\documentclass{article}

\usepackage{arxiv}

\usepackage[utf8]{inputenc} 
\usepackage[T1]{fontenc}    
\usepackage{hyperref}       
\usepackage{url}            
\usepackage{booktabs}       
\usepackage{amsfonts}       
\usepackage{nicefrac}       
\usepackage{microtype}      
\usepackage{lipsum} 

\usepackage{graphicx}
\usepackage{amsmath,amsthm,bm,mathrsfs, amssymb, amscd, amsthm, amsfonts}
\usepackage{caption}
\usepackage{subcaption}

\usepackage[nottoc]{tocbibind}
\usepackage{hyperref}
\usepackage[british]{babel}
\usepackage{afterpage}
\usepackage{caption}
\usepackage[square, comma, sort&compress]{natbib}
\usepackage[ruled,vlined, linesnumbered]{algorithm2e}
\usepackage{algorithmic}

\usepackage{booktabs}
\usepackage{multirow}

\usepackage{algorithmic}

\title{Fair Insurance Premium Level in Connected SIR Model under Epidemic Outbreak}

\author{ {\sc Aleksandr A. Shemendyuk, Alexey A. Chernov, Mark Y Kelbert}\\[2pt]
National Research University Higher School of Economics, \\
 Myasnitskaya 20, Moscow, Russia.\\[6pt]}

\newcommand{\mrm}{\mathrm}
\newcommand{\mbb}{\mathbb}

\begin{document}
\maketitle

\begin{abstract}
{In this paper we aim to study an optimal insurance premium level for health-care in a deterministic and stochastic SIR models with migration fluxes and vaccination of population. The studied model considers two standard SIR centres connected via links and continuous migration fluxes. The premium is calculated using the basic equivalence principle. Even in this simple setup there are non-intuitive results that illustrate how the premium depends on migration rates, severeness of a disease and initial distribution of healthy and infected individuals through the centres. We investigate how the vaccination program effects the insurance costs by comparing the savings in benefits with the expenses for vaccination. We compare the results of deterministic and stochastic models.}

\end{abstract}




\clearpage
\setcounter{page}{2}

    

\section{Introduction}
Epidemics cause severe damage to social welfare and can result in a massive loss of working days. Many models are developed to estimate the dynamics of epidemics, e.g. ~(see pioneer works \cite{bernoulli_1760, ross_1910, kermack_1927, macdonald_1957} and later works \cite{HETHCOTE_1978, GLEISSNER_1988, bolker_1995, bailey_2019, burton_etall_2012, lee_2012}). One of the most fundamental works with strict mathematical and medical approaches is \cite{daley_gani_1999}. Investigation of these models can help understanding the key points of the phenomena and to determine the optimal vaccination strategy, which could stop the spread of the pandemic and reduce the economical costs.

Urgent decisions are required when the epidemic starts to spread. The public health-care should provide the medicine for citizens to insure they get immunity. The optimal vaccine allocation in connected SIR centres was discussed in~\cite{Chernov_2019}. The vaccine might not always be free of charge. On the contrary, for an average person it could be too expensive to afford. If a person does not agree to purchase the vaccination, he is forced to buy a mandatory health insurance policy that covers treatment expenses and (possibly) pays a lump sum when the policyholder becomes healthy again after disease. An insurance company collects premiums from susceptible group and covers the health-care benefits to infected policyholders. The optimal premium level is then calculated according to the classical equivalence principle
\begin{equation}
\label{eq:equiv_principle}
    \mathbb{E}[\text{benefit outgo}] = \mathbb{E}[\text{premium income}].
\end{equation}

In this paper we consider deterministic and stochastic SIR models with two connected centres and constant migration fluxes. The model considers three population groups: Susceptible to a disease $S_t$, Infected $I_t$ and Removed $R_t$ persons. The last group consist of subjects who either got an immunity and will not become ill again. The centres could represent cities, countries, health-care facilities, etc. We calculate health-care premium level for different parameters of model and investigate insurance costs.

Epidemic model have been proposed in different actuarial methods in \cite{feng_garrido_2011}, ~\cite{lefevre_picard_simon_2017}. \cite{feng_garrido_2011} consider constant infection and recovery rates in isolated SIR model without migration. Authors present calculations of annuity premiums and annuity benefits, lump-sum benefits and death benefits. They also propose power series solutions for evaluating the dynamic of the model. Then the premium adjustments are introduced and the method is applied to real data sets: great plague in Eyam in 1665, SARS epidemic in Honk Kong in 2003. In~\cite{lefevre_picard_simon_2017} authors consider time-continuous Markov model of epidemic spread with one centre and different epidemic scenarios: general and fatal epidemics, case with exponentially-dependent rates, and a single-time change of infection rate model.

The aim of this work is to investigate the optimal health-care premium level $\pi$ in deterministic and stochastic connected SIR centres. We consider different scenarios (different centres' characteristics), investigate the dependence of premium $\pi$ from the amount of vaccine available, infection rates, migration intensities. Further, we compare the result of different scenarios for stochastic and deterministic models.

The paper is organised as follows. In Section~2 we consider deterministic SIR model and premium calculation in this case. In Section~3 we study stochastic SIR centres and describe simulation methods by Markov chain. Section~4 is dedicated numerical simulations of deterministic and stochastic SIR models for different initial parameters and comparing results. Conclusion and an outlook for a further work is given in Section 5.

\section{Deterministic model}
\subsection{Connected SIR model}
\label{sect:sir-model-det}
The classical SIR model for $n$ centres with migration fluxes is governed by the following ODEs:
\begin{equation}
\label{eq:SIR_model_det}
    \begin{aligned}
        \frac{\mrm{d} S_i}{\mrm{d} t} &= -\beta(R_i) \, S_i I_i - \sum_{j \neq i} k_{i j} \, S_i + \sum_{j \neq i} k_{j i} \, S_j \\
        \frac{\mrm{d} I_i}{\mrm{d} t} &= \beta(R_i) \, S_i I_i - \mu(R_i) \, I_i - \sum_{j \neq i} l_{i j} \, I_i + \sum_{j \neq i} l_{j i} \, I_j \\
        \frac{\mrm{d} R_i}{\mrm{d} t} &= \mu(R_i) \, I_i \\
        S_i(0) &= S_{i, 0}, \, I_i(0) = I_{i, 0}, \, R_i(0) = R_{i, 0}
    \end{aligned}
\end{equation}
where $i = 1, 2, \ldots, n$, infection rate $\beta(R_i)$ and recovery (fatality) rate $\mu(R_i)$ are dependent on the number of removed people in $i$-th centre; $k_{i j}$ and $l_{i j}$ are the migration rates of susceptible and infected groups, respectively, from $i$-th to $j$-th centre; $S_{i, 0}$, $I_{i, 0}$ and $R_{i, 0}$ are initial numbers of susceptible, infected and removed persons. The total population in centre $i$ is then $N_i(t) = S_i(t) + I_i(t) + R_i(t)$. Total number of population is $N(t) = \sum_{i=1}^n N_i(t)$. The analysis of SIR model with migration is given in~\cite{sazonov_kelbert_migration_2015}.

Infection and recovery rates depending on the number of removed persons allow us to represent different epidemic scenarios by choosing corresponding response to the dynamic of the epidemic. For instance, in case of fast spreading disease (i.e. flue, influenza, measles) intensity of infection reduces as the number of immune persons increases, yielding to decreasing $\beta(R_i)$. On the other hand, in case of sexually transmitted diseases, as the efficiency of treatment methods improves, people are tending to have riskier sexual connections, leading to increasing function $\beta(R_i)$. Moreover, treatment of patients implies examination of the disease, following increasing function $\mu(R_i)$. On contrary, if a disease is more persistent as the number of infected people increases (i.e. number of removed goes down), then the function $\mu(R_i)$ is decreasing.

Usually, when an infected individual is transferred into removed class, he is considered to get an immunity and total number of the population remains unchanged. Hence, the natural assumption is $\beta(R_i) \equiv \beta_i = \alpha / N_i$. In case of fatal epidemic (see~\cite{GLEISSNER_1988}) the removed group consist of dead people, and the natural assumption is to let $\beta(R_i) \equiv \alpha / (N_i - R_i)$. So, the dependence of the rates from number of removals is very diverse an can describe different scenarios on practice. In this case, the basic reproduction number is defined as $R_0 = \alpha / \mu$.

Another key object of the model is the time moment $T$ when the epidemic stops. It is widely known that nontrivial solutions of \eqref{eq:SIR_model_det} have exponential character. Therefore, we can never obtain pure zero number of infectives by integrating the system. It is natural to say that if there are less infectives then a certain value, then the epidemic is assumed to be suppressed. We therefore set the threshold percentage $\theta$ of living population at time $T$. In case of general epidemic define $T$ as

\begin{equation} \label{eq:T_def_general}
    T = \inf\left\{t: \sum_{i = 1,\ldots, n} I_i(t) < \theta \cdot N(t)\right\},
\end{equation}
and in case of fatal epidemic as follows:
\begin{equation} \label{eq:T_def_fatal}
    T = \inf\left\{t: \sum_{i = 1,\ldots, n} I_i(t) < \theta \cdot \left(N(t) - \sum_{i=1}^n R_i(t)\right)\right\}.
\end{equation}

\subsection{Infectivity and susceptibility times}
One of the key functionals is
\begin{equation*}
    \label{eq:A_i(T)}
    A_i(T) = \int_0^T I_i(t) \, \mrm{d}t, \quad i = 1, \ldots, n,
\end{equation*}
which describes the total number of lost working days during the epidemic in $i$-th centre. This functional also allows us to measure the medical expenses for treating infected persons.

Since we are interested in total number of lost working days in the network, we sum over $i$ and define
\begin{equation}
    \label{eq:A_T}
    A_T = \sum\limits_{i=1}^nA_i(T) =\sum\limits_{i=1}^n \int_0^T I_i(t) \, \mrm{d}t.
\end{equation}

During the epidemic, an insurance company collects premiums from healthy individuals with some rate (will be introduced in Sect.~\ref{subsec:premium}). The total exposure to premiums of
all susceptible persons at $i$-th centre equals
\begin{equation*}
    \label{eq:B_i(T)}
    B_i(T) = \int_0^T S_i(t) \, \mrm{d}t, \quad i = 1, \ldots, n .
\end{equation*}

By summing over $i$, we define
\begin{equation}
    \label{eq:B_T}
    B_T = \sum\limits_{i=1}^nB_i(T) =\sum_{i=1}^n \int_0^T S_i(t) \, \mrm{d}t.
\end{equation}

\subsection{Premium calculation}
\label{subsec:premium}
A premium is calculated according to the standard equivalence principle \eqref{eq:equiv_principle}. On the liability side, the benefits are payed by insurance company to infected persons with constant intensity $c_1$. For a person transferred to the removed group (i.e. got the immunity or has died due to a fatal disease) the company pays a lump sum $c_2$ to every beneficiary. In line with \cite{lefevre_picard_simon_2017}, we assume that all claims are settled at time $T$ when the epidemic is extinct. 

We assume that the insurance company has a permission to resell the vaccine to the population. The amount of vaccine $\mathcal{V}$ could be bought from other centre (i.e. city, country), ordered to be developed / manufactured, etc. In any case, there is a liability part and we let $c_3$ be a cost of one vaccine unit. Therefore, the expected liability is
\begin{equation}
\label{eq:premia_outgo}
    \mathbb{E}[\text{benefit outgo}] = c_1 A_T + c_2 \sum_{i=1}^n R_i(T) + c_3 \mathcal{V}.
\end{equation}

On the other hand, the insurance company collects constant premiums $\pi$ from susceptible population until they become ill or the epidemic has ended. Also, we let $\mathcal{V}_{\text{sold}}$ be a number of sold vaccine. So, the expected income of the company is
\begin{equation}
\label{eq:premia_income}
    \mathbb{E}[\text{income}] = \pi B_T + c_4 \mathcal{V}_{\text{sold}}.
\end{equation}

By equivalence principle \eqref{eq:equiv_principle} and expressions \eqref{eq:premia_outgo} -- \eqref{eq:premia_income} we get
\begin{equation}
    \label{eq:premium}
    \pi = \frac{1}{B_T} \left( c_1 A_T + c_2 \sum_{i=1}^n R_i(T) + c_3 \mathcal{V}  - c_4 \mathcal{V}_{\text{sold}}\right).
\end{equation}

We also consider a constant time-continuous discounting factor $\delta$ to calculate the present value of benefits and incomes. Therefore, we complete formulae \eqref{eq:A_T} and \eqref{eq:B_T} as follows:
\begin{equation}
\label{eq:A_T_B_T_disc}
    \begin{aligned}
        A_{\delta, T} &= \sum_{i=1}^n \int_0^T \exp(-\delta t)\, I_i(t) \, \mrm{d}t \\
        B_{\delta, T} &= \sum_{i=1}^n \int_0^T \exp(-\delta t)\, S_i(t) \, \mrm{d}t
    \end{aligned}
\end{equation}

In the end, we also enhance the premium calculation as
\begin{equation}
    \label{eq:premium_discount}
    \pi_\delta = \frac{1}{B_{\delta, T}} \left( c_1 A_{\delta, T} + c_2 \exp(-\delta T) \sum_{i=1}^n R_i(T) + c_3 \mathcal{V} - c_4 \mathcal{V}_{\text{sold}} \right).
\end{equation}

\subsection{Optimal vaccine allocation}
\label{subsec:opt_vaccine}

Once the epidemic starts and the company possesses vaccine amount $\mathcal{V}$, it is important to wisely allocate the vaccine in the centres. We assume that the vaccination is instantaneous and people prefer to vaccinate rather than risk their health. Let $w_i$ be the shares of vaccine stock $\mathcal{V}$ allocated in $i$-th centre, $i = 1,\ldots,n$. In~\cite{Chernov_2019} it is shown that the best vaccination time is $t = 0$, since any delay in vaccination let the infection to spread. Therefore, the SIR model~\eqref{eq:SIR_model_det} with vaccine allocation $(w_1, \ldots, w_n): w_i \geq 0, \, w_1 + \ldots + w_n = 1$ takes the form
\begin{equation}
    \label{eq:SIR_model_det_vaccine}
    \begin{aligned}
        &\frac{\mrm{d} S_i}{\mrm{d} t} = -\beta(R_i) \, S_i I_i - \sum_{j \neq i} k_{i, j} \, S_i + \sum_{j \neq i} k_{j, i} \, S_j - \min\{S_{i, 0}, w_i \mathcal{V}\} \delta(t), \\
        &\frac{\mrm{d} I_i}{\mrm{d} t} = \beta(R_i) \, S_i I_i - \mu(R_i) \, I_i - \sum_{j \neq i} l_{i, j} \, I_i + \sum_{j \neq i} l_{j, i} \, I_j, \\
        &\frac{\mrm{d} R_i}{\mrm{d} t} = \mu(R_i) \, I_i, \\
        &S_i(0) = S_{i, 0}, \, I_i(0) = I_{i, 0}, \, R_i(0) = R_{i, 0},
    \end{aligned}
\end{equation}
where $\delta(t)$ is Dirac delta function.

One natural way to optimally allocate the vaccine is to minimise the resulting premium \eqref{eq:premium} or \eqref{eq:premium_discount}. Since the health-care is assumed to be mandatory, we would like to reduce the premium payed by susceptibles as much as possible. More formally, for given vaccine stock and model parameters we get
\begin{equation}
    \label{eq:pi_minimisation}
    \pi^* = \min_{\substack{(w_1,\ldots,w_n): w_i \geq 0, \\ w_1 + \ldots + w_n = 1}} \frac{1}{B_T} \left( c_1 A_T + c_2 \sum_{i=1}^n R_i(T) + c_3 \mathcal{V}  - c_4 \mathcal{V}_{\text{sold}}\right).
\end{equation}

Allocation $(w_1, \ldots, w_n)$ affects the initial number of susceptible persons, which affect $A_T$, $B_T$ and $R_i(T)$. Here we allow cases when the vaccine amount is not fully utilised. It is clearly not optimal to ``waste'' vaccine, but those cases are negligible. Therefore, the vaccine amount $\mathcal{V}$ and the vaccine sold $\mathcal{V}_{\text{sold}}$ are not always the same.

Alternatively, we could set as a goal to keep low epidemic costs, i.e. have as less lost working days as possible. Therefore, optimal vaccine allocation could be such that the functional $A_T$ is minimised:
\begin{equation}
    \label{eq:A_minimisation}
    A_T^* = \min_{\substack{(w_1,\ldots,w_n): w_i \geq 0, \\ w_1 + \ldots + w_n = 1}} \sum_{i=1}^n \int_0^T I_i(t) \, \mrm{d}t.
\end{equation}

It is interesting to see the relation between two ``optimality" points of view. Both have sensible reasons of treating epidemic -- first is financially oriented, whilst the second prevents population from being ill.
\section{Stochastic model}

As in previous chapters, we denote number of susceptible persons at time $t$ as $S(t)$, infected persons -- $I(t)$ and recovered / removed persons -- $R(t)$. Here the main interest is to model the infection spread among population assuming small number of initially infected persons.

If number of infectives is small, the infection can get naturally suppressed, not causing any epidemic outbreak. A stochastic model is generally devoted for such cases, whereas deterministic differential models describe evolution of epidemic in high population with large number of infected individuals. The algorithm of merging two approaches is proposed in \cite{SAZONOV2011_twostage}.

All stochastic models are considered to have discrete state space in continuous time. Also, models have Markov property, i.e. probability of transition from one state to other does not depend on history. These assumptions are natural because in real epidemics the number of people getting infection only depends on the number of contacts with infected persons. We do not study cases when a disease can infect people from other sources (water, air breeze, unsanitary conditions, intentional release of a disease, etc.). Similarly, recovery / death probability only depends on the host of the infection.

\subsection{Markov chain for one centre}

In case of one centre model the transition probabilities of standard stochastic model with constant coefficients $\beta$ and $\mu$ are presented in Table~\ref{tab:MC_1}.

We denote the time of $j$-th jump (switch of state) as $t_j$. Corresponding waiting time (i.e. time between $j-1$-th and $j$-th jumps) as $s_j$.

\begin{table}[ht]
    \centering
    \caption{Transition probabilities of time continuous Markov chain in case of one centre.}
    \begin{tabular}{lll}
        \toprule
        Event & Rate & Condition \\
        \midrule
        $S \to S-1$, $I \to I + 1$ & $\beta S I$ & $S > 0$ \\
        $I \to I - 1$, $R \to R + 1$ & $\mu I$ & $I > 0$ \\
        Absorbing state & $0$ & $I = 0$ \\
        \bottomrule
    \end{tabular}
    \label{tab:MC_1}
\end{table}

The final time of epidemic is a random variable that depends on the chain evolution. We define the end of epidemic as
\begin{equation}
    \label{def:T_1_stoch}
    T = \inf\{t: \; I(t) = 0\}.
\end{equation}

Simulation algorithm is presented in Algorithm~\ref{algo:MC_1}. The general approach is to calculate all transition rates at time $t_j$ and simulate the waiting time $s_j$ (line 4) as exponential random variable with inverse mean equal to sum of all rates. Then generate a standard uniform random variable $u_j$ (line 6) to decide which event took place. Calculate the probability of an event as its rate divided by sum of all rates (line 7). Finally, choose a corresponding event (lines 8--12).

\begin{algorithm}[H]
\label{algo:MC_1}
    \SetAlgoLined
    \KwResult{$T, S(T), I(T), R(T)$}
    set $S = S_0$, $I = I_0$, $R = 0$, $t_0 = 0$, $j = 1$\;
    calculate $N = S + I + R$\;
    \While{$I > 0$}{
        generate $s_j \sim \mrm{Exp}(\beta S I + \mu I)$\;
        set $t_j = t_{j-1} + s_j$\;
        generate $u_j \sim \mrm{Unif}(0, 1)$\;
        set probabilities $p_1 = \frac{\beta S I}{\beta S I + \mu I}$ and $p_2 = \frac{\mu I}{\beta S I + \mu I}$\;
        \eIf{$u_j < p_1$}{
            set $S = S - 1$, $I = I + 1$\;
        }{
            set $I = I - 1$, $R = R + 1$\;
        }
        $j \gets j + 1$
    }
    set $T = t_j$
    \caption{Markov chain simulation for one centre}
\end{algorithm}

\subsection{Markov chain for multiple centres}
\label{sect:mc_n}

Here we consider a stochastic model for epidemic in multiple centres. Denote a number of susceptibles in $i$-th centre at time $t$ as $S_i(t)$, infectives -- $I_i(t)$ and recovered / removed -- $R_i(t)$ for $i = 1, \ldots, n$. We let coefficients $\beta(R_i)$ and $\mu(R_i)$ now be dependent on the number of removals (see Sect.~\ref{sect:sir-model-det}). Then the transition rates for $i, j = 1 \ldots n$ are presented in Table~\ref{tab:MC_n}. The end of epidemic is defined as
\begin{equation*}
    \label{def:T_n_stoch}
    T = \inf \left\{t: \; \sum_{i=1}^n I_i(t) = 0\right\}.
\end{equation*}

\begin{table}[ht]
    \centering
    \caption{Transition probabilities of time continuous Markov chain in case of multiple centres}
    \begin{tabular}{lll}
        \toprule
        Event & Rate & Condition \\
        \midrule
        $S_i \to S_i - 1$, $I_i \to I_i + 1$ & $\beta(R_i) S_i I_i$ & $S_i > 0$ \\
        $I_i \to I_i - 1$, $R_i \to R_i + 1$ & $\mu(R_i) I_i$ & $I_i > 0$ \\
        $S_i \to S_i - 1$, $S_j \to S_j + 1$ & $k_{i j} S_i$ & $S_i > 0$, $i \neq j$ \\
        $I_i \to I_i - 1$, $I_j \to I_j + 1$ & $k_{i j} I_i$ & $I_i > 0$, $i \neq j$ \\
        Absorbing state & $0$ & $\sum_{i=1}^n I_i = 0$ \\
        \bottomrule
    \end{tabular}
    \label{tab:MC_n}
\end{table}

The simulation algorithm of one Markov chain for $n$ centres is similar to Algorithm~\ref{algo:MC_1}. However, here the dimension of one state is $3n - 1$, where we have three numbers for susceptible-infected-recovered triplet and $n$ centres with constant population throughout the network. Latter reduces the dimension by one.

\subsection{Vaccination in the random setup}

Introduction of additional centres makes definition of optimal vaccine allocation more complicated. We calculate the health-care premium according to the equivalence principle~\eqref{eq:equiv_principle}. However, unlike~\eqref{eq:premium}, the formula for premium in stochastic model has the form
\begin{equation}
    \label{eq:premium_stoch}
    \pi = \frac{1}{\mbb{E}[B_T]} \left( c_1 \mbb{E}[A_T] + c_2 \sum_{i=1}^n \mbb{E}[R_i(T)] + c_3 V  - c_4 V_{\text{sold}}\right),
\end{equation}
and discounted premium is
\begin{equation}
    \label{eq:premium_discount_stoch}
    \pi_\delta = \frac{1}{\mbb{E}[B_{\delta, T}]} \left( c_1 \mbb{E}[A_{\delta, T}] + c_2 \sum_{i=1}^n \mbb{E}[\exp(-\delta T) R_i(T)] + c_3 V - c_4 V_{\text{sold}} \right).
\end{equation}

Unlike deterministic model~\eqref{eq:SIR_model_det_vaccine} with vaccine stock $\mathcal{V}$, in stochastic setup we need to clarify what is optimal vaccine allocation with integers $(w_1, \ldots, w_n): w_i \geq 0, \sum_{i=1}^n w_i = V$. In stochastic model we let $V \equiv \mathcal{V}$ Therefore, we apply the following procedure:

\begin{algorithm}
\caption{Calculation of optimal vaccine allocation for stochastic model for $n$ centres}
\label{alg:optalloc}
    \SetAlgoLined
    \SetKwInOut{Input}{input}
    \Input{amount of vaccine $V$,\\
        minimisation option $Q \in \{\pi, \pi_\delta, A_T, A_{\delta, T} \}$, \\
        number of simulations $N_{\mrm{sim}}$}
    \KwResult{$w^* = (w_1, w_2, \ldots, w_n)$}
    
    \For{all possible $w = (w_1, w_2, \ldots, w_n): w_i \geq 0, \sum_{i=1}^n w_i = V$} {
        simulate $N_{\mrm{sim}}$ Markov Chains, computing $Q_k(w)$, $k = 1,\ldots N_{\mrm{sim}}$\;
        calculate average $Q = \sum_{k=1}^{N_{\mrm{sim}}} Q_k(w)$\;
    }
    choose $w^* = (w_1, w_2, \ldots, w_n)$ such that $Q(w)$ is minimal\;
    i.e. $w^* = \mrm{arg}\min_w Q(w)$ \;
\end{algorithm}

The idea of Algorithm~\ref{alg:optalloc} is to consider all possible vaccine allocations and simulate $N_{\mrm{sim}}$ Markov chains for each allocation (i.e. perform Monte Carlo method). Then we choose such allocation $w^* = (w_1, w_2, \ldots, w_n)$ that minimises the desired functional (resulting health-care premium $\pi$, lost working days $A_T$, or corresponding discounted values).

After computing optimal vaccine allocation $w^*$, we calculate corresponding optimal premium $\pi^*, \pi^*_{\delta}, \pi(A^*_T), \pi(A_{\delta, T})$, i.e. use formulae~\eqref{eq:premium_stoch} -- \eqref{eq:premium_discount_stoch}.
\section{Numerical experiments}

Let us consider models \eqref{eq:SIR_model_det} and \eqref{sect:mc_n} with two centres. Fix the removal rate $\mu(R_i) \equiv \mu = 1$, and consider functions $\beta(R_i)$ for different scenarios (general and fatal epidemic), and vary basic reproduction number to investigate the optimal premiums. Let initial number of removed persons be $R_{i, 0} = 0$ for $i = 1, \, 2$.

In line with \cite{lefevre_picard_simon_2017}, let the constant payment intensity $c_1 = 1$ and a lump sum $c_2 = 2$. We let the cost of one vaccine unit $\mathcal{V}$ be $c_3 = 4$. The selling price of one vaccine unit $c_4$ is usually assumed to be higher than the unit cost.

In all figures (in deterministic section) horizontal axis corresponds to the vaccine amount required to vaccinate $V$ percents of susceptible individuals, i.e. $\mathcal{V} = V S$. For example, if $S = S_1 + S_2 = 100$, then $V = 0.2$ means that the company can vaccinate $20\%$ of susceptibles, i.e. $\mathcal{V} = 20$ individuals. We only consider $V \in [0, \, 0.5]$ since it becomes extremely hard to vaccinate more than a half of population. Moreover, since people have a choice to buy the vaccine or not, it becomes less likely that the company can sell entire vaccine stock.

\textit{General epidemic.}
In general epidemic setup we assume that after being infected a person gets immunity and will not get ill for the rest of epidemic period. Therefore, the number of total population remains constant and a natural assumption is to let $\beta(R_i) = \alpha / N_i$.

\textit{Fatal epidemic.}
Here we assume that the disease is fatal and removed class corresponds to the number of dead people. In this case the population is decreasing over time and a natural assumption is to let $\beta(R_i) = \alpha / (N_i - R_i)$, see \cite{GLEISSNER_1988}.

In both epidemic options the basic reproduction number becomes $R_0 \equiv \frac{\alpha}{\mu}$. We investigate some cases of $R_0 \in [2, 12]$ that correspond to reproduction numbers of different diseases. Further, we investigate the optimal premium level with and without discounting factor $\delta = \ln 1.01$.

\subsection{Deterministic model}
\label{sect:numerical_deterministic}

Assume that the end of epidemic $T$ is the moment when the total number of infectives is smaller than $\theta = 0.5\%$ of living population at time $T$, i.e. use formulae \eqref{eq:T_def_general} -- \eqref{eq:T_def_fatal}.

\subsubsection{Basic deterministic scenario} \label{sub:basic_scenario2}
Here we consider a basic scenario, in which we have similar migration fluxes and both centres are equivalent. We let the initial populations be $S_{i, 0} = 100$, $I_{i, 0} = 10$, and let migration flows be $k_{1, 2} = k_{2, 1} = 0.5$ and $l_{1, 2} = l_{1, 2} = 0.1$.

\begin{figure}[ht]
    \centering

    \begin{subfigure}[t]{.45\linewidth}
        \centering
        \includegraphics[width = \linewidth]{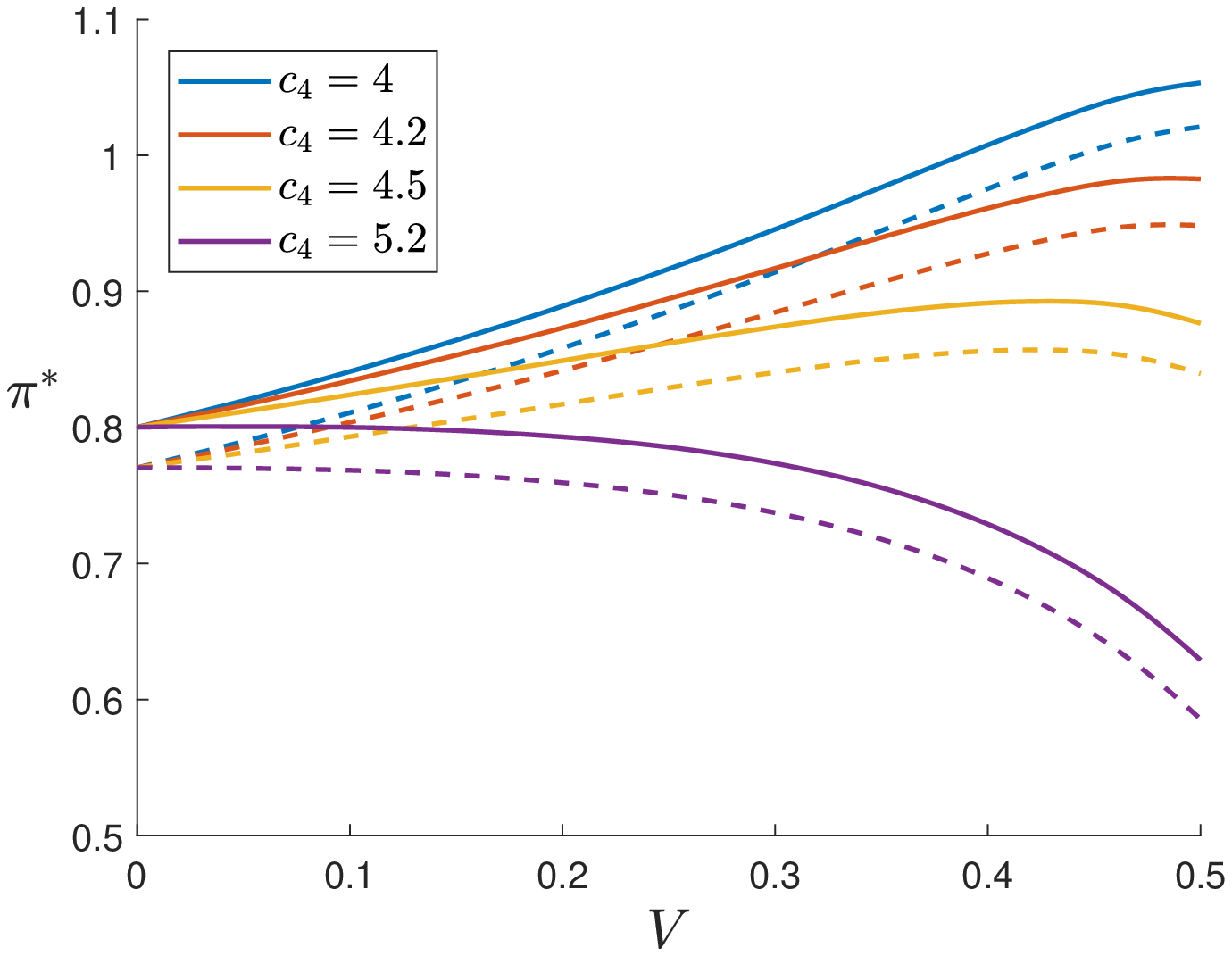}
        \caption{General epidemic, premium minimisation}
        \label{fig:1a}
    \end{subfigure}
    \begin{subfigure}[t]{.45\linewidth}
        \centering
        \includegraphics[width = \linewidth]{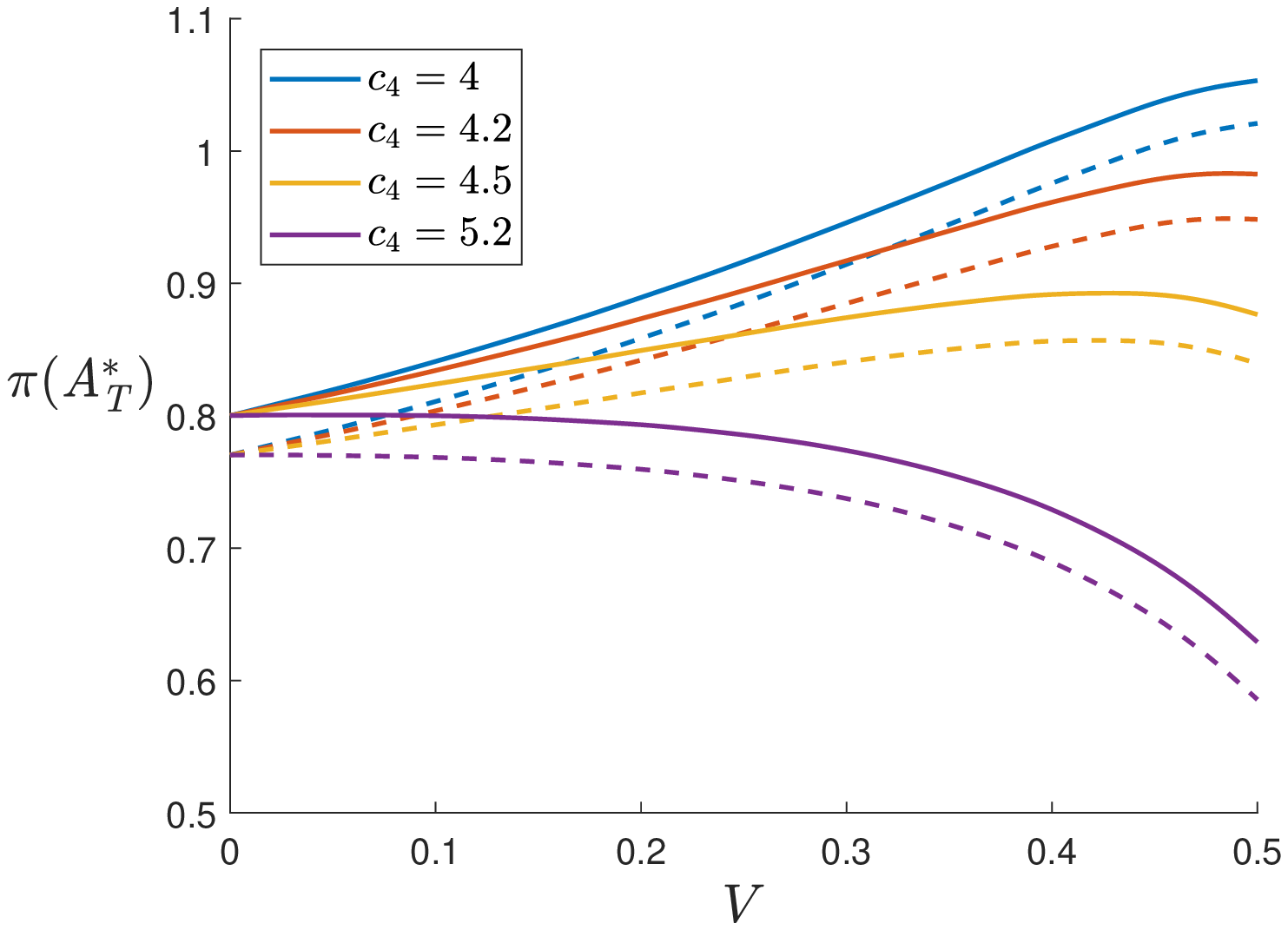}
        \caption{General epidemic, $A_T$ minimisation}
        \label{fig:1b}
    \end{subfigure}
    \begin{subfigure}[t]{.45\linewidth}
        \centering
        \includegraphics[width = \linewidth]{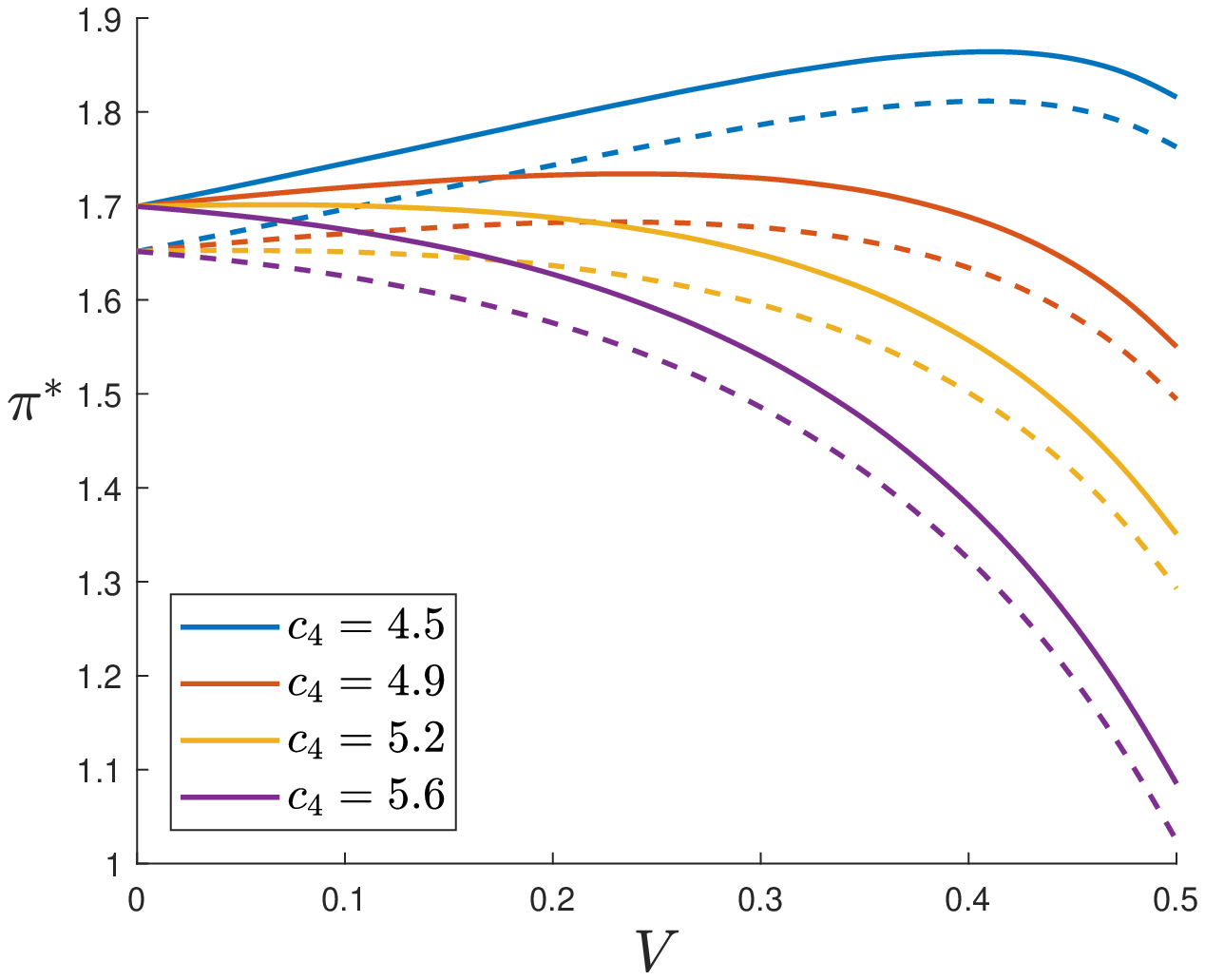}
        \caption{Fatal epidemic, premium minimisation}
        \label{fig:1c}
    \end{subfigure}
    \begin{subfigure}[t]{.45\linewidth}
        \centering
        \includegraphics[width = \linewidth]{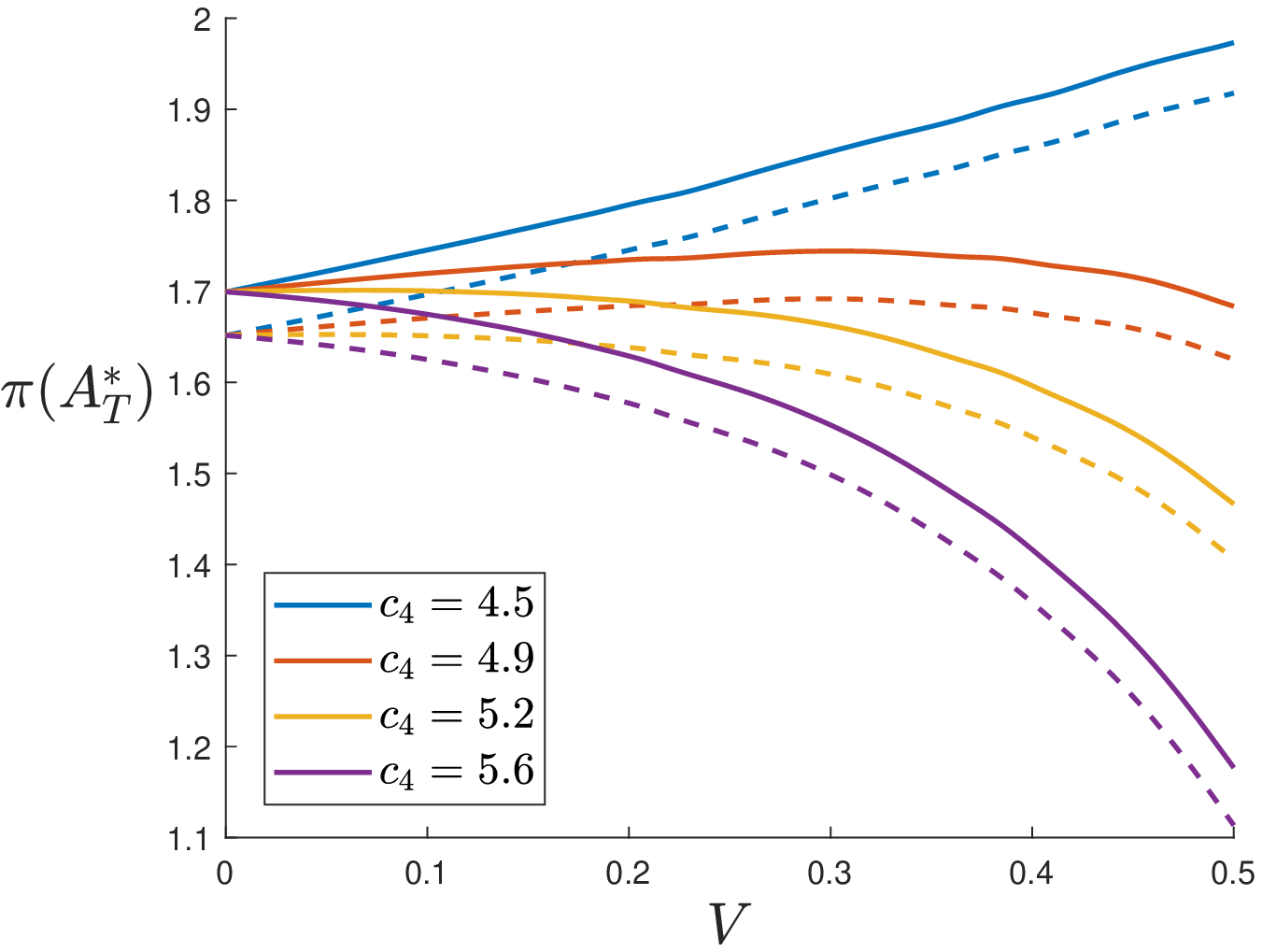}
        \caption{Fatal epidemic, $A_T$ minimisation}
        \label{fig:1d}
    \end{subfigure}
    
    \caption{Premium level for general and fatal epidemics in basic scenario for Ebola, influenza. Dashed line -- corresponding premium with discounting factor $\delta = \ln 1.01$. Basic reproduction number $R_0 = 2$.}
    \label{fig:1}
\end{figure}

This scenario represents the simplest case when both centres are the same. The interest here is to investigate the vaccination options and detect some noticeable dependencies.

\begin{figure}[ht]
    \centering

    \begin{subfigure}[t]{.45\linewidth}
        \centering
        \includegraphics[width = \linewidth]{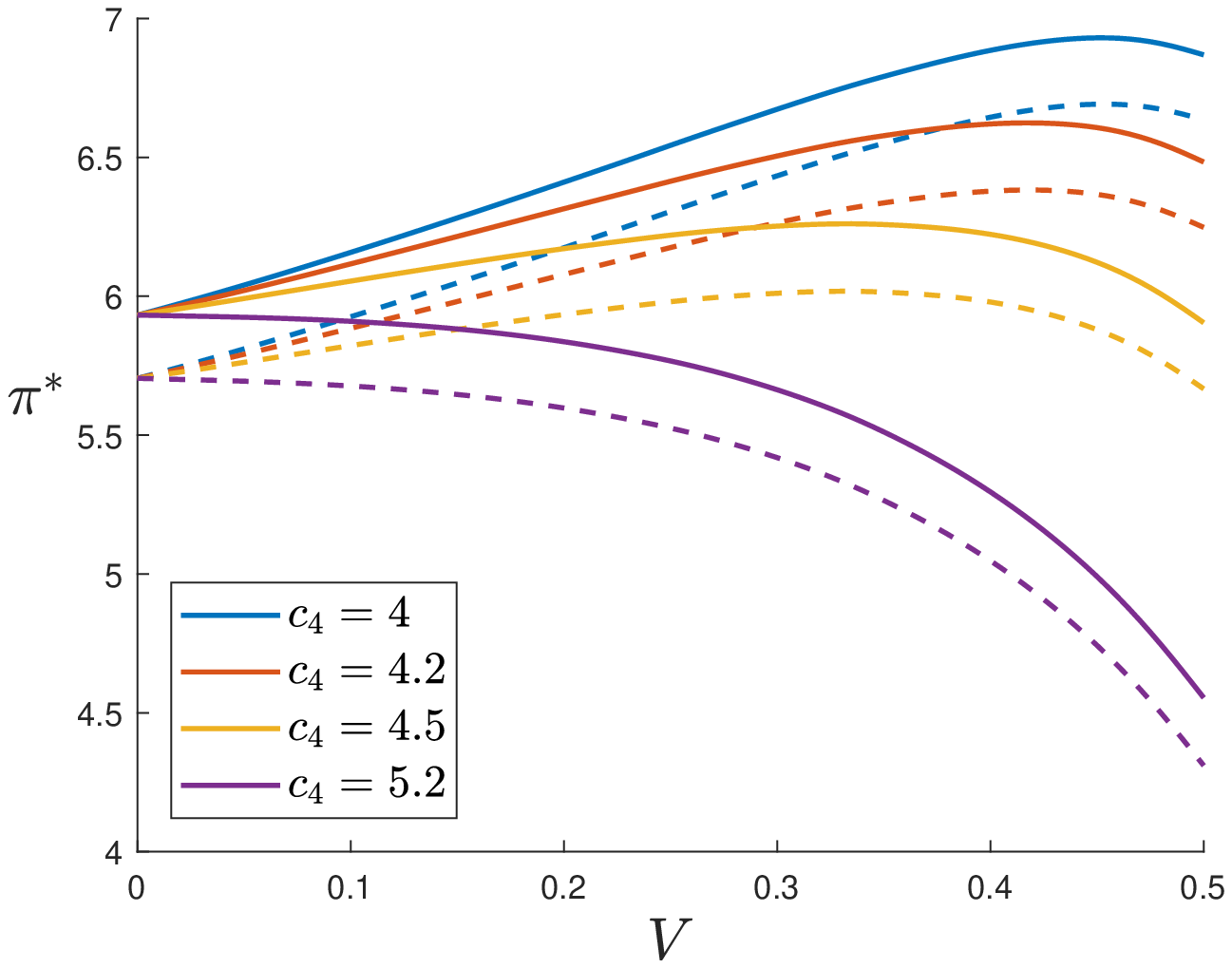}
        \caption{General epidemic, premium minimisation}
        \label{fig:2a}
    \end{subfigure}
    \begin{subfigure}[t]{.45\linewidth}
        \centering
        \includegraphics[width = \linewidth]{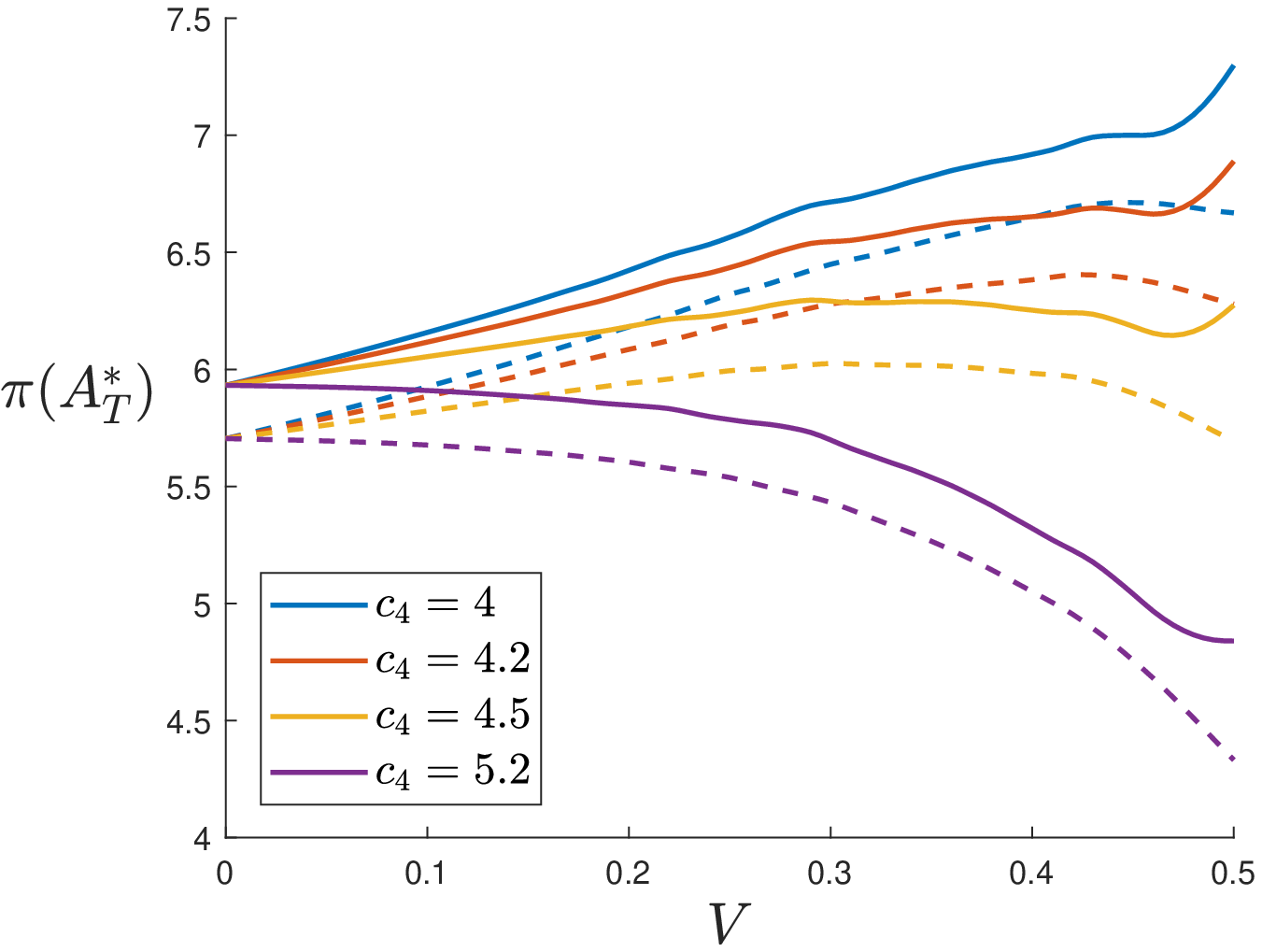}
        \caption{General epidemic, $A_T$ minimisation}
        \label{fig:2b}
    \end{subfigure}
    \begin{subfigure}[t]{.45\linewidth}
        \centering
        \includegraphics[width = \linewidth]{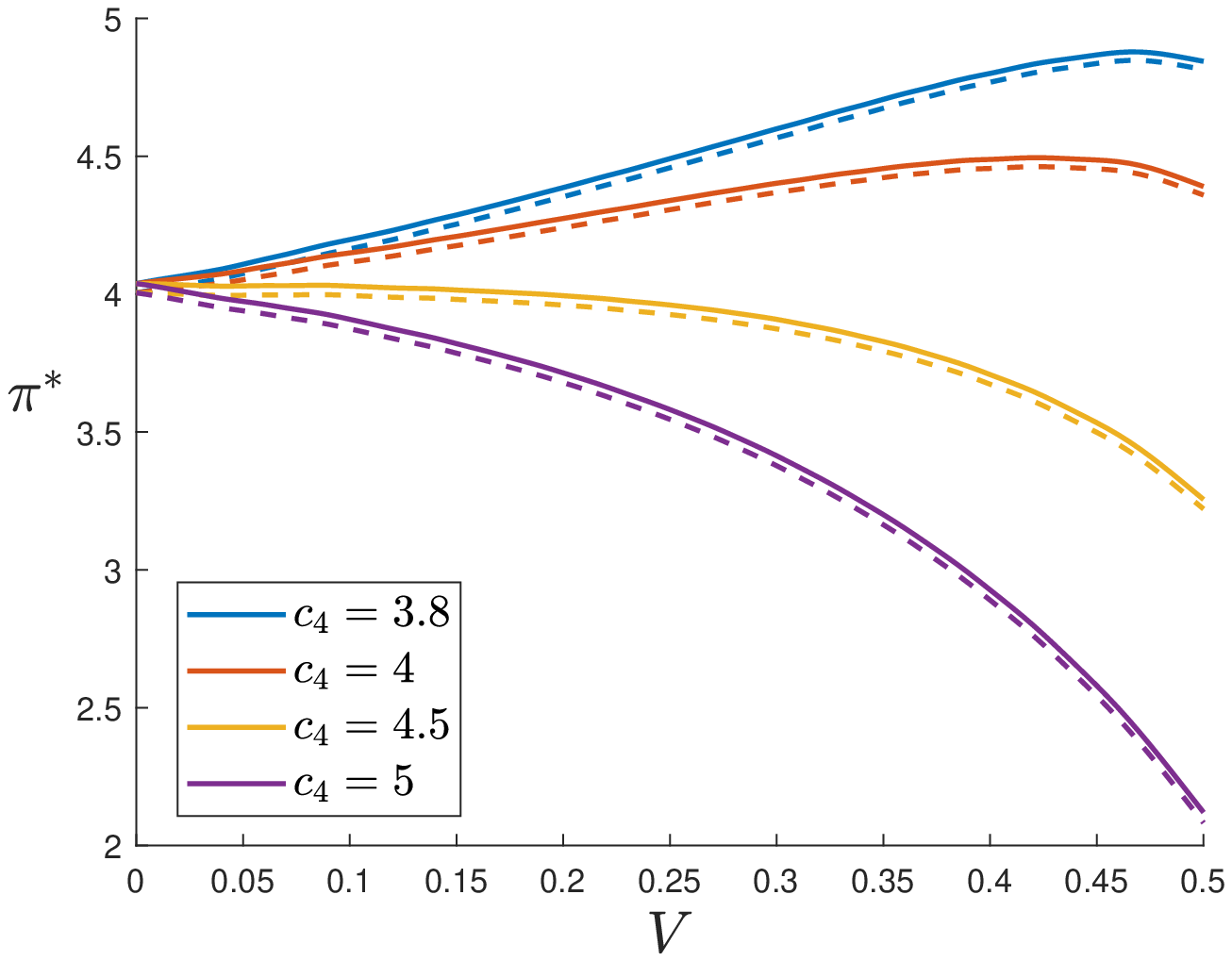}
        \caption{Fatal epidemic, premium minimisation}
        \label{fig:2c}
    \end{subfigure}
    \begin{subfigure}[t]{.45\linewidth}
        \centering
        \includegraphics[width = \linewidth]{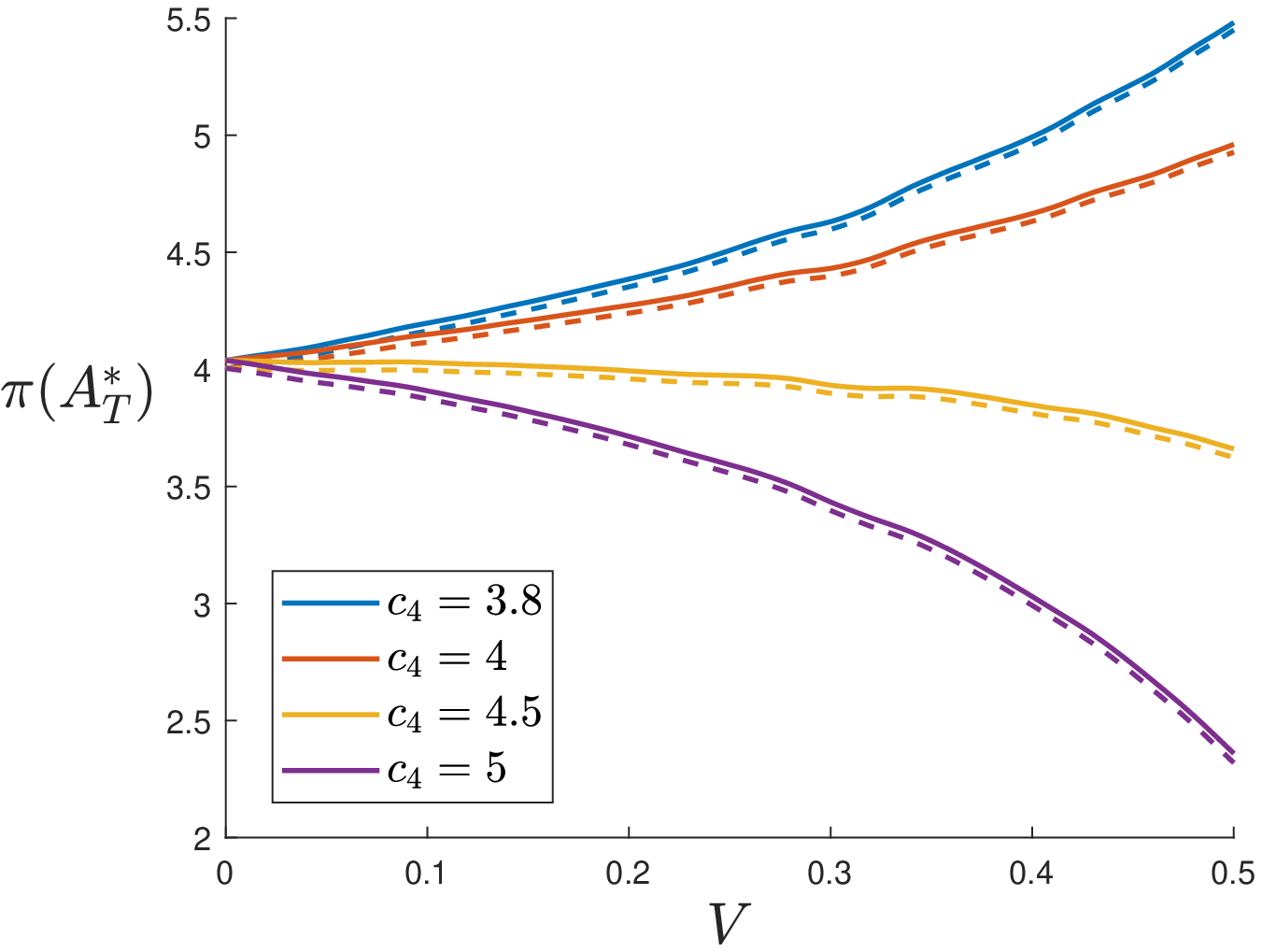}
        \caption{Fatal epidemic, $A_T$ minimisation}
        \label{fig:2d}
    \end{subfigure}
    
    \caption{Premium level for general and fatal epidemics in basic scenario for diphtheria, mumps, polio, smallpox. Dashed line -- corresponding premium with discounting factor $\delta = \ln 1.01$. Basic reproduction number $R_0 = 6$.}
    \label{fig:2}
\end{figure}

\begin{figure}[ht]
    \centering

    \begin{subfigure}[t]{.45\linewidth}
        \centering
        \includegraphics[width = \linewidth]{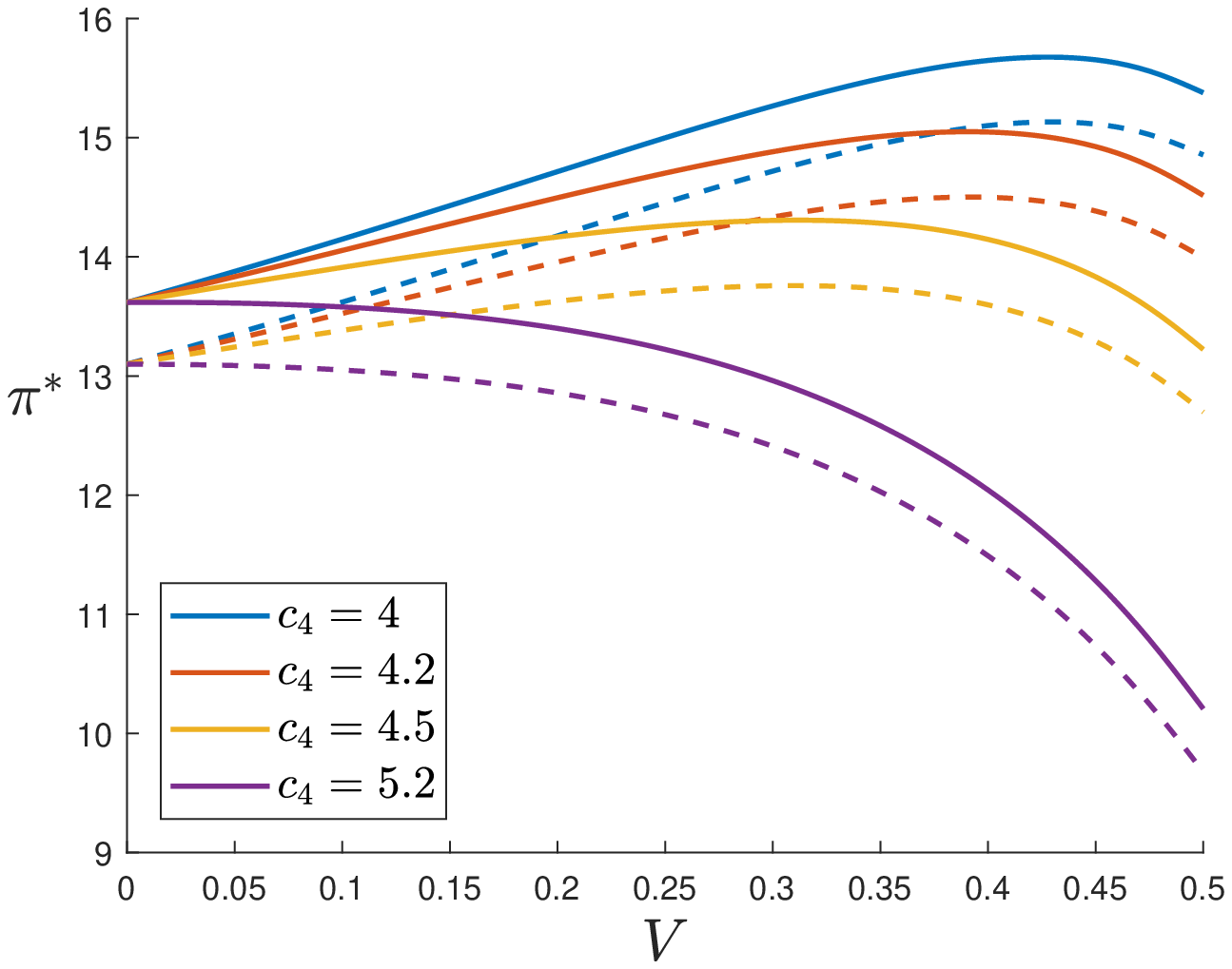}
        \caption{General epidemic, premium minimisation}
        \label{fig:3a}
    \end{subfigure}
    \begin{subfigure}[t]{.45\linewidth}
        \centering
        \includegraphics[width = \linewidth]{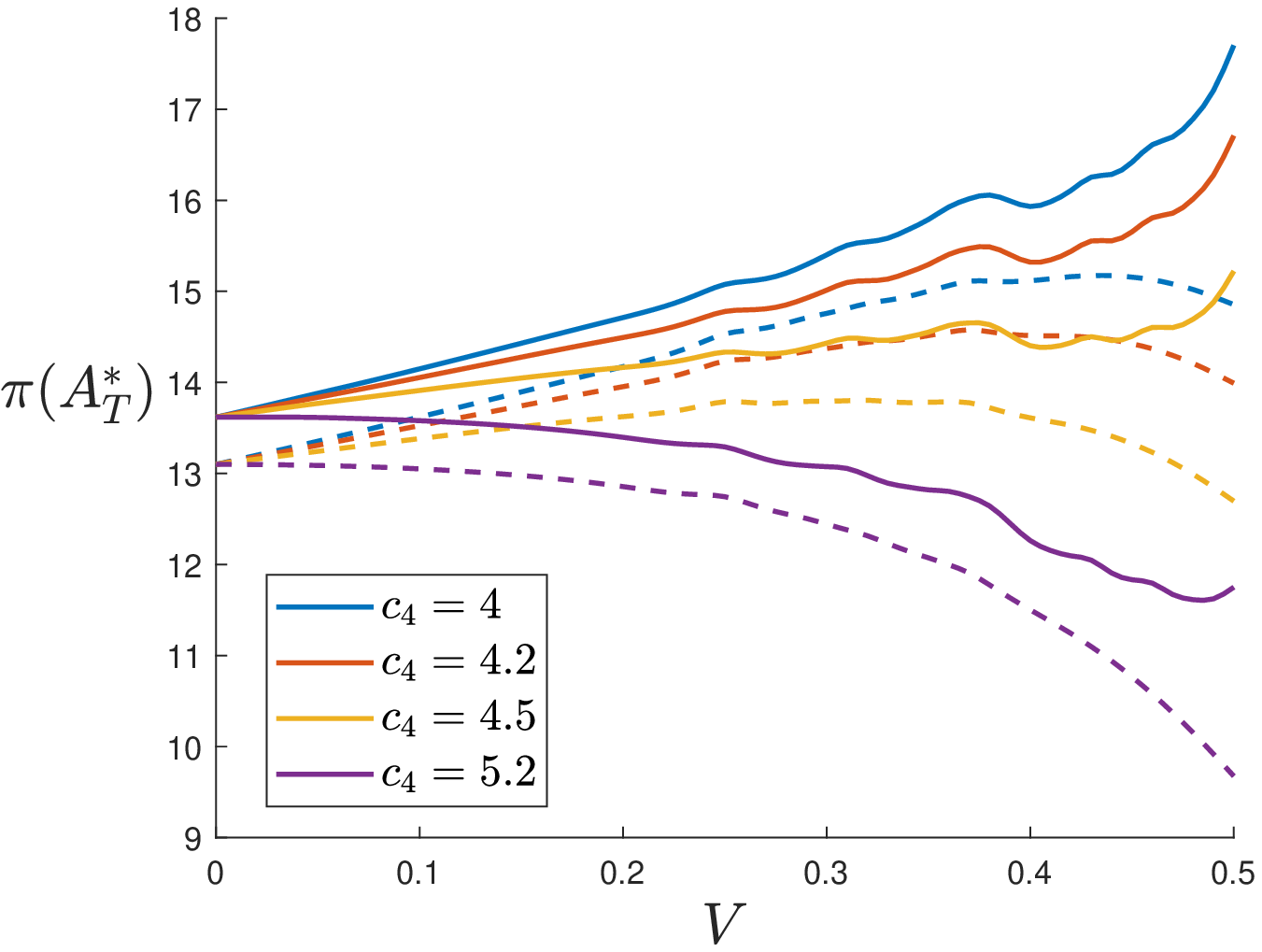}
        \caption{General epidemic, $A_T$ minimisation}
        \label{fig:3b}
    \end{subfigure}
    \begin{subfigure}[t]{.45\linewidth}
        \centering
        \includegraphics[width = \linewidth]{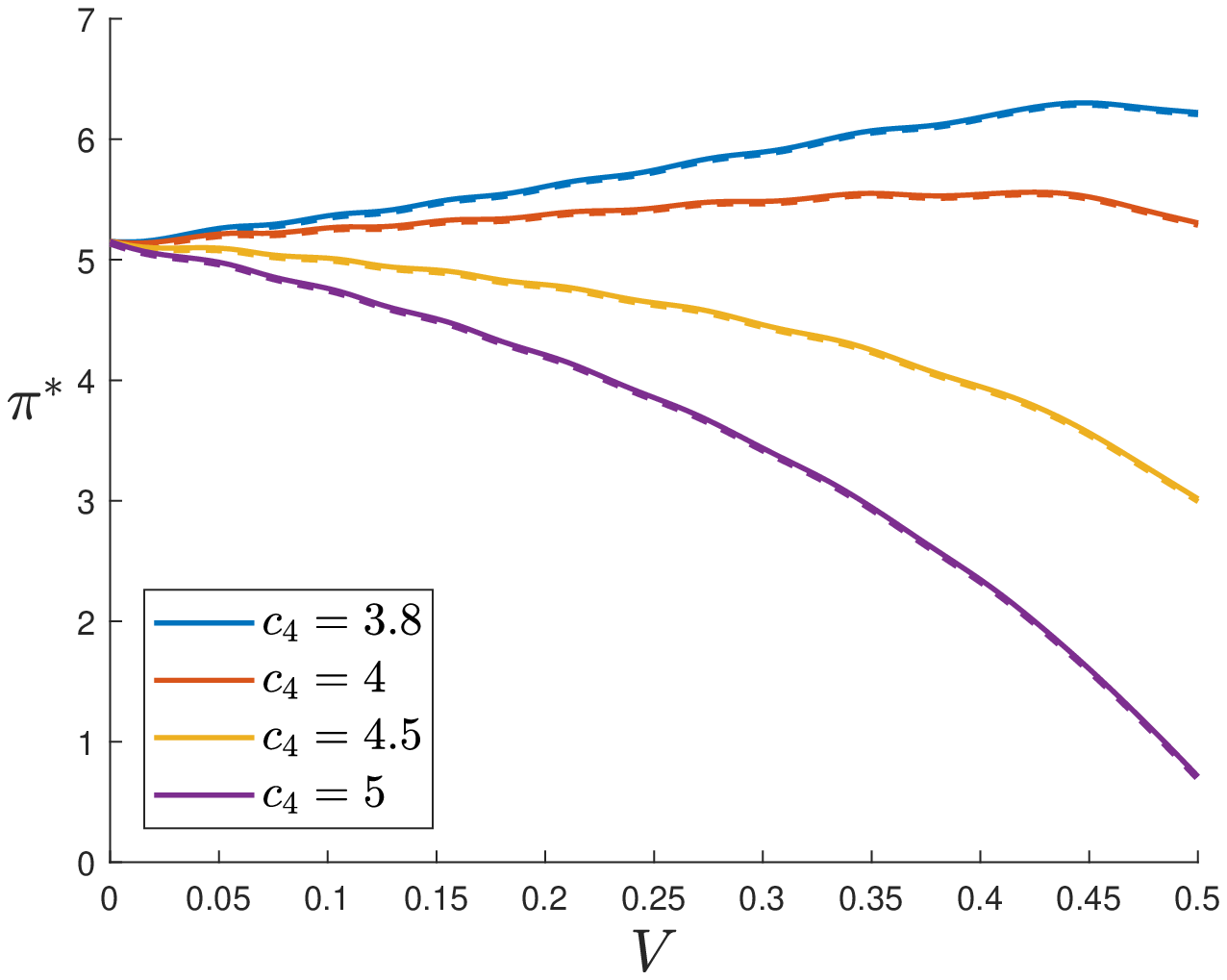}
        \caption{Fatal epidemic, premium minimisation}
        \label{fig:3c}
    \end{subfigure}
    \begin{subfigure}[t]{.45\linewidth}
        \centering
        \includegraphics[width = \linewidth]{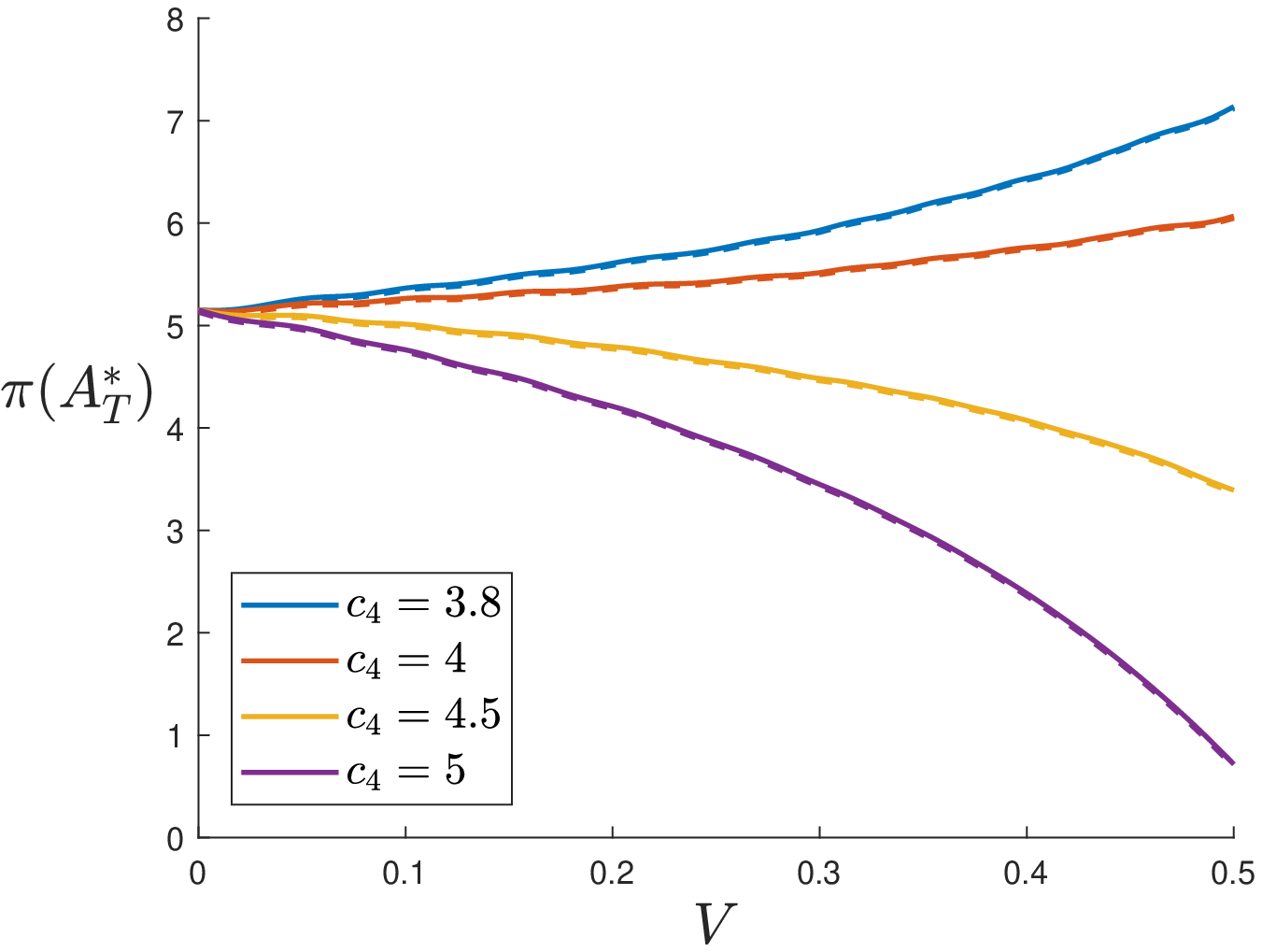}
        \caption{Fatal epidemic, $A_T$ minimisation}
        \label{fig:3d}
    \end{subfigure}
    
    \caption{Premium level for general and fatal epidemics in basic scenario for malaria and measles. Dashed line -- corresponding premium with discounting factor $\delta = \ln 1.01$. Basic reproduction number $R_0 = 12$.}
    \label{fig:3}
\end{figure}

For infections that have low basic reproduction number (see Fig.~\ref{fig:1}) both ``optimality'' approaches have similar results (compare Fig.~\ref{fig:1a} with Fig.~\ref{fig:1b}, and Fig.~\ref{fig:1c} with Fig.~\ref{fig:1d}). The resulting premium depends on selling price $c_4$. On Fig.~\ref{fig:1}, \ref{fig:2} and \ref{fig:3} we clearly see that increase in selling price $c_4$ result in reducing premium level.

From Fig.~\ref{fig:1} we see that the discounted premium \eqref{eq:premium_discount} is lower than the premium without discount \eqref{eq:premium}. In general (Fig.~\ref{fig:1a}--\ref{fig:1b}) and fatal (Fig.~\ref{fig:1c}--\ref{fig:1d}) epidemics discounted premiums noticeably differ from non-discounted.

Due to concavity of the premium level (Fig.~\ref{fig:1}) with respect to vaccine amount, it is better either not to purchase any vaccine at all, or alternatively purchase as much as possible, if the company is confident that it will be sold to the population at price $c_4$.

In case of more serious diseases (see Fig.~\ref{fig:2}) the difference between two ``optimality'' approaches is only seen in case of general epidemic (Fig.~\ref{fig:2a}--\ref{fig:2b}) for vaccine amount close to $50\%$ of susceptibles. This is the point when vaccine allocation for lost working days minimisation problem (equations~\eqref{eq:pi_minimisation}-\eqref{eq:A_minimisation}) switches strategy from putting all vaccine available into one centre to dividing equally between two centres (see \cite{Chernov_2019}). For larger $V$ the effect of vaccination is not that noticeable. Therefore, the vaccine becomes not self-sustaining.

In case of fatal epidemic (Fig.~\ref{fig:2c}--\ref{fig:2d}) we do not see the same behaviour. According to small difference between discounted and non-discounted premium levels, the epidemic is taking over, so that all susceptible persons die quite fast, and only vaccinated people pay money for the vaccine. Very similar results are for severe infections, see Fig.~\ref{fig:3}.

\subsubsection{Health-centre}\label{sub:health_centre2}
Assume that the epidemic has started in two centres, but only one of them has a medical facility that can cure the disease effectively. In this case, infected persons from the second centre are willing to get to the first centre to cure the disease, while infected persons from the first centre stay at home. We also assume that the second centre has a medical institution, but it is not that effective.

It is natural to assume that the first centre, which has the medical facility, is technologically more advanced and has higher population. Let the population of the first centre be $S_{1, 0} = 5000$ and $I_{1, 0} = 800$, and the population of the second centre be $S_{2, 0} = 1000$ and $I_{2, 0} = 200$.

\begin{figure}[ht]
    \centering

    \begin{subfigure}[t]{.45\linewidth}
        \centering
        \includegraphics[width = \linewidth]{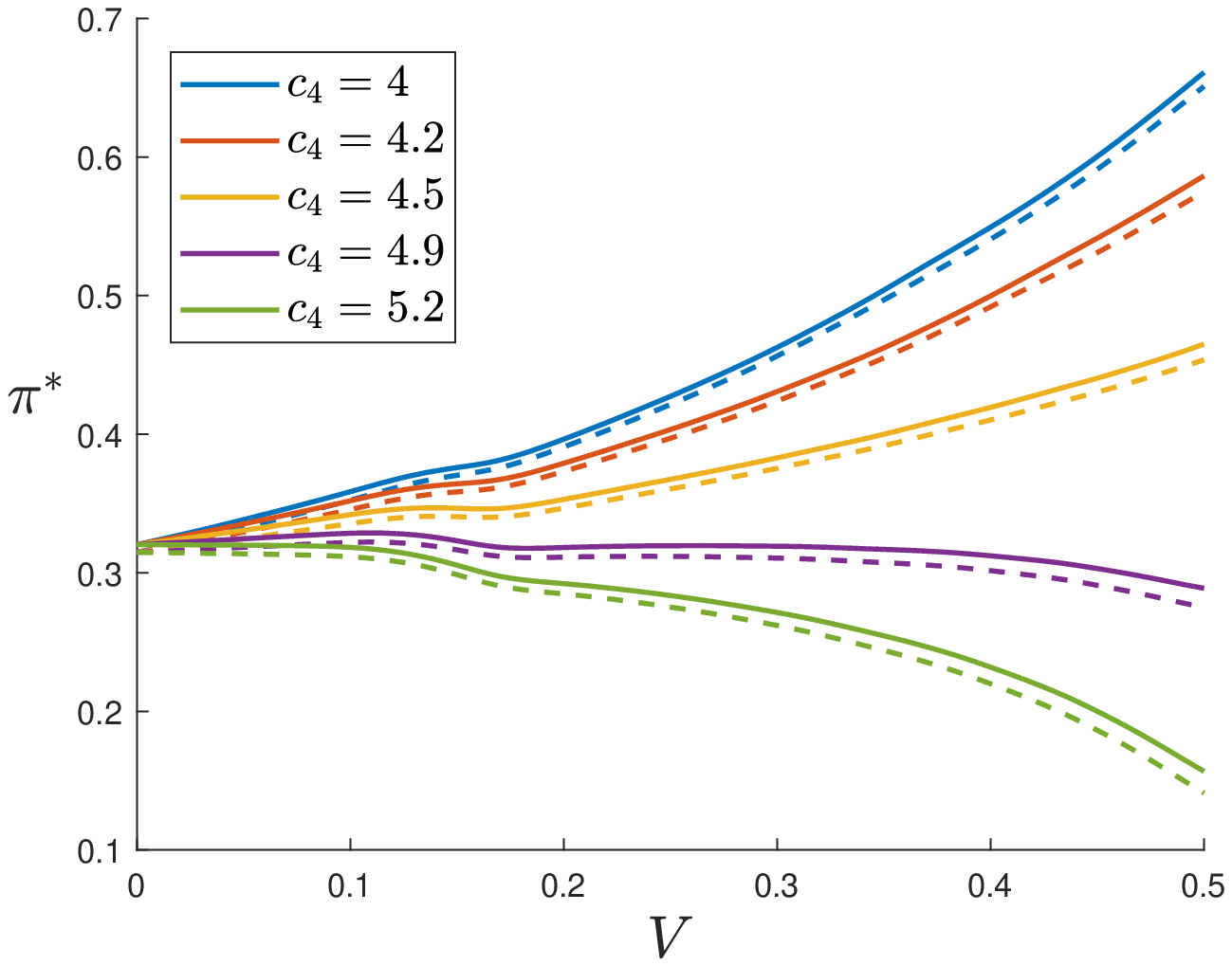}
        \caption{General epidemic, premium minimisation}
        \label{fig:4a}
    \end{subfigure}
    \begin{subfigure}[t]{.45\linewidth}
        \centering
        \includegraphics[width = \linewidth]{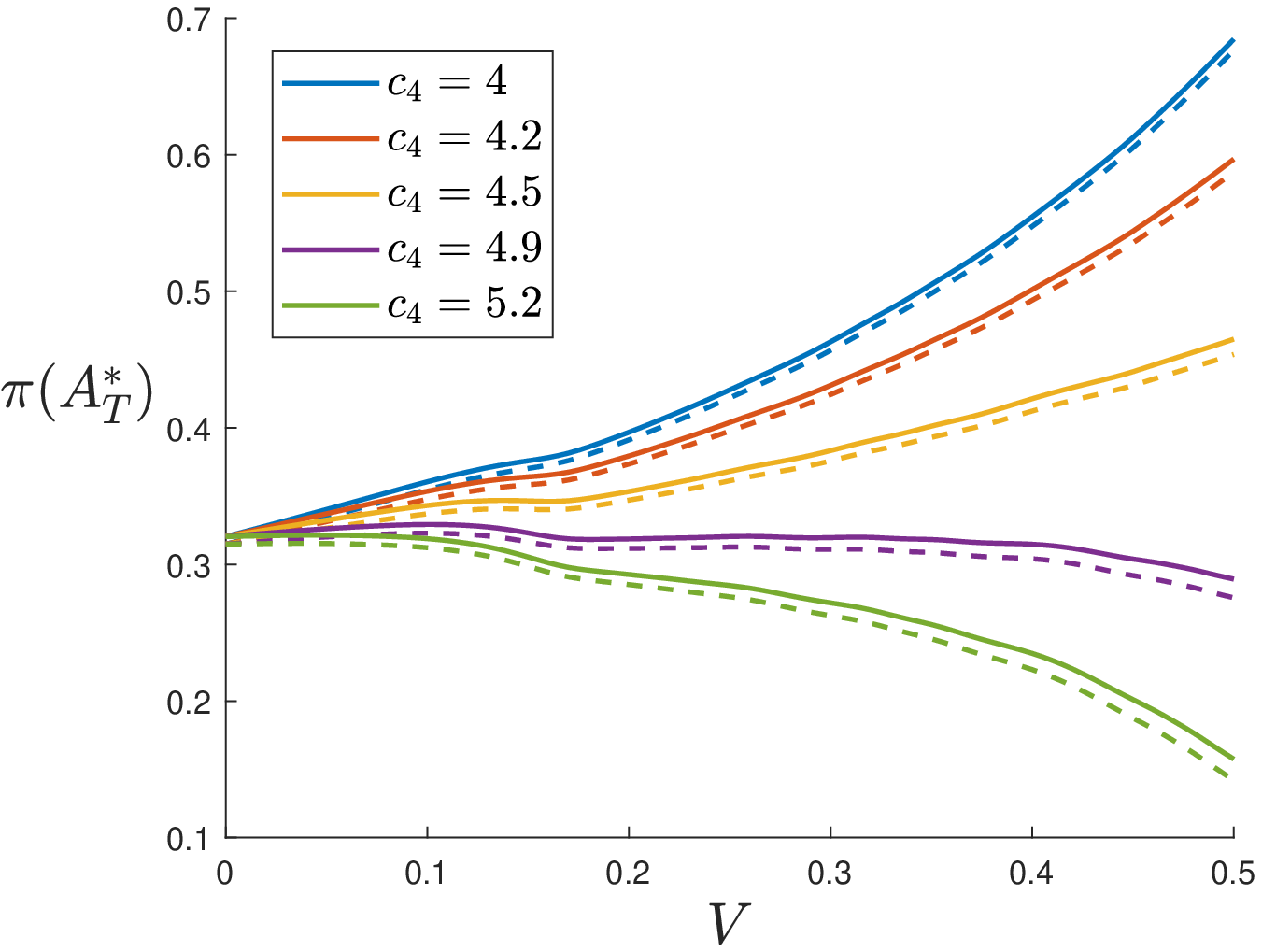}
        \caption{General epidemic, $A_T$ minimisation}
        \label{fig:4b}
    \end{subfigure}
    \begin{subfigure}[t]{.45\linewidth}
        \centering
        \includegraphics[width = \linewidth]{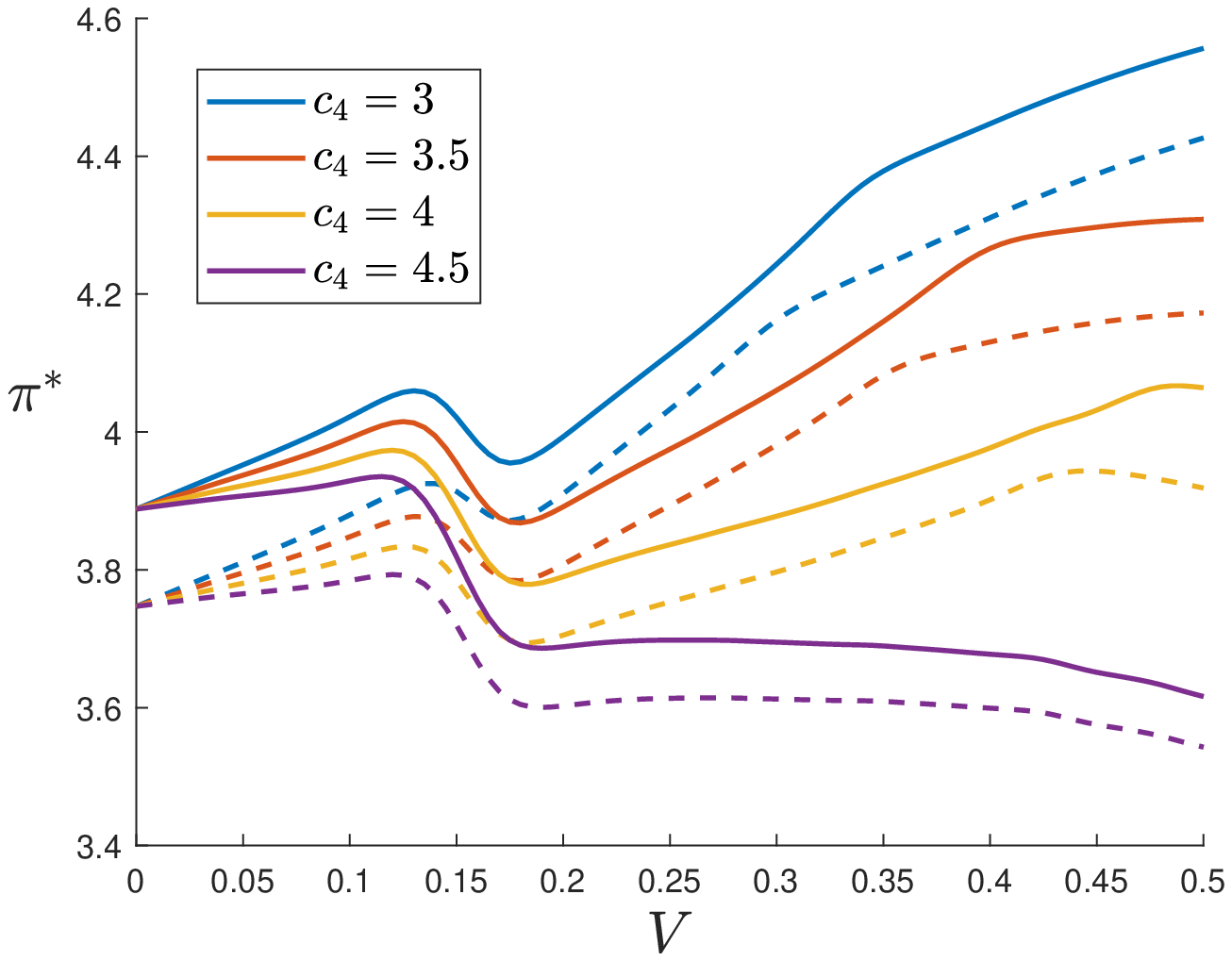}
        \caption{Fatal epidemic, premium minimisation}
        \label{fig:4c}
    \end{subfigure}
    \begin{subfigure}[t]{.45\linewidth}
        \centering
        \includegraphics[width = \linewidth]{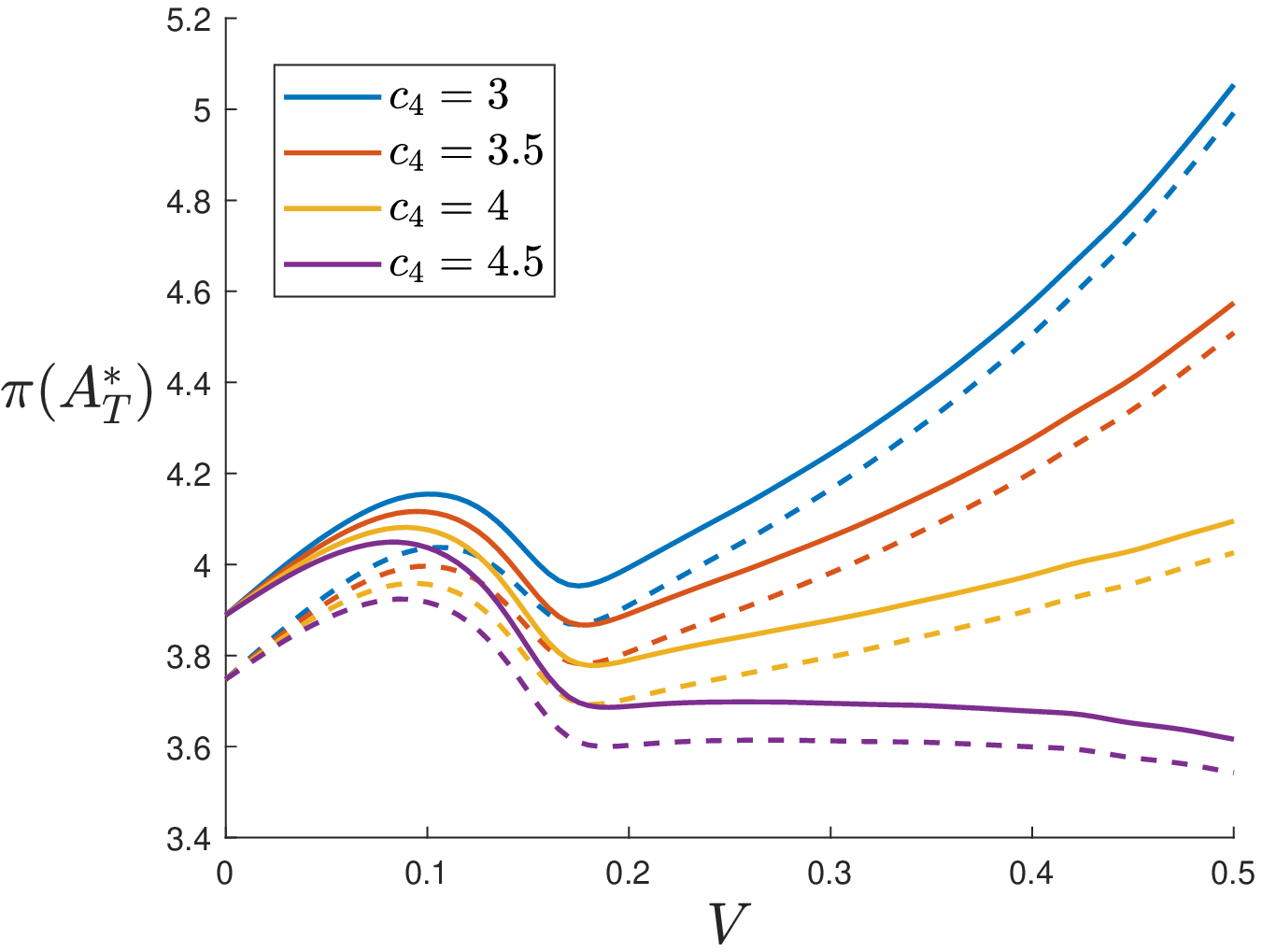}
        \caption{Fatal epidemic, $A_T$ minimisation}
        \label{fig:4d}
    \end{subfigure}
    
    \caption{Premium level for general and fatal epidemics in health-centre scenario for Ebola, influenza. Dashed line -- corresponding premium with discounting factor $\delta = \ln 1.01$. Basic reproduction number $R_0 = 2$.}
    \label{fig:4}
\end{figure}

In this scenario, we assume that susceptible groups in both centres are travelling with rates $k_{12} = 0.1$ and $k_{21} = 0.15$, which represent the well-being of both centres. On contrary, the infected group in the first centre almost do not travel $l_{12} = 0.05$, while infectives from the second centre are coming to the health-centre with rate $l_{21} = 2.5$.

We also assume that a health-centre is affecting curing time, while the infection rate of a disease remains the same (basic reproduction number is unchanged with $\mu = 1$). Hence, for a general epidemic we let $\mu_1 = 2$ and $\mu_2 = -0.9$ (see~\eqref{eq:SIR_hc}), which corresponds to faster curing at the first centre and lower recovery rate at the second. In case of fatal epidemic, the first centre provide better treatment, so that a person lives longer, and it is opposite at the second centre. Therefore, for a fatal epidemic we consider symmetric case: $\mu_1 = -0.9$ and $\mu_2 = 2$.

\begin{figure}[ht]
    \centering

    \begin{subfigure}[t]{.45\linewidth}
        \centering
        \includegraphics[width = \linewidth]{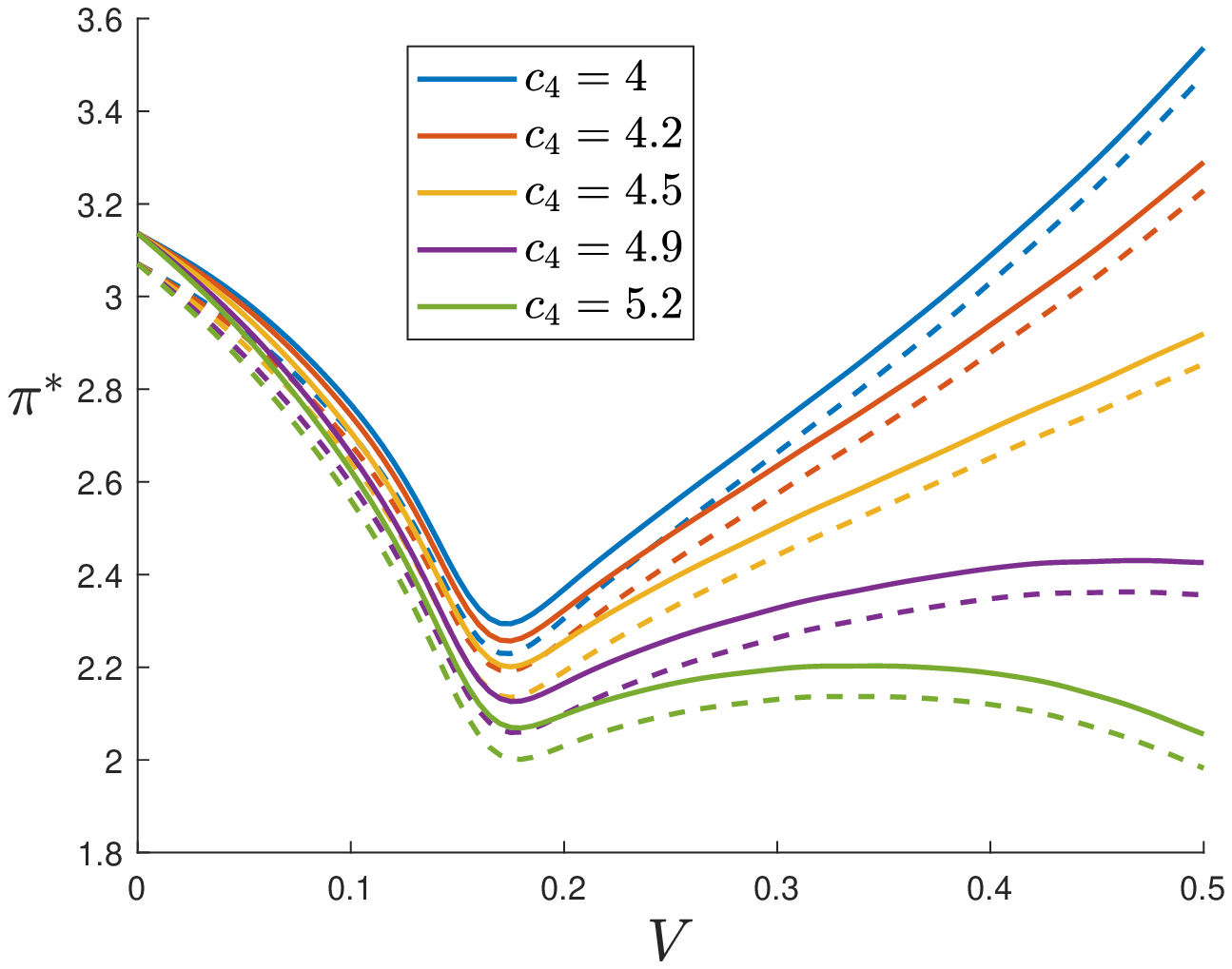}
        \caption{General epidemic, premium minimisation}
        \label{fig:5a}
    \end{subfigure}
    \begin{subfigure}[t]{.45\linewidth}
        \centering
        \includegraphics[width = \linewidth]{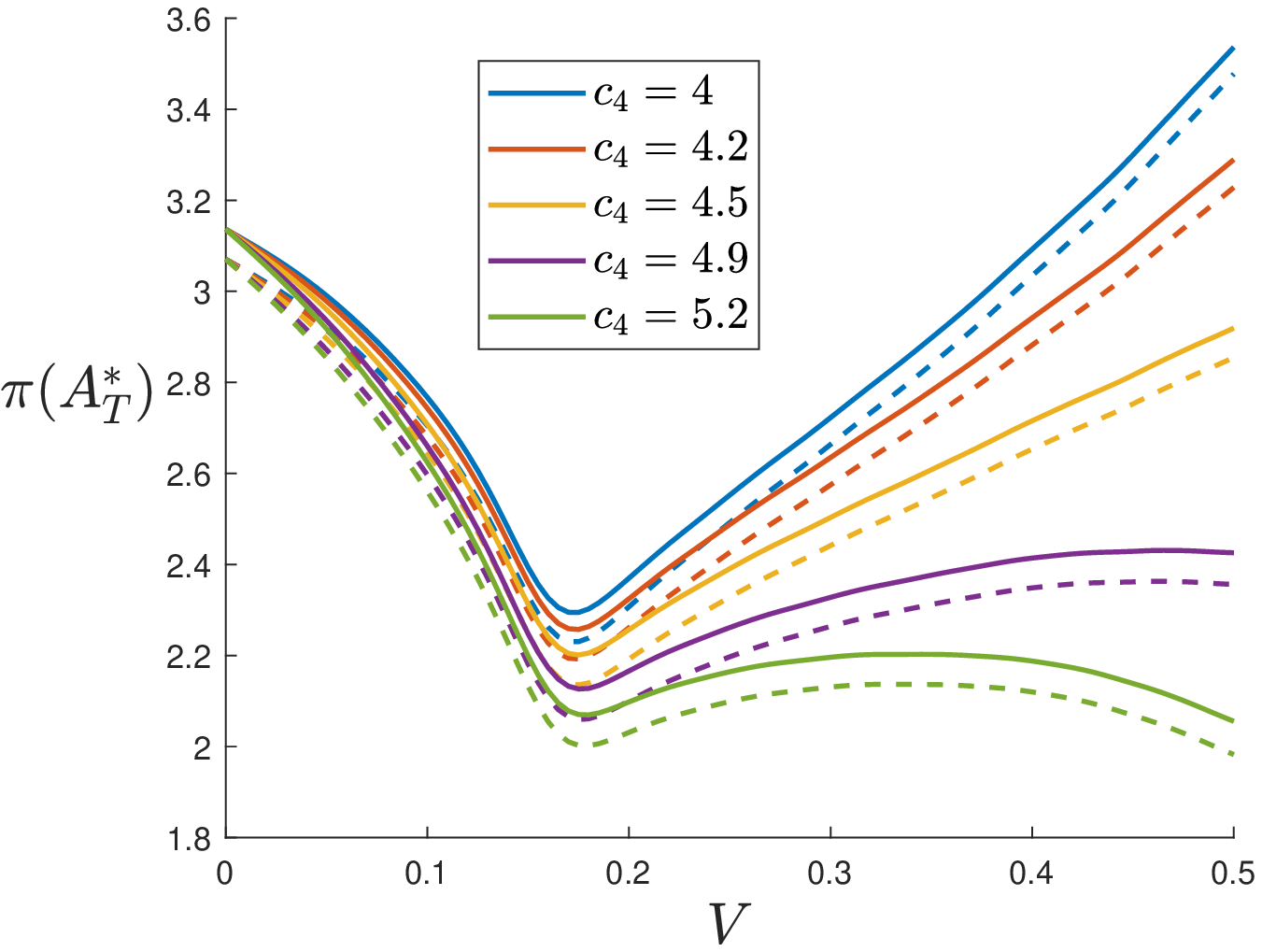}
        \caption{General epidemic, $A_T$ minimisation}
        \label{fig:5b}
    \end{subfigure}
    \begin{subfigure}[t]{.45\linewidth}
        \centering
        \includegraphics[width = \linewidth]{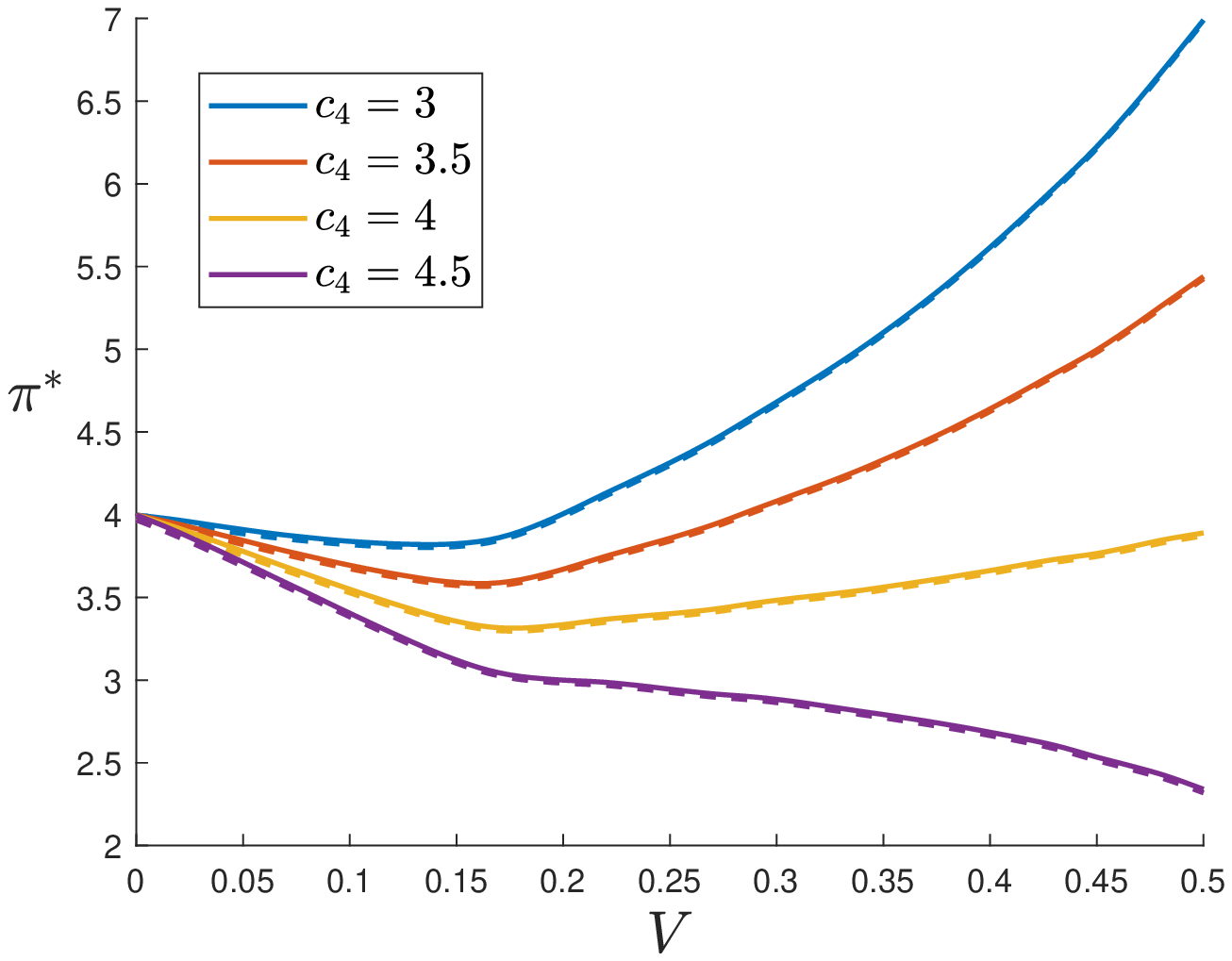}
        \caption{Fatal epidemic, premium minimisation}
        \label{fig:5c}
    \end{subfigure}
    \begin{subfigure}[t]{.45\linewidth}
        \centering
        \includegraphics[width = \linewidth]{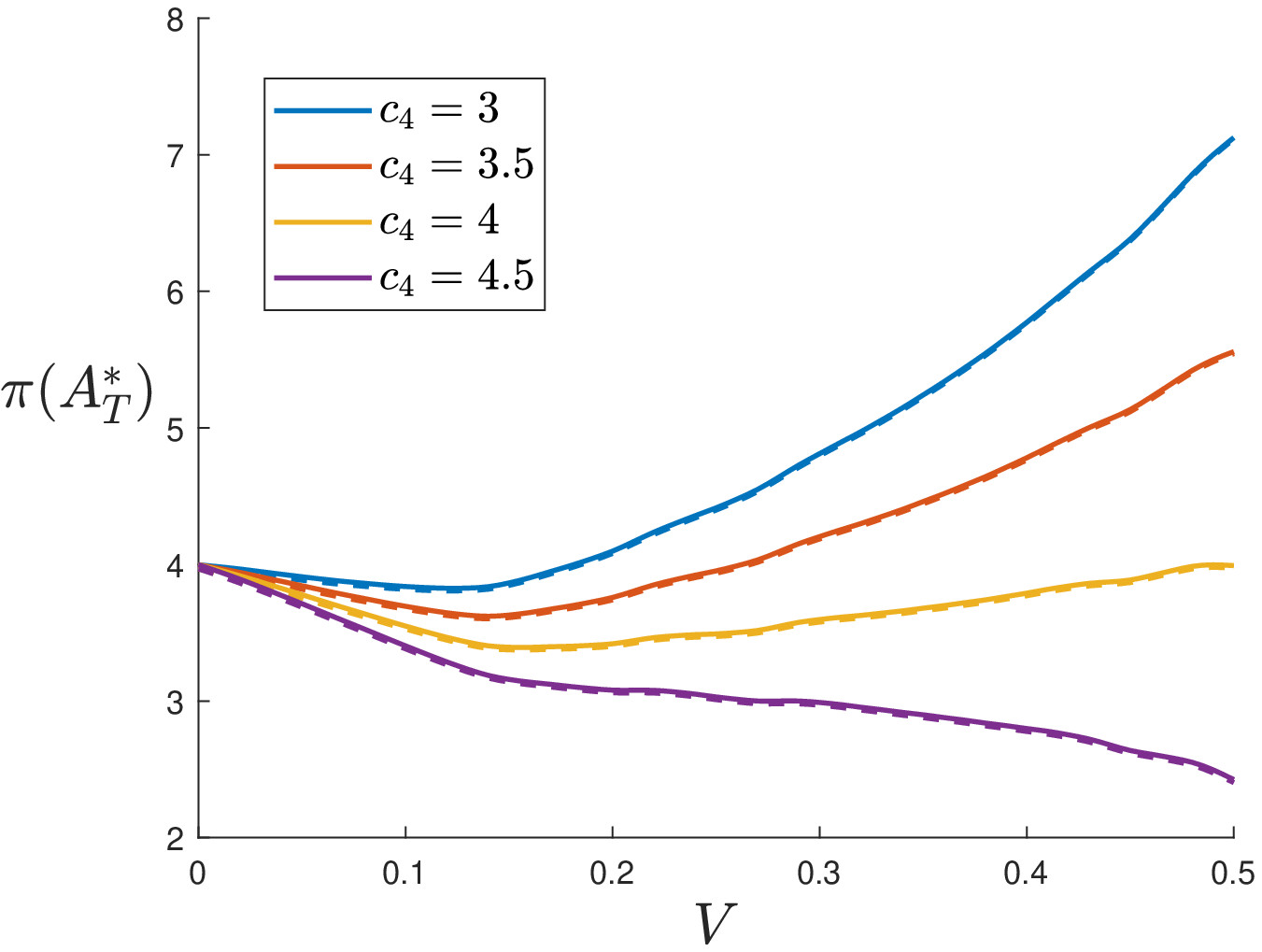}
        \caption{Fatal epidemic, $A_T$ minimisation}
        \label{fig:5d}
    \end{subfigure}
    
    \caption{Premium level for general and fatal epidemics in basic scenario for diphtheria, mumps, polio, smallpox. Dashed line -- corresponding premium with discounting factor $\delta = \ln 1.01$. Basic reproduction number $R_0 = 6$.}
    \label{fig:5}
\end{figure}

\begin{equation}
\label{eq:SIR_hc}
    \begin{aligned}
        \frac{\mrm{d} S_i}{\mrm{d} t} &= -\beta(R_i) \, S_i I_i - \sum_{j \neq i} k_{i j} \, S_i + \sum_{j \neq i} k_{j i} \, S_j \\
        \frac{\mrm{d} I_i}{\mrm{d} t} &= \beta(R_i) \, S_i I_i - (\mu + \mu_i) \, I_i - \sum_{j \neq i} l_{i j} \, I_i + \sum_{j \neq i} l_{j i} \, I_j \\
        \frac{\mrm{d} R_i}{\mrm{d} t} &= (\mu + \mu_i) \, I_i \\
        S_i(0) &= S_{i, 0}, \, I_i(0) = I_{i, 0}, \, R_i(0) = R_{i, 0}
    \end{aligned}
\end{equation}

\begin{figure}[ht]
    \centering

    \begin{subfigure}[t]{.45\linewidth}
        \centering
        \includegraphics[width = \linewidth]{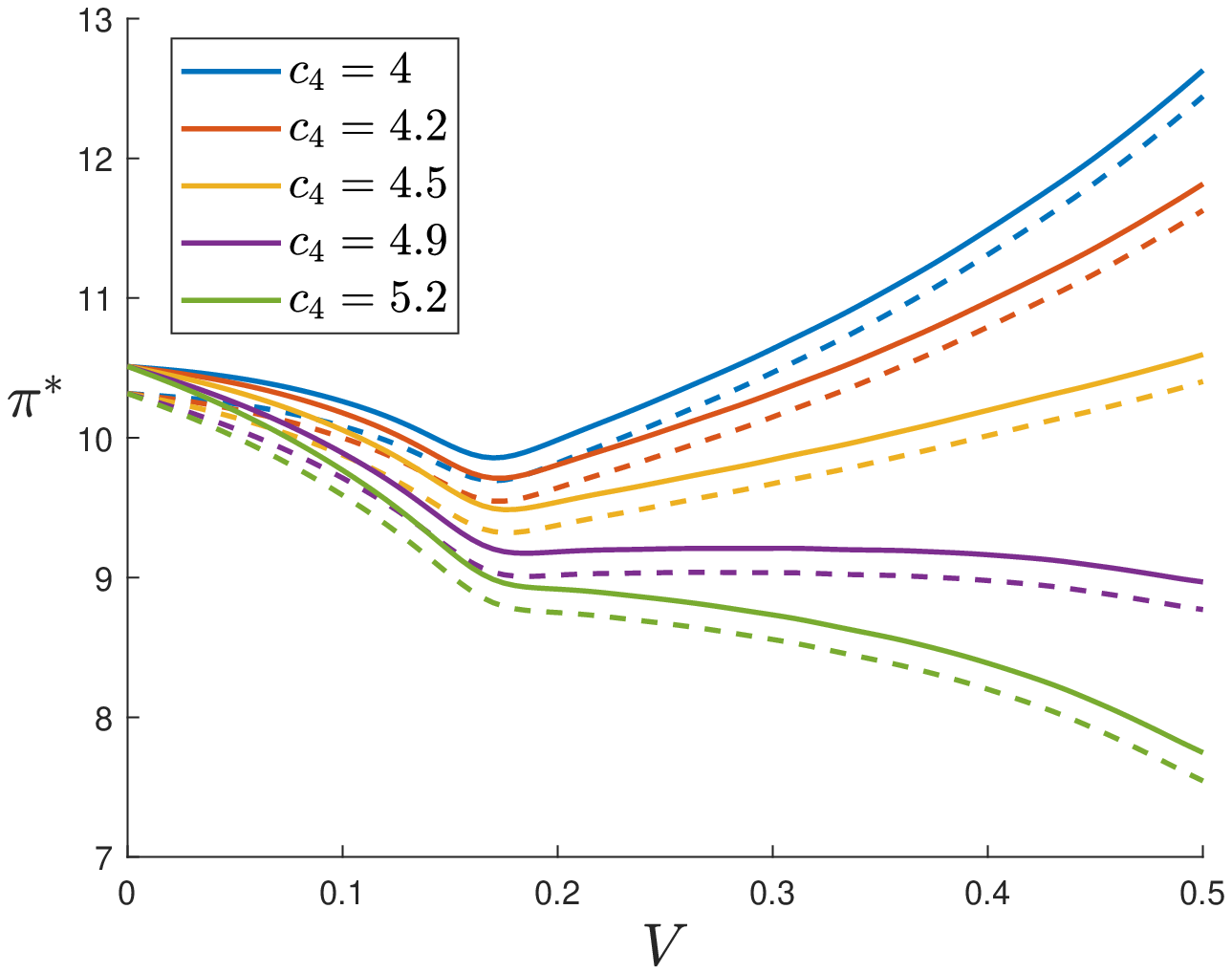}
        \caption{General epidemic, premium minimisation}
        \label{fig:6a}
    \end{subfigure}
    \begin{subfigure}[t]{.45\linewidth}
        \centering
        \includegraphics[width = \linewidth]{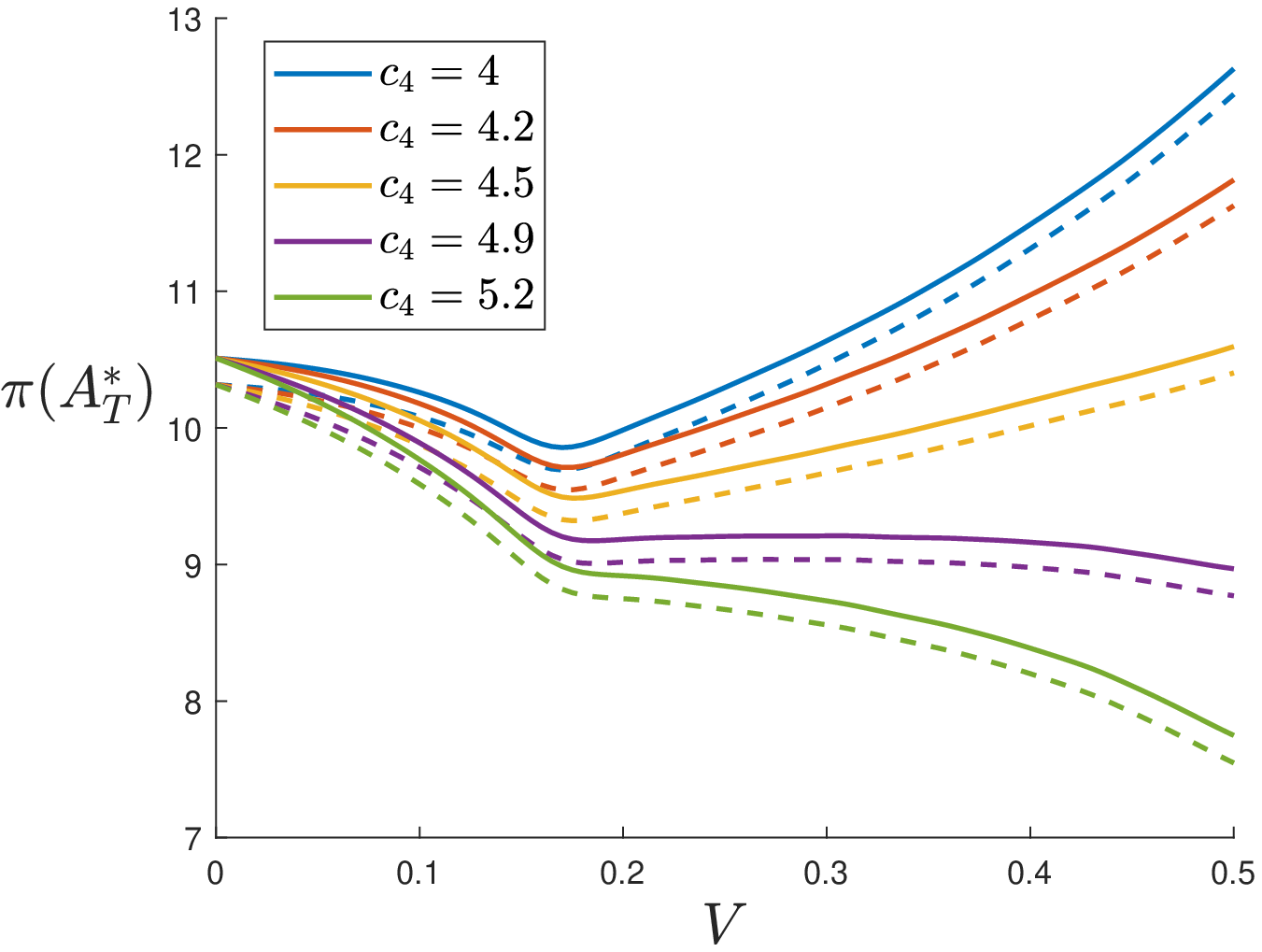}
        \caption{General epidemic, $A_T$ minimisation}
        \label{fig:6b}
    \end{subfigure}
    \begin{subfigure}[t]{.45\linewidth}
        \centering
        \includegraphics[width = \linewidth]{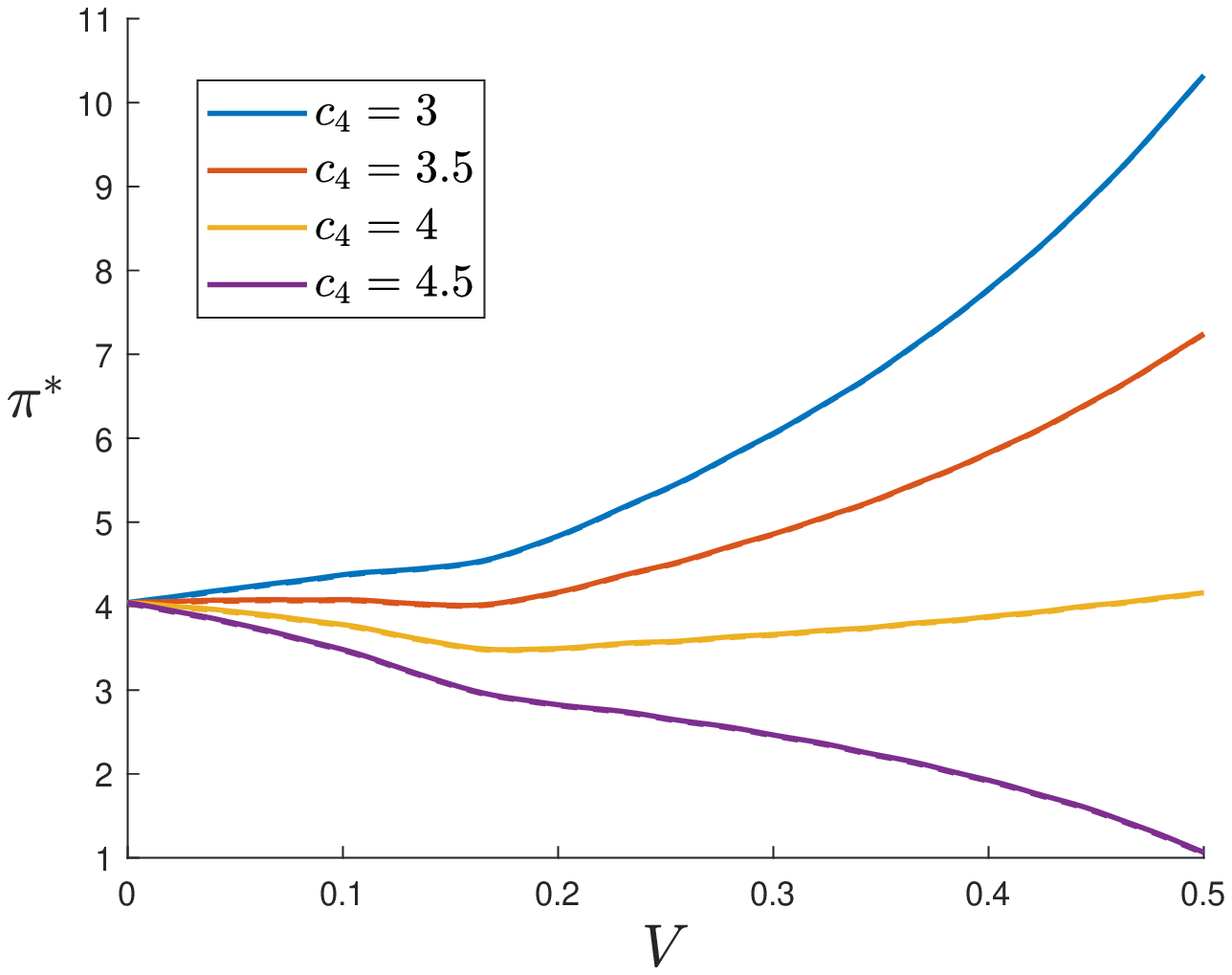}
        \caption{Fatal epidemic, premium minimisation}
        \label{fig:6c}
    \end{subfigure}
    \begin{subfigure}[t]{.45\linewidth}
        \centering
        \includegraphics[width = \linewidth]{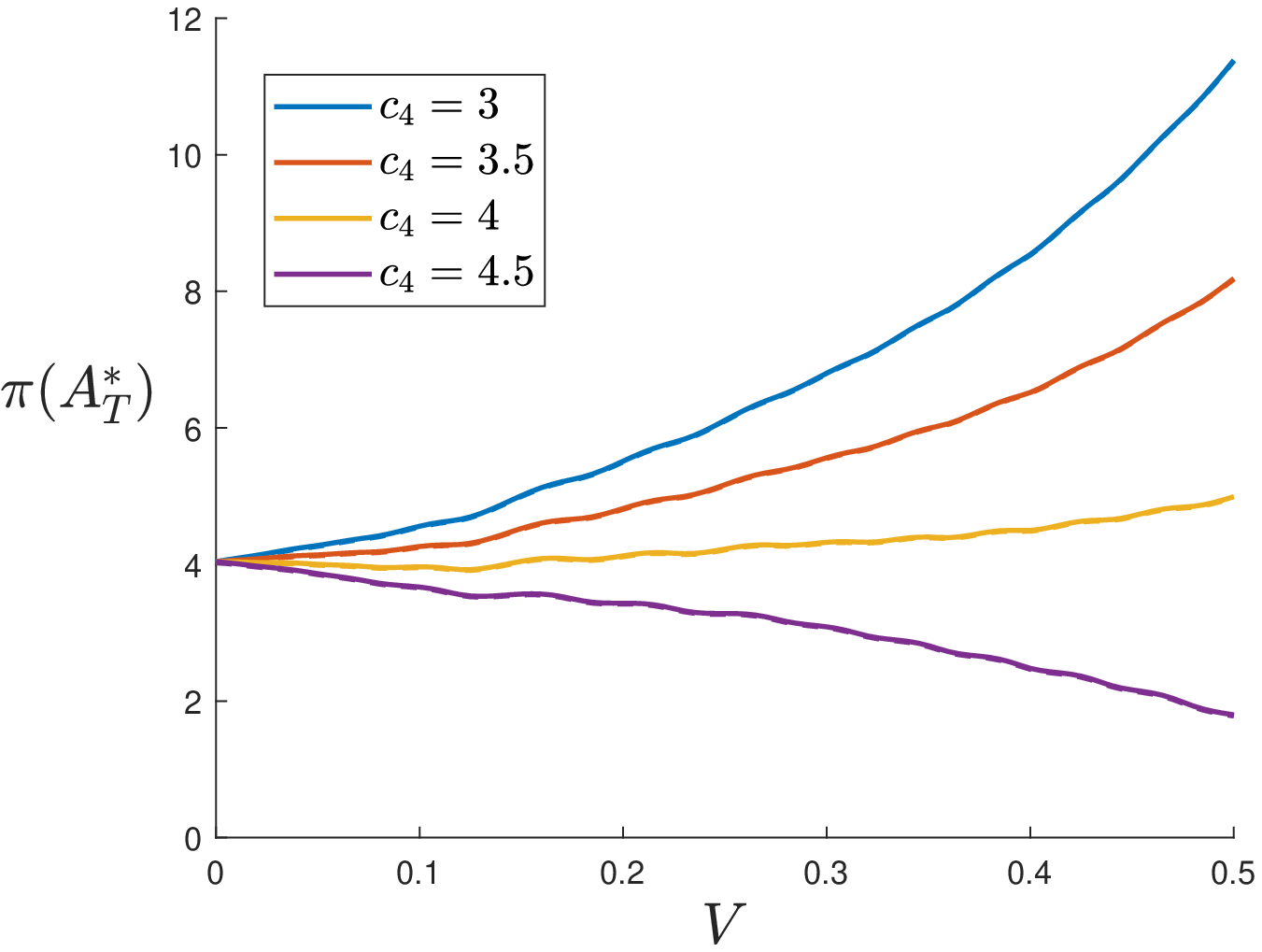}
        \caption{Fatal epidemic, $A_T$ minimisation}
        \label{fig:6d}
    \end{subfigure}
    
    \caption{Premium level for general and fatal epidemics in basic scenario for malaria and measles. Dashed line -- corresponding premium with discounting factor $\delta = \ln 1.01$. Basic reproduction number $R_0 = 12$.}
    \label{fig:6}
\end{figure}

Fig.~\ref{fig:4}, \ref{fig:5} and \ref{fig:6} show results for mild, harmful and severe diseases, correspondingly. In general epidemic setup the results are completely different comparing to the basic scenario \ref{sub:basic_scenario2}. One of the main differences is the shape of the graphs -- for smaller selling prices $c_4$ premiums in basic scenario \ref{sub:basic_scenario2} mostly have concave shape, while corresponding graphs in health-centre scenario \ref{sub:health_centre2} have convex shape. Moreover, here we see characteristic threshold (local minimum of premium) for all reproduction numbers.

\begin{table}[t]
    \centering
    \caption{Optimal vaccine share in health-centre scenario harmful $R_0 = 6$ and severe $R_0 = 12$ diseases for fatal epidemics.}
    \begin{tabular}{c|cccc}
        \toprule
        $c_4$ & $3$ & $3.5$ & $4$ & $4.5$ \\
        \midrule
        $V^*$ & $0.14$ & $0.16$ & $0.17$ & $0.18$\\
        \bottomrule
    \end{tabular}
    \label{tab:hc-opt-V}
\end{table}

In case of mild diseases (see Fig.~\ref{fig:4}) there is a noticeable vaccine share $V^* \approx 0.18$, which corresponds to local minimum of both ``optimal'' premiums for fatal epidemics (see Fig.~\ref{fig:4c} and \ref{fig:4d}) and less distinct local minimum for general epidemics (see Fig.~\ref{fig:4a} and \ref{fig:4b}). Therefore, for slowly spreading infections there is a universal strategy of vaccinating $18\%$ of susceptible population -- fully vaccinate the second centre (i.e. $1000$ susceptibles) and partly vaccinate the health-centre (i.e. $80$ individuals).

Fig.~\ref{fig:5} shows results for harmful diseases. As it is seen in Fig.~\ref{fig:5a} and \ref{fig:5b}, there is a more sharp minimum of both ``optimal'' premiums in case of general epidemic, which occurs in $V^* \approx 0.18$ uniformly with respect to selling price $c_4$. Hence, for general epidemic of infection with basic reproduction number $R_0 = 6$ it is enough to vaccinate $18\%$ of population, allocating the vaccine in the same way: $80$ individuals in the first centre and $1000$ -- in the second. The selling price should be decided such that the population can afford to buy it (i.e. if $c_4$ is too high, people might prefer to buy a health-care policy instead of vaccine, which will shift the local minimum point).

For fatal diseases (see Fig.~\ref{fig:5c}, \ref{fig:5d}, \ref{fig:6c} and \ref{fig:6d}) the local minimum point is not that distinct as in Fig.~\ref{fig:4c} and \ref{fig:4d}, and depends on selling price $c_4$. 
Particularly, for both ``optimal'' premiums the vaccine shares $V^*$ are presented in Table~\ref{tab:hc-opt-V}.


\subsection{Stochastic model}
\label{sect:numerical_stochastic}

In this section we abandon notation $\mathcal{V}$ and let $V$ denote the available vaccine stock. It is a natural number (including zero), since the stochastic model concentrates on particular individuals, so the vaccine amount cannot be a real number.

According to proposed Algorithm~\ref{alg:optalloc} for computation of optimal vaccine allocation, for small vaccine number $V$ we need to look over small number of possible allocations. In fact, in case of two centres the number $N_V$ of all possible allocations of vaccine amount $V$ is
\begin{equation*}
    N_V = V+1
\end{equation*}
Therefore, we can afford to run a lot of Monte Carlo simulations $N_{\mrm{sim}}$ for small $V$, and we try to keep $N_{\mrm{sim}}$ low for larger vaccine amounts. So, we propose the following number of simulations depending on $V$:
\begin{equation*}
    N_{\mrm{sim}} = \lceil 100 \cdot (1 + 29e^{-V/5}) \rceil.
\end{equation*}
This will allow us to calculate average premiums with higher accuracy (i.e. with lower variance) for lower $V$, while maintaining the precision for higher vaccine stocks.

In all graphs in this section we present premiums with $95\%$ confidence intervals
\begin{equation*}
    \pi^* \pm z_{0.975} \sqrt{\frac{\mrm{var}(\pi^*)}{N_{\mrm{sim}}}},
\end{equation*}
where $z_{0.975}$ is quantile of standard normal distribution and $\mrm{var}(\pi^*)$ is calculated during Algorithm~\ref{alg:optalloc} as sample variance. This form of confidence interval follows directly from Monte Carlo estimator properties.

Stochastic model is generally used to describe the infection spread throughout the population for small numbers of infected persons. The key point is that the infection can be extinct, i.e. get naturally suppressed, not causing any outbreak. For this reason we complement Algorithm~\ref{alg:optalloc} with calculation of average number of extinct epidemics. For this cause we set percentage $\eta = 0.1$ of population, and say that there was no outbreak, if maximal number of infected individuals at any point of time is smaller than $10\%$ of total susceptible population at time $t=0$:
\begin{equation*}
    \max_{0 \leq t \leq T} \sum_{i=1}^2 I_i(t) < \eta \cdot (S_{1, 0} + S_{2, 0}).
\end{equation*}

In general, many computationally expensive simulations should be performed to obtain precise result, which is usually done on supercomputers. Therefore, the following results are not reliable enough to conclude anything about optimal vaccine allocation and precise premium calculations. However, they provide a satisfactory behaviour of premium dependencies.

\begin{figure}[ht]
    \centering
    
    \begin{subfigure}[t]{.45\linewidth}
        \includegraphics[width = \linewidth]{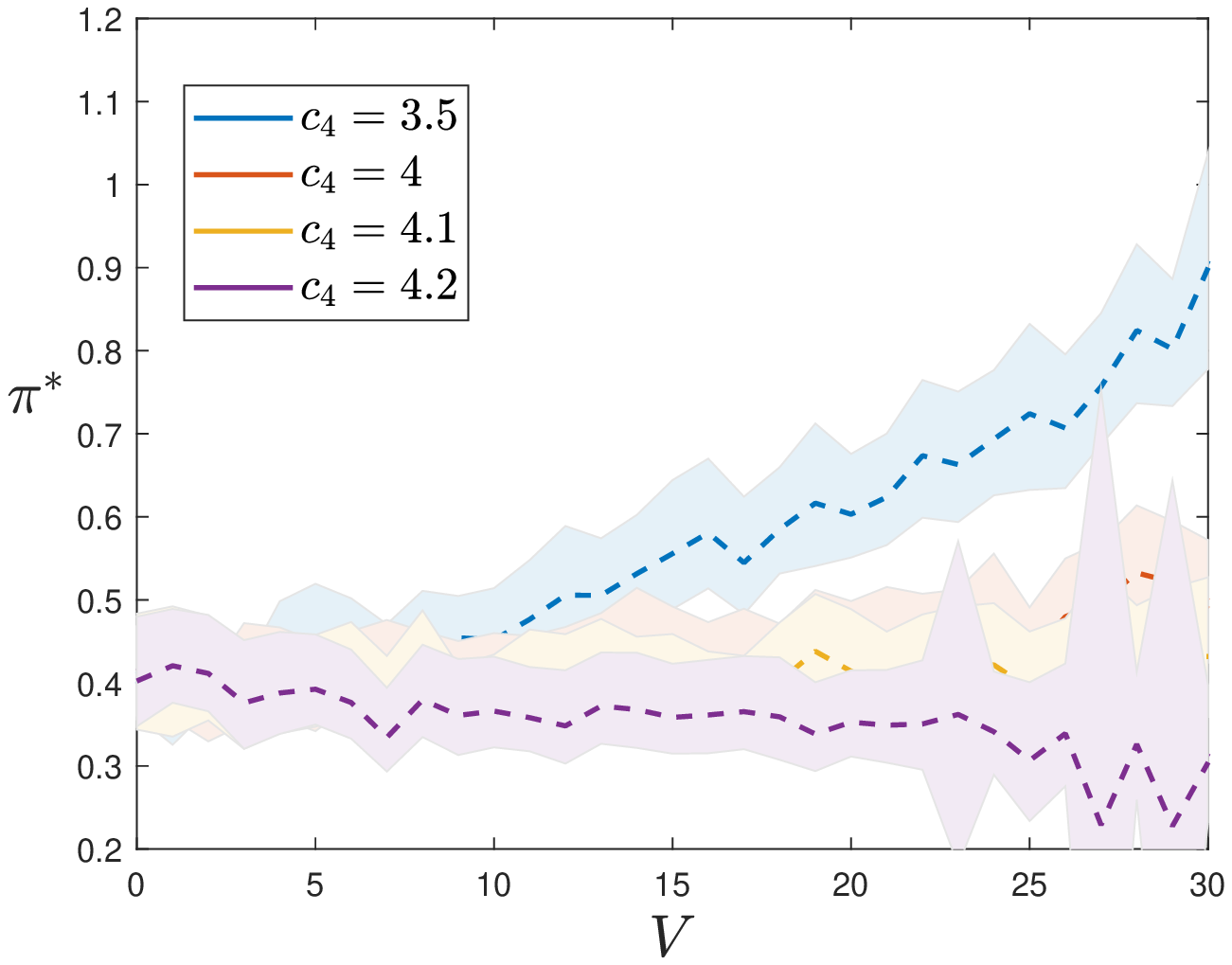}
        \caption{General epidemic, premium minimisation}
        \label{fig:7a}
    \end{subfigure}
    \begin{subfigure}[t]{.45\linewidth}
        \includegraphics[width = \linewidth]{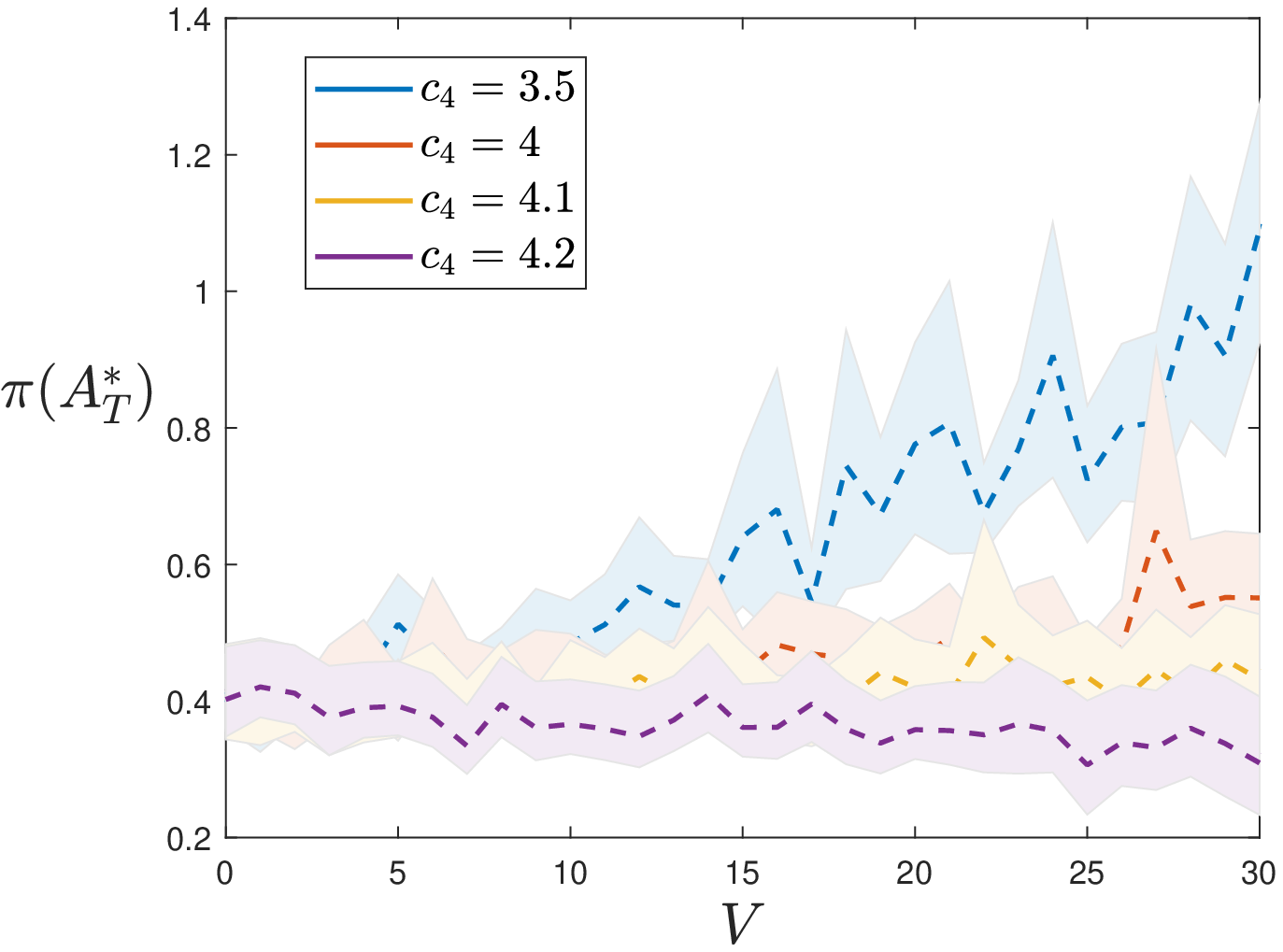}
        \caption{General epidemic, $A_T$ minimisation}
        \label{fig:7b}
    \end{subfigure}
    \begin{subfigure}[t]{.45\linewidth}
        \includegraphics[width = \linewidth]{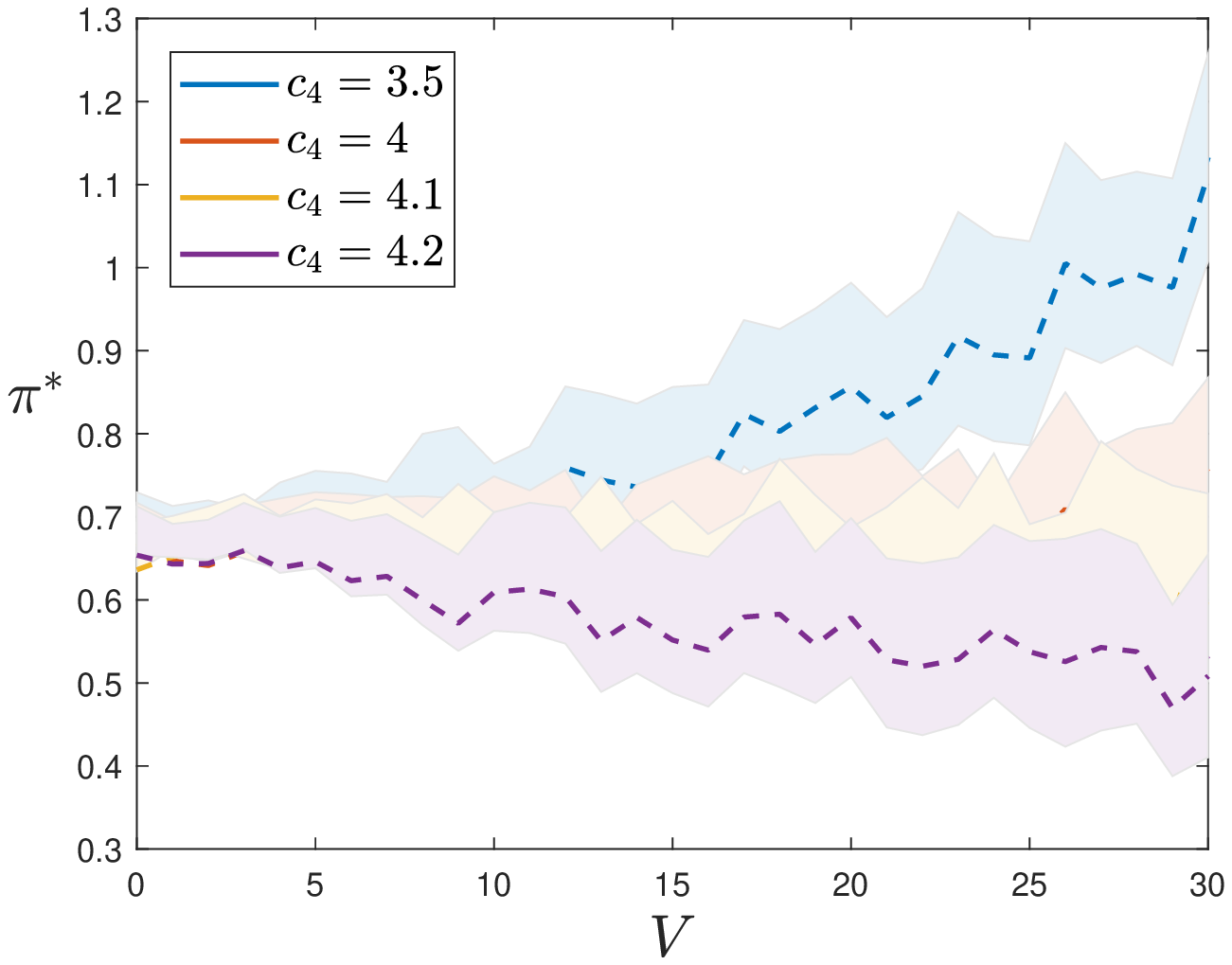}
        \caption{Fatal epidemic, premium minimisation}
        \label{fig:7c}
    \end{subfigure}
    \begin{subfigure}[t]{.45\linewidth}
        \includegraphics[width = \linewidth]{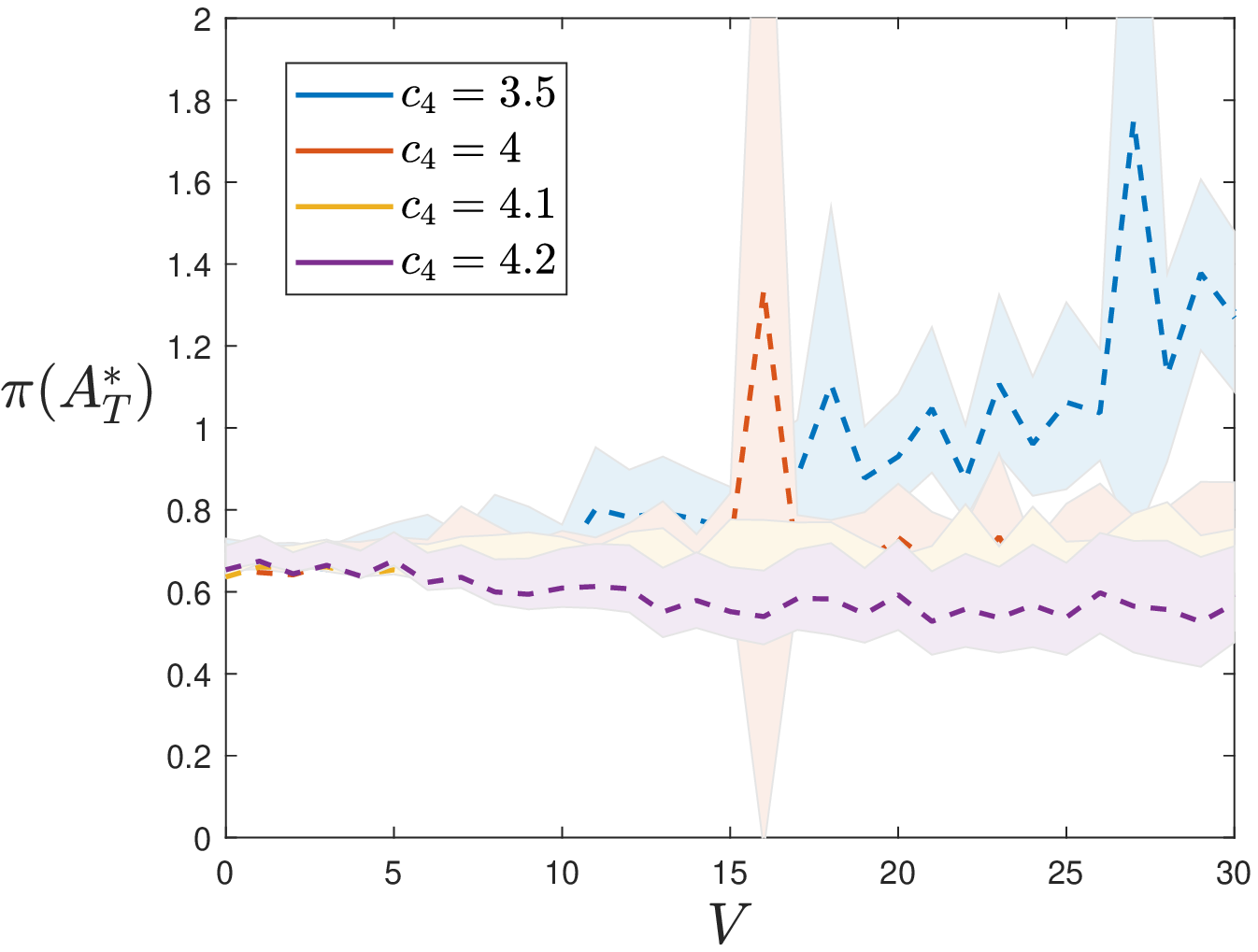}
        \caption{Fatal epidemic, $A_T$ minimisation}
        \label{fig:7d}
    \end{subfigure}
    
    \caption{Premium level for general and fatal epidemics in basic stochastic scenario for Ebola, influenza. Dashed line -- corresponding premium with discounting factor $\delta = \ln 1.01$. Shaded area -- $95\%$ confidence interval of $\pi$. Basic reproduction number $R_0 = 2$.}
    \label{fig:7}
\end{figure}

\subsubsection{Basic stochastic scenario}
Here we consider a basic scenario that is analogous to deterministic case Sect.~\ref{sub:basic_scenario2}. In line with \cite{lefevre_picard_simon_2017}, we let initial number of susceptible individuals be $S_{i, 0} = 30$, however, we consider less infected people: $I_{i, 0} = 1$ for $i = 1, 2$. Further, we let migration fluxes be $k_{12} = k_{21} = 0.1$ and $l_{12} = l_{21} = 0.05$.

In stochastic model for minor diseases in general epidemic case (see Fig.~\ref{fig:7}) all premium levels are increasing for selling prices $c_4$ less than vaccine cost $c_3 = 4$. If epidemic is fatal, the premium level becomes decreasing for $c_4 > c_3$.

During numerical simulations we assess number of epidemics that are naturally suppressed and calculate their fractions, see Table~\ref{tab:stoch_extinct_R2} for mild diseases ($R_0 = 2$), Table~\ref{tab:stoch_extinct_R6} for harmful diseases ($R_0 = 6$) and Table~\ref{tab:stoch_extinct_R12} for severe diseases ($R_0 = 12$).

\begin{table}[ht]
    \centering
    \caption{Basic scenario. Extinct epidemic rates for mild diseases $R_0 = 2$.}
    \begin{tabular}{l | rlrlrlrl}
        \toprule
        \multirow{2}{*}[-3pt]{Minimisation} & \multicolumn{8}{l}{$\qquad\; V = 0 \qquad\qquad\; V = 7 \qquad\qquad\; V = 15 \qquad\qquad\; V = 30$} \\
          & General & Fatal & General & Fatal & General & Fatal & General & Fatal\\
         \cmidrule{1-9}
        $\pi$ & $30.2\%$ & $26.0\%$ & $32.1\%$ & $28.7\%$ & $32.2\%$ & $35.9\%$ & $37.0\%$ & $54.6\%$ \\
        $A_T$ & $30.2\%$ & $26.0\%$ & $32.1\%$ & $28.7\%$ & $32.2\%$ & $36.7\%$ & $55.6\%$ & $56.5\%$ \\
        \bottomrule
    \end{tabular}
    \label{tab:stoch_extinct_R2}
\end{table}

\begin{table}[ht]
    \centering
    \caption{Basic scenario. Extinct epidemic rates for harmful diseases $R_0 = 6$.}
    \begin{tabular}{l | rlrlrlrl}
        \toprule
        \multirow{2}{*}[-3pt]{Minimisation} & \multicolumn{8}{l}{$\qquad\; V = 0 \qquad\qquad\; V = 7 \qquad\qquad\; V = 15 \qquad\qquad\; V = 30$} \\
          & General & Fatal & General & Fatal & General & Fatal & General & Fatal\\
         \cmidrule{1-9}
        $\pi$ & $2.3\%$ & $2.3\%$ & $3.3\%$ & $3.2\%$ & $3.4\%$ & $3.6\%$ & $14.0\%$ & $15.0\%$ \\
        $A_T$ & $2.3\%$ & $2.3\%$ & $3.3\%$ & $3.6\%$ & $3.4\%$ & $3.8\%$ & $14.0\%$ & $15.0\%$ \\
        \bottomrule
    \end{tabular}
    \label{tab:stoch_extinct_R6}
\end{table}

\begin{table}[ht]
    \centering
    \caption{Basic scenario. Extinct epidemic rates for severe diseases $R_0 = 12$.}
    \begin{tabular}{l | rlrlrlrl}
        \toprule
        \multirow{2}{*}[-3pt]{Minimisation} & \multicolumn{8}{l}{$\qquad\; V = 0 \qquad\qquad\; V = 7 \qquad\qquad\; V = 15 \qquad\qquad\; V = 30$} \\
          & General & Fatal & General & Fatal & General & Fatal & General & Fatal\\
         \cmidrule{1-9}
        $\pi$ & $0.5\%$ & $1.2\%$ & $0.7\%$ & $0.8\%$ & $0.6\%$ & $0.9\%$ & $6.3\%$ & $6.5\%$ \\
        $A_T$ & $0.5\%$ & $1.2\%$ & $0.7\%$ & $0.8\%$ & $0.8\%$ & $0.9\%$ & $6.3\%$ & $6.5\%$ \\
        \bottomrule
    \end{tabular}
    \label{tab:stoch_extinct_R12}
\end{table}

\newpage
Table~\ref{tab:stoch_extinct_R2} shows that even when we do not vaccinate the population, there is still relatively high chance ($30.2\%$ in general and $26\%$ in fatal case) to surpass the infection. When we vaccinate half of the population, i.e. $V = 30$, in case of fatal epidemic both ``optimality'' options lead to similar extinction probability. However, if the infection is not fatal, the premium minimisation problem \eqref{eq:premium_stoch} leads to considerably lower chances or suppressing the infection, while the minimisation of lost working days still provides high escaping chance. The first option happens because the insurance company tries to maintain the epidemic being active in order to collect premiums from susceptibles by allocating vaccine evenly between centres. In this case, the epidemic has higher chance to outbreak than in case, when one centre is completely vaccinated.

Note that some sharp peaks (e.g. Fig.~\ref{fig:7d}) are rather due to poor number of simulations, rather than specifics of certain vaccine allocation (statement supported by high variance, i.e. big confidence interval).

\begin{figure}[t]
    \centering
    
    \begin{subfigure}[t]{.45\linewidth}
        \includegraphics[width = \linewidth]{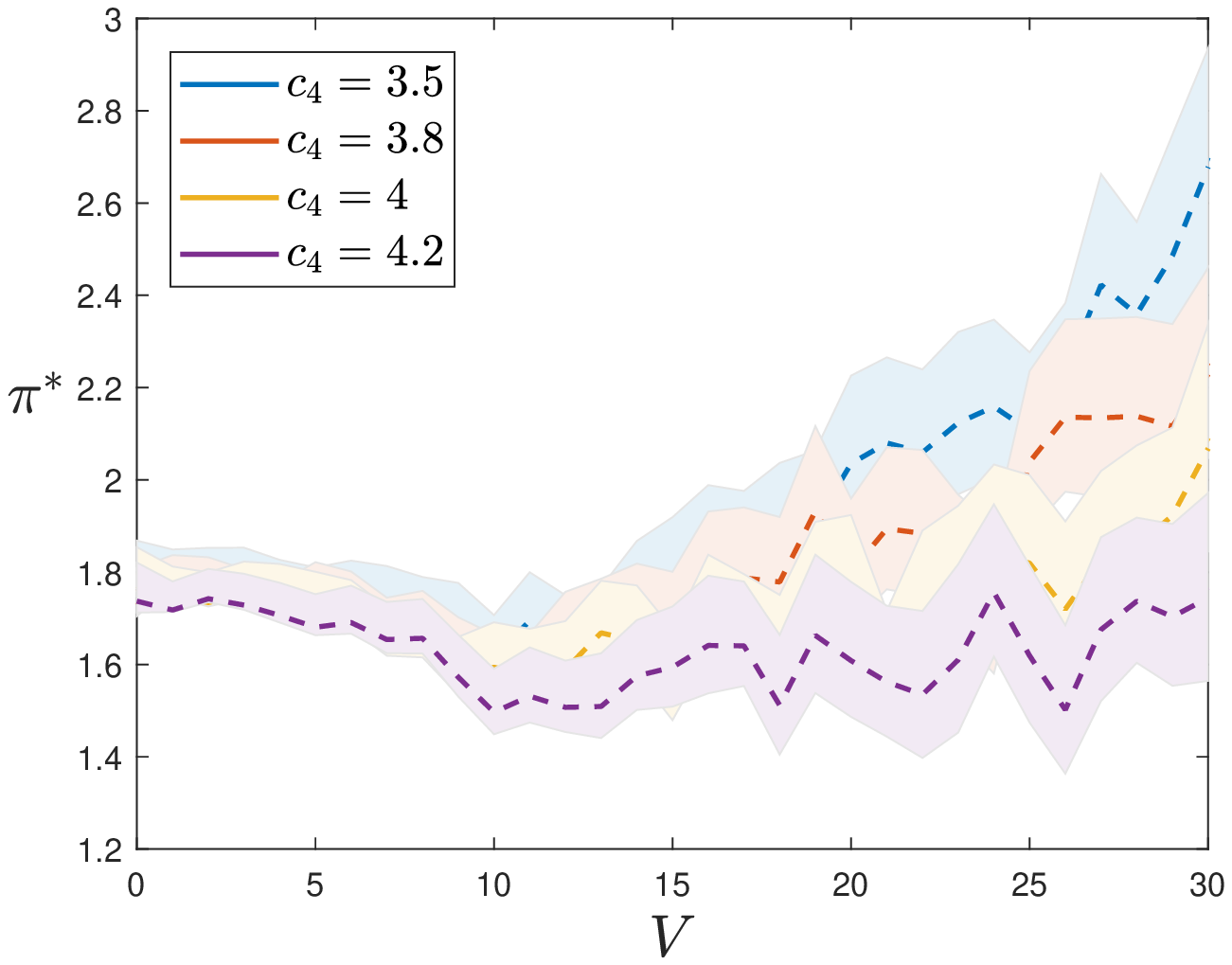}
        \caption{General epidemic, premium minimisation}
        \label{fig:8a}
    \end{subfigure}
    \begin{subfigure}[t]{.45\linewidth}
        \includegraphics[width = \linewidth]{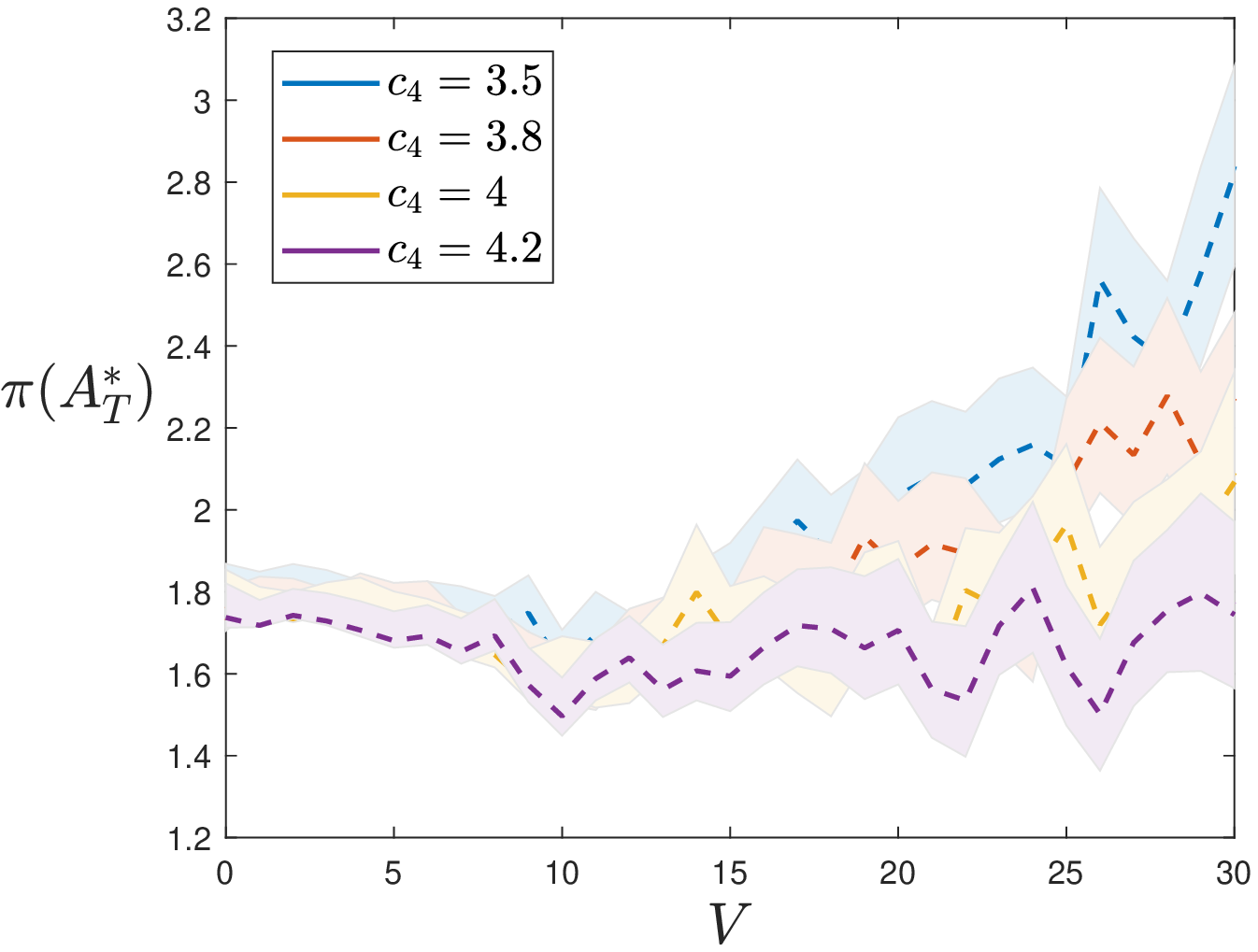}
        \caption{General epidemic, $A_T$ minimisation}
        \label{fig:8b}
    \end{subfigure}
    \begin{subfigure}[t]{.45\linewidth}
        \includegraphics[width = \linewidth]{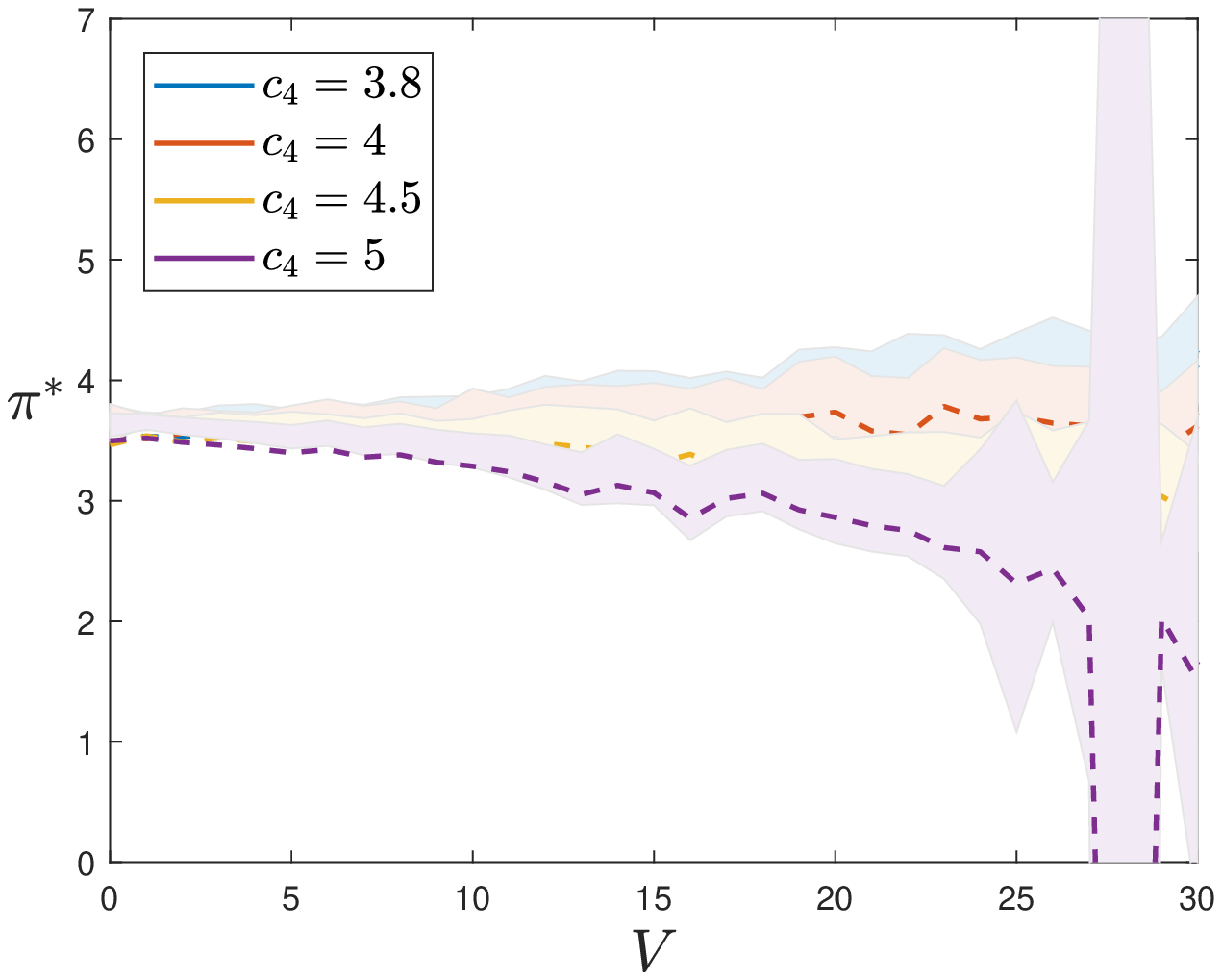}
        \caption{Fatal epidemic, premium minimisation}
        \label{fig:8c}
    \end{subfigure}
    \begin{subfigure}[t]{.45\linewidth}
        \includegraphics[width = \linewidth]{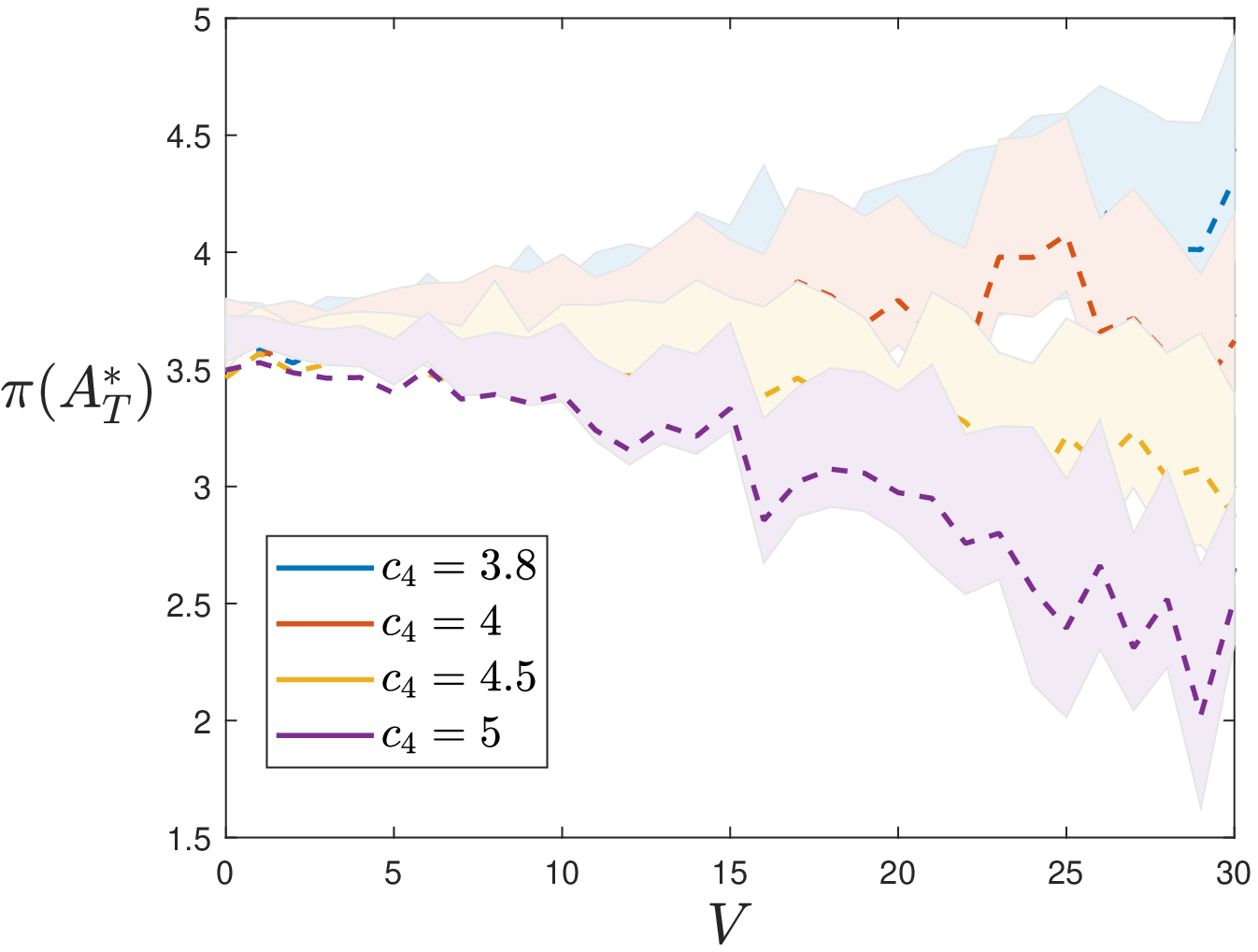}
        \caption{Fatal epidemic, $A_T$ minimisation}
        \label{fig:8d}
    \end{subfigure}
    
    \caption{Premium level for general and fatal epidemics in basic stochastic scenario for diphtheria, mumps, polio, smallpox. Dashed line -- corresponding premium with discounting factor $\delta = \ln 1.01$. Shaded area -- $95\%$ confidence interval of $\pi$. Basic reproduction number $R_0 = 6$.}
    \label{fig:8}
\end{figure}

\newpage
For harmful diseases with reproduction number $R_0 = 6$ in non-fatal case there is characteristic minimum for both premiums at point $V = 10$, after which (in average) the premiums start to increase. In Fig.~\ref{fig:8c} there is another example of poor stochastic simulation results.

\begin{figure}[t]
    \centering
    
    \begin{subfigure}[t]{.45\linewidth}
        \includegraphics[width = \linewidth]{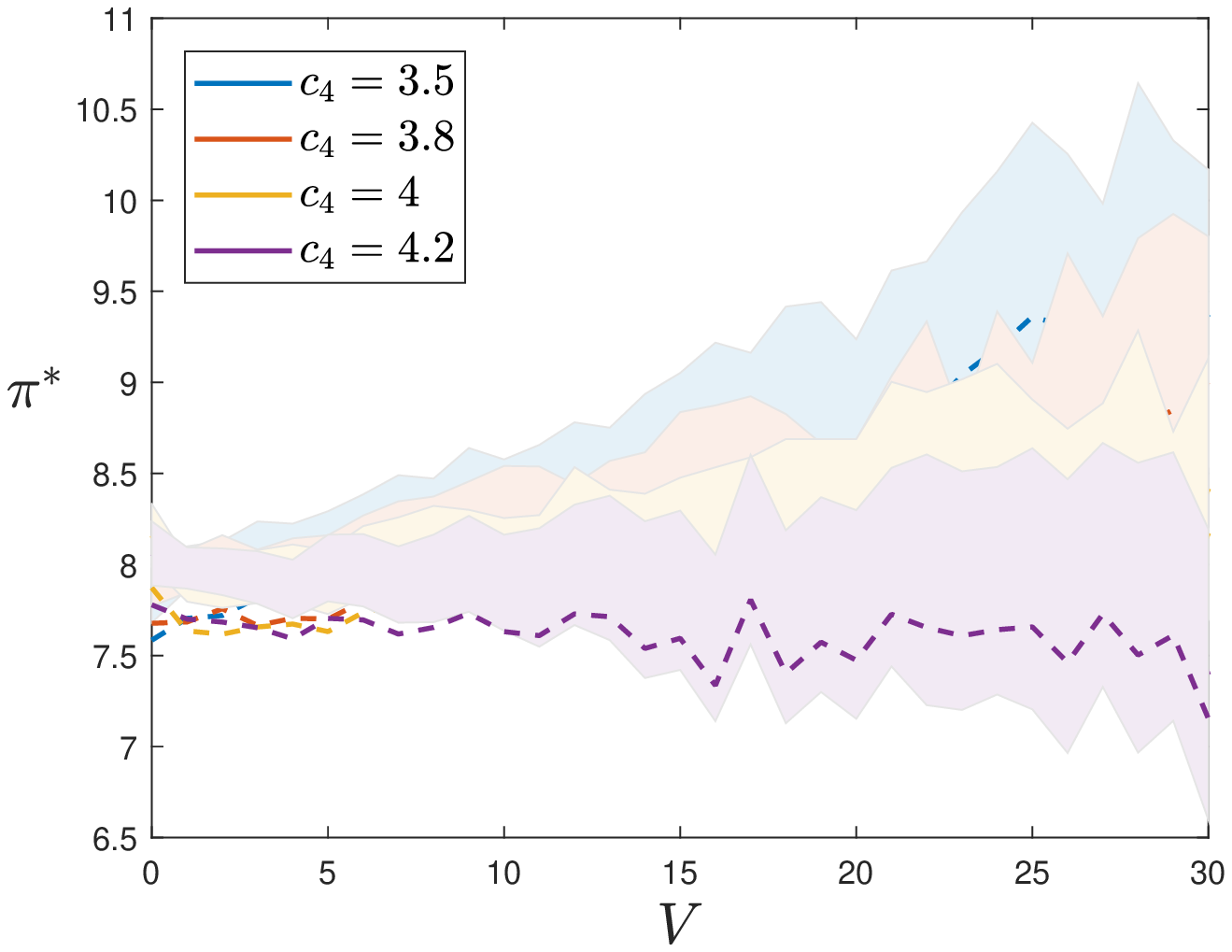}
        \caption{General epidemic, premium minimisation}
        \label{fig:9a}
    \end{subfigure}
    \begin{subfigure}[t]{.45\linewidth}
        \includegraphics[width = \linewidth]{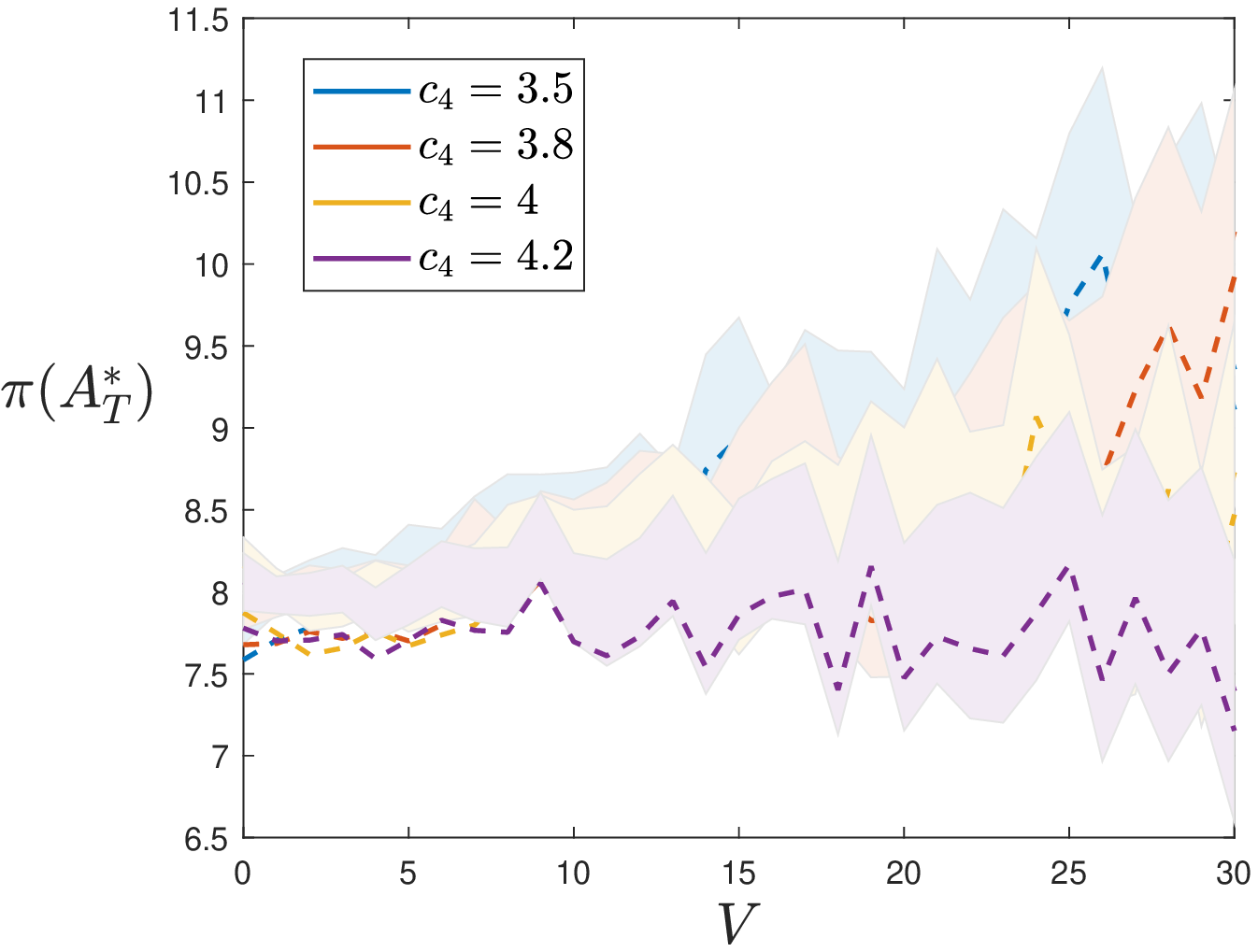}
        \caption{General epidemic, $A_T$ minimisation}
        \label{fig:9b}
    \end{subfigure}
    \begin{subfigure}[t]{.45\linewidth}
        \includegraphics[width = \linewidth]{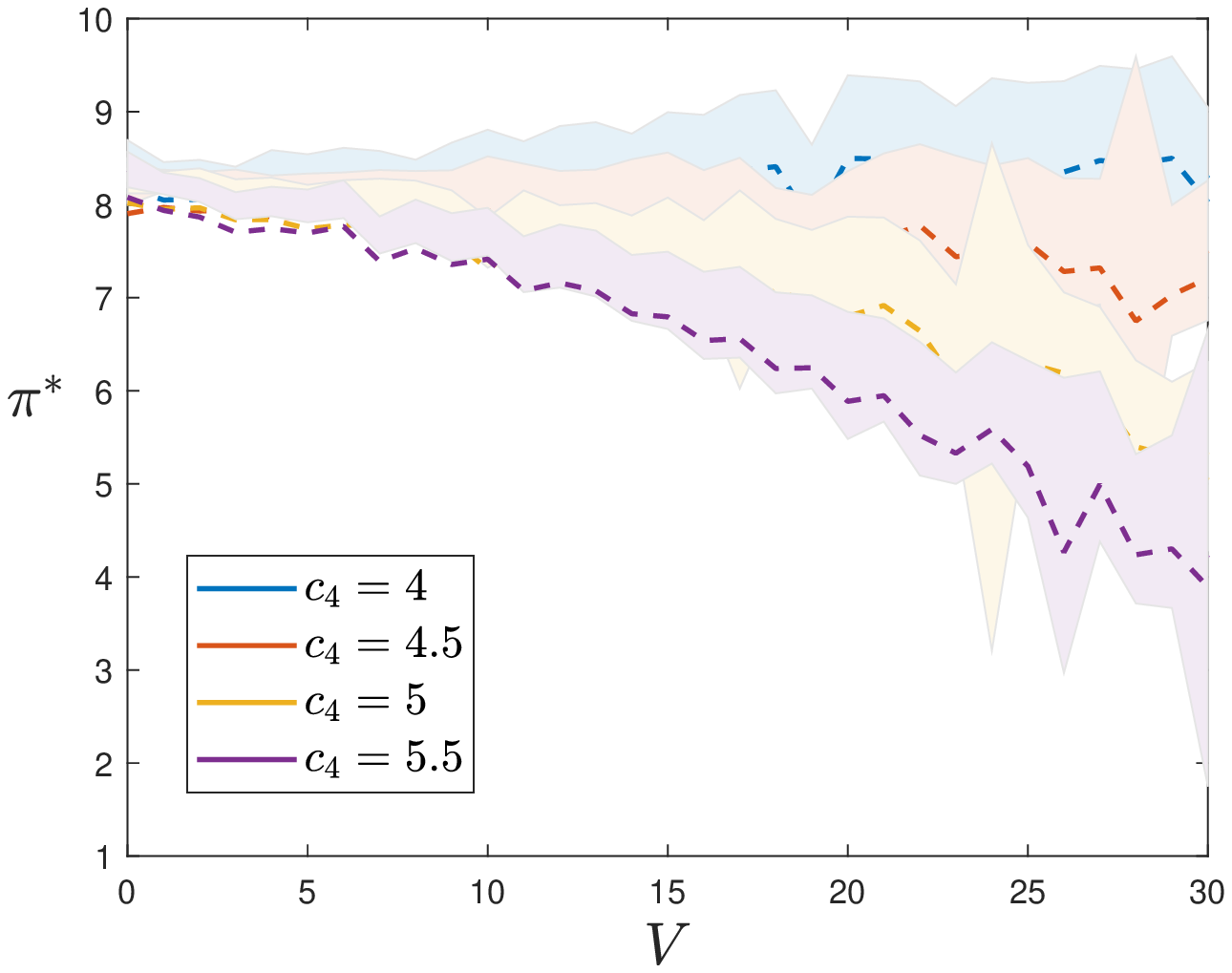}
        \caption{Fatal epidemic, premium minimisation}
        \label{fig:9c}
    \end{subfigure}
    \begin{subfigure}[t]{.45\linewidth}
        \includegraphics[width = \linewidth]{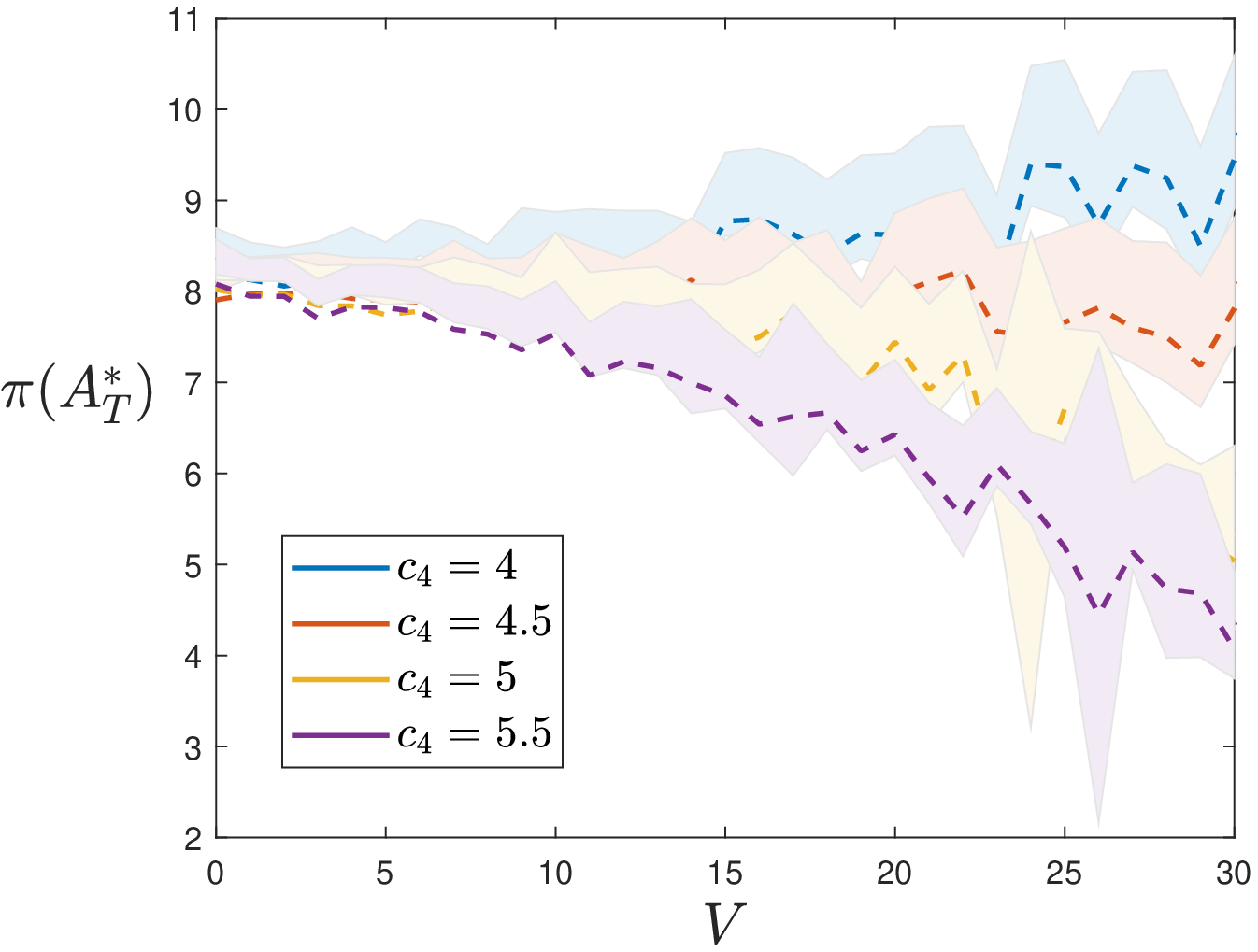}
        \caption{Fatal epidemic, $A_T$ minimisation}
        \label{fig:9d}
    \end{subfigure}
    
    \caption{Premium level for general and fatal epidemics in basic stochastic scenario for malaria and measles. Dashed line -- corresponding premium with discounting factor $\delta = \ln 1.01$. Shaded area -- $95\%$ confidence interval of $\pi$. Basic reproduction number $R_0 = 12$.}
    \label{fig:9}
\end{figure}

Usually, for severe diseases most of the population becomes infected, therefore there is almost no difference between premiums in both ``optimality'' approaches. We can see on Fig.~\ref{fig:9} that the behaviour of premiums is similar to mild disease case (see. Fig~\ref{fig:7}) -- they are increasing for small selling prices $c_4$, and decreasing as selling prices becomes higher than a certain threshold. However, the case of harmful diseases (see Fig.~\ref{fig:8}) has more distinct vaccine amount $V \approx 10$ when the premium has local minimum.

\subsubsection{Big and small centres}
In this scenario we consider two cities, that resemble health-centre scenario, considered in Sect.~\ref{sub:health_centre2}. Here we assume that we have a big centre with treatment facility, and a small centre with poor treatment. All infected persons try to migrate to the big centre, while infected in the first centre generally stay at home. Susceptible group is travelling as usual.

We define initial number of susceptibles as $S_{1, 0} = 50$ and $S_{2, 0} = 10$, and infectives as $I_{1, 0} = 2$, $I_{2, 0} = 1$. Migration fluxes are $k_{12} = 0.1$, $k_{21} = 0.15$, $l_{12} = 0.05$ and $l_{21} = 2.5$. Similarly to Sect.~\ref{sub:health_centre2}, in non-fatal case we let treatment effect in the first centre be $\mu_1 = 2$ ($\mu_1 = -0.9$ in case of fatal case), and mistreatment in the second centre be $\mu_2 = -0.9$ ($\mu_2 = 2$ in fatal case). The natural immunisation is $\mu(R_i) = 1$.

\begin{figure}[ht]
    \centering
    \begin{subfigure}[t]{.45\linewidth}
        \includegraphics[width = \linewidth]{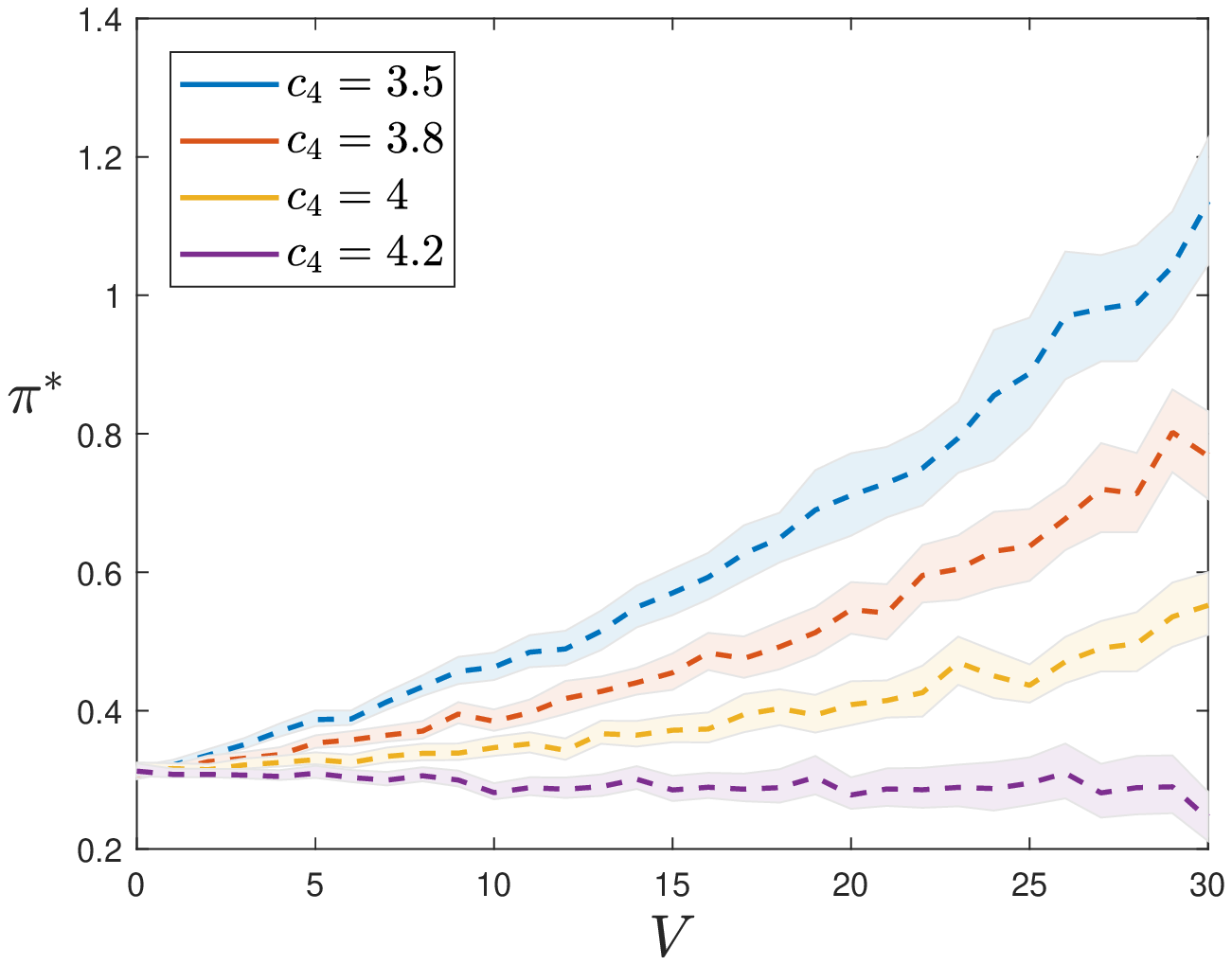}
        \caption{General epidemic, premium minimisation}
        \label{fig:10a}
    \end{subfigure}
    \begin{subfigure}[t]{.45\linewidth}
        \includegraphics[width = \linewidth]{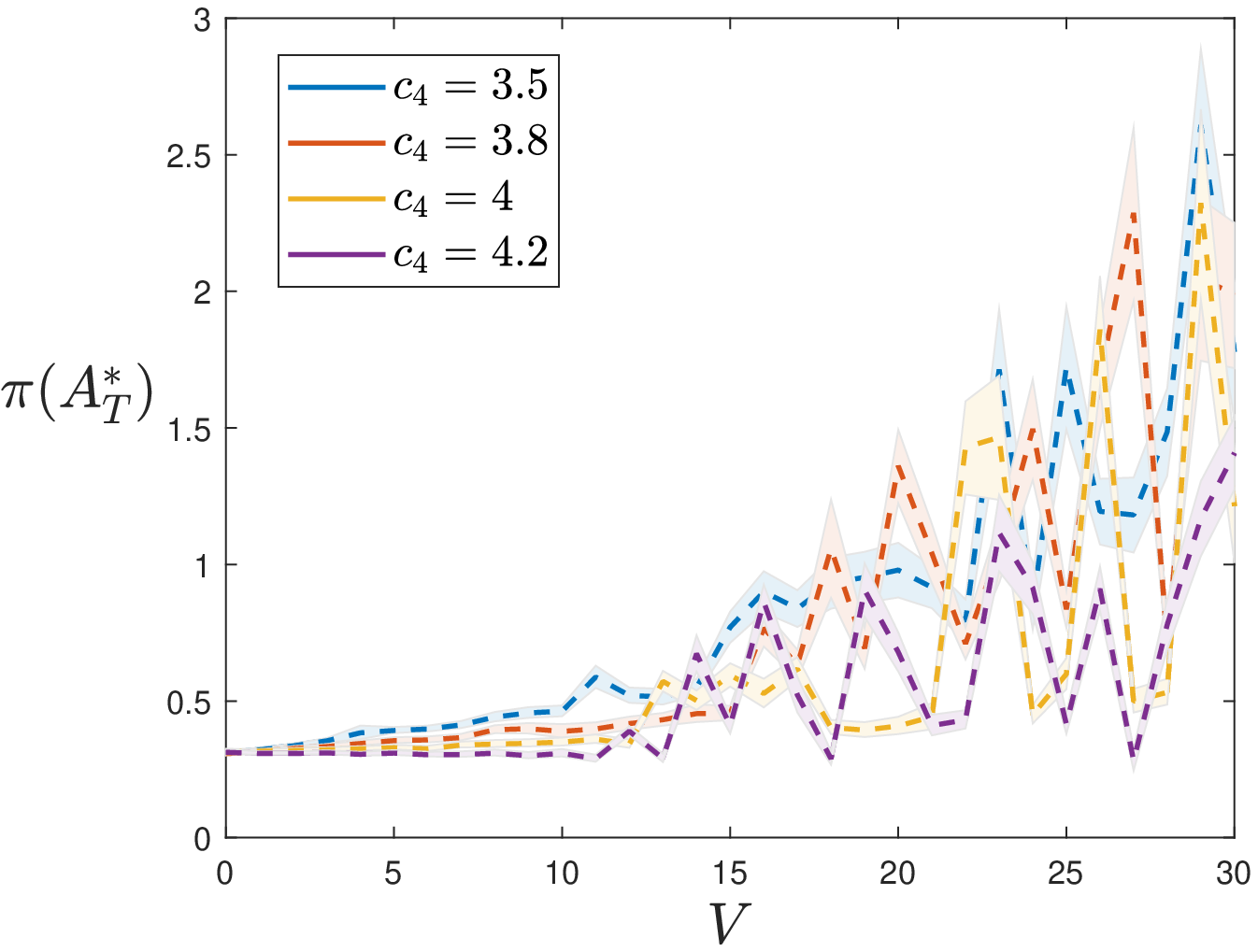}
        \caption{General epidemic, $A_T$ minimisation}
        \label{fig:10b}
    \end{subfigure}
    \begin{subfigure}[t]{.45\linewidth}
        \includegraphics[width = \linewidth]{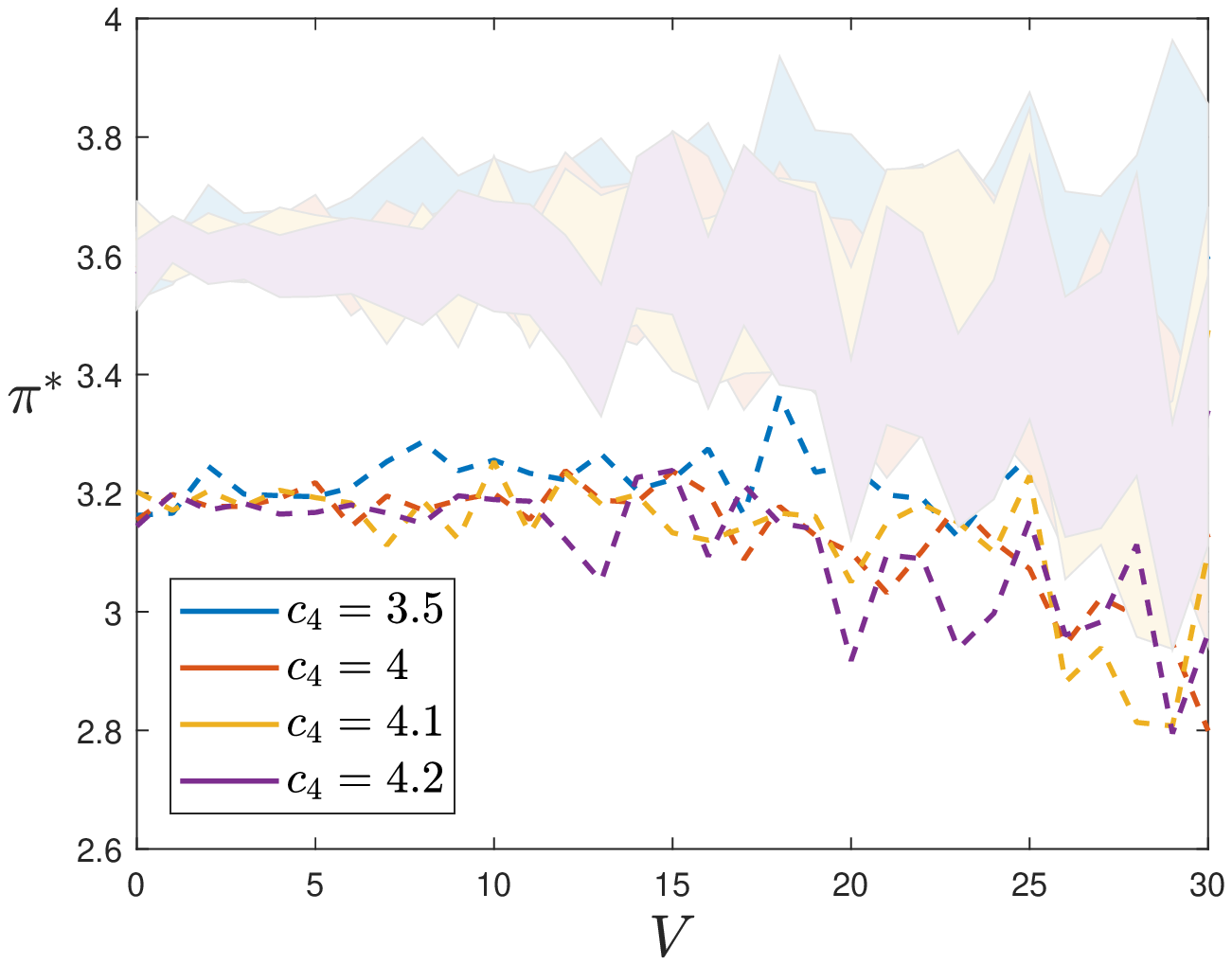}
        \caption{Fatal epidemic, premium minimisation}
        \label{fig:10c}
    \end{subfigure}
    \begin{subfigure}[t]{.45\linewidth}
        \includegraphics[width = \linewidth]{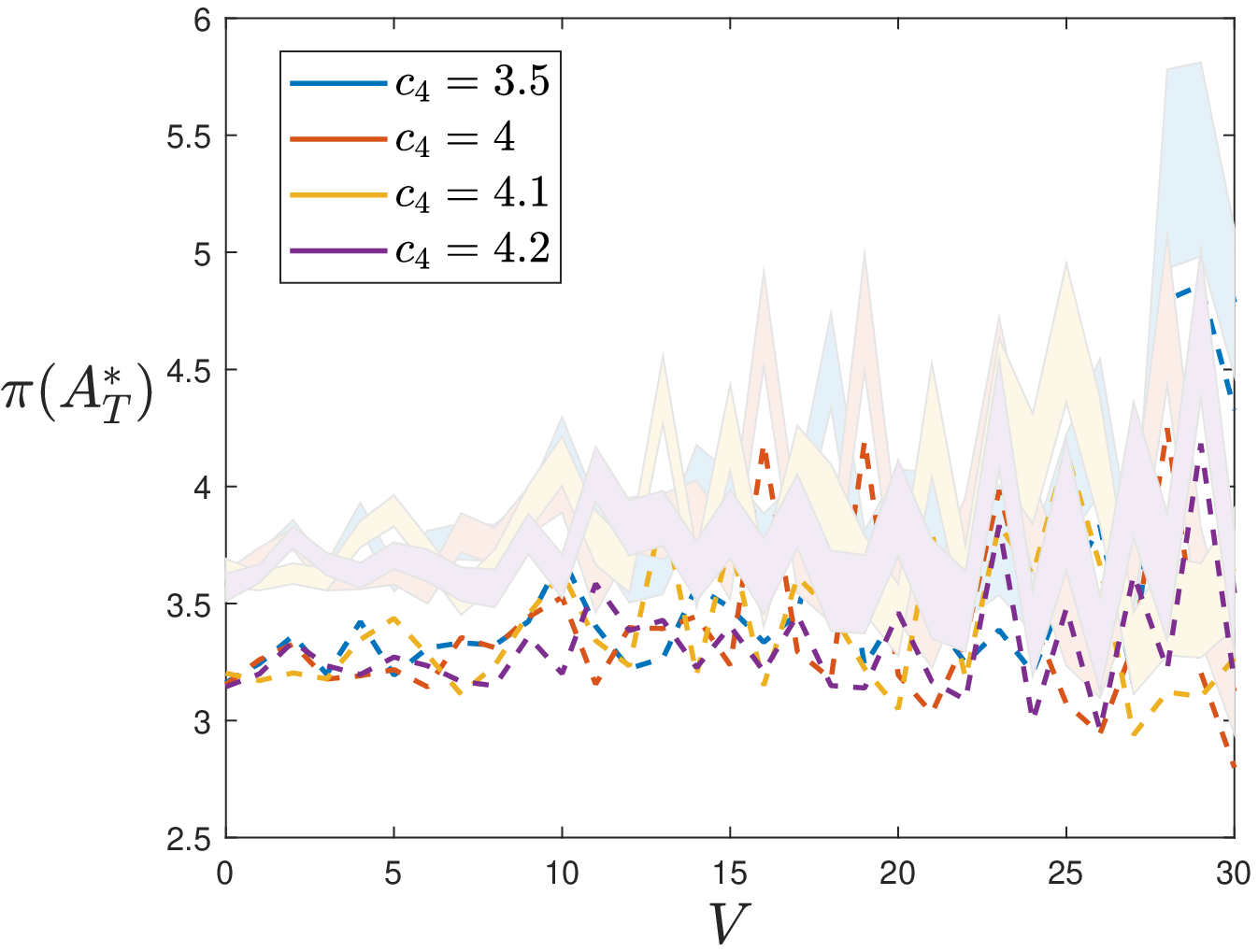}
        \caption{Fatal epidemic, $A_T$ minimisation}
        \label{fig:10d}
    \end{subfigure}
    
    \caption{Premium level for general and fatal epidemics in stochastic health-care scenario for Ebola, influenza. Dashed line -- corresponding premium with discounting factor $\delta = \ln 1.01$. Shaded area -- $95\%$ confidence interval of $\pi$. Basic reproduction number $R_0 = 2$.}
    \label{fig:10}
\end{figure}

Fig.~\ref{fig:10} shows the dependence of premiums from vaccine stock for the health-centre scenario in stochastic model for mild diseases. In non-fatal case (see Fig.~\ref{fig:10a} and \ref{fig:10b}) the number of simulations is sufficient to provide narrow confidence intervals. In Fig.~\ref{fig:10b} after a threshold $V = 10$ there is a switch of strategy from vaccinating the second centre to vaccination of both. Due to randomness of simulations, the minimisation of function $A_T$ becomes less reliable. When vaccine amount is less than $10$, the behaviour is similar to Fig.~\ref{fig:10a}, where we observe precise steady grow (decline) of premium, similar to deterministic results.

\begin{table}[ht]
    \centering
    \caption{Big and small centres. Extinct epidemic rates for mild diseases $R_0 = 2$.}
    \begin{tabular}{l | rlrlrlrl}
        \toprule
        \multirow{2}{*}[-3pt]{Minimisation} & \multicolumn{8}{l}{$\qquad\; V = 0 \qquad\qquad\; V = 7 \qquad\qquad\; V = 15 \qquad\qquad\; V = 30$} \\
          & General & Fatal & General & Fatal & General & Fatal & General & Fatal\\
         \cmidrule{1-9}
        $\pi$ & $65.1\%$ & $0.3\%$ & $70.4\%$ & $0.1\%$ & $79.1\%$ & $0.2\%$ & $80.9\%$ & $0.2\%$ \\
        $A_T$ & $65.1\%$ & $0.3\%$ & $70.4\%$ & $0.3\%$ & $79.3\%$ & $0.3\%$ & $84.2\%$ & $0.2\%$ \\
        \bottomrule
    \end{tabular}
    \label{tab:stoch_extinct_hc_R2}
\end{table}

\begin{table}[ht]
    \centering
    \caption{Big and small centres. Extinct epidemic rates for harmful diseases $R_0 = 6$.}
    \begin{tabular}{l | rlrlrlrl}
        \toprule
        \multirow{2}{*}[-3pt]{Minimisation} & \multicolumn{8}{l}{$\qquad\; V = 0 \qquad\qquad\; V = 7 \qquad\qquad\; V = 15 \qquad\qquad\; V = 30$} \\
          & General & Fatal & General & Fatal & General & Fatal & General & Fatal\\
         \cmidrule{1-9}
        $\pi$ & $6.4\%$ & $0\%$ & $7.0\%$ & $\approx 0\%$ & $10.6\%$ & $\approx 0\%$ & $15.9\%$ & $\approx 0\%$ \\
        $A_T$ & $6.4\%$ & $0\%$ & $7.0\%$ & $\approx 0\%$ & $10.4\%$ & $\approx 0\%$ & $15.9\%$ & $\approx 0\%$ \\
        \bottomrule
    \end{tabular}
    \label{tab:stoch_extinct_hc_R6}
\end{table}

\begin{figure}[t]
    \centering
    
    \begin{subfigure}[t]{.45\linewidth}
        \includegraphics[width = \linewidth]{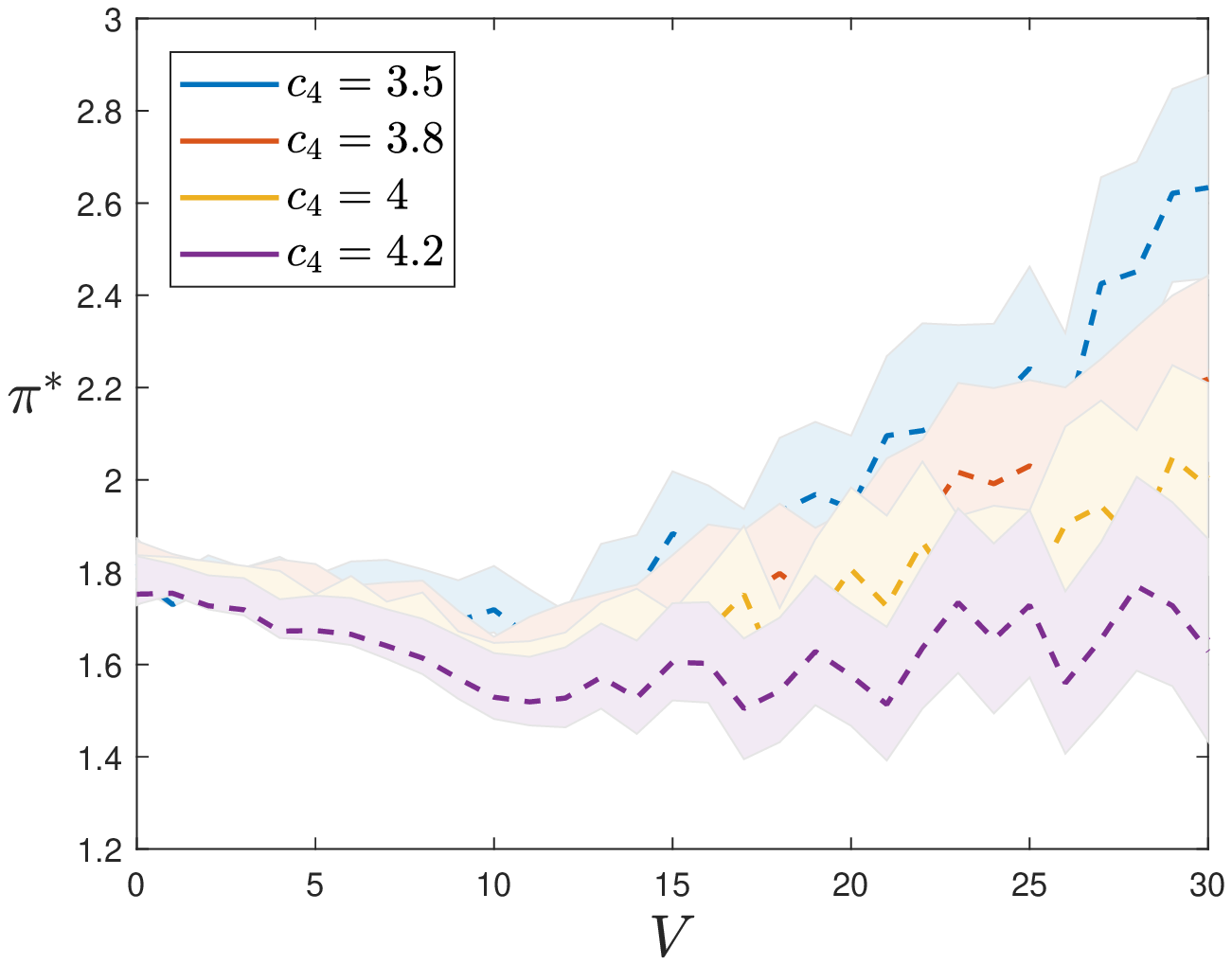}
        \caption{General epidemic, premium minimisation}
        \label{fig:11a}
    \end{subfigure}
    \begin{subfigure}[t]{.45\linewidth}
        \includegraphics[width = \linewidth]{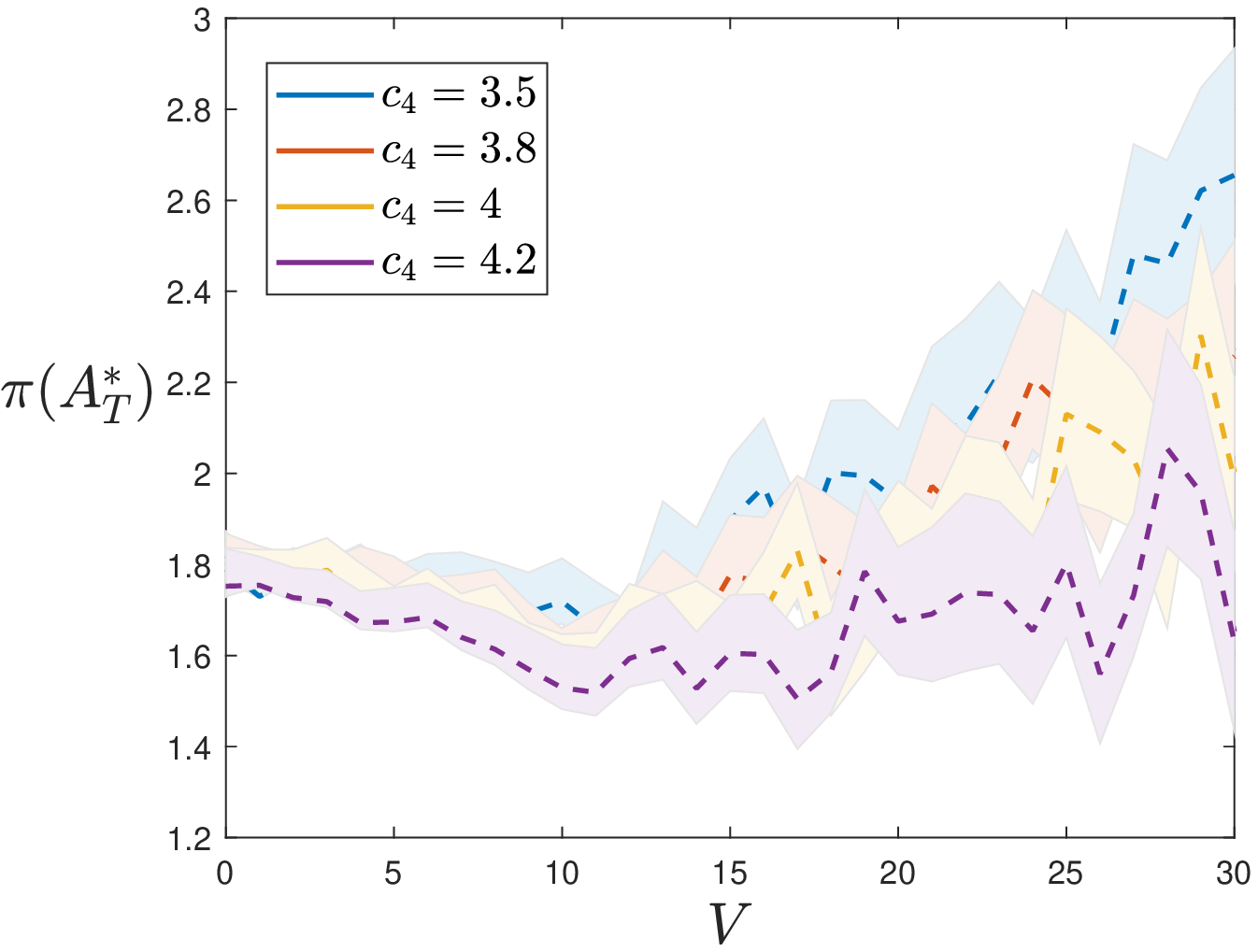}
        \caption{General epidemic, $A_T$ minimisation}
        \label{fig:11b}
    \end{subfigure}
    \begin{subfigure}[t]{.45\linewidth}
        \includegraphics[width = \linewidth]{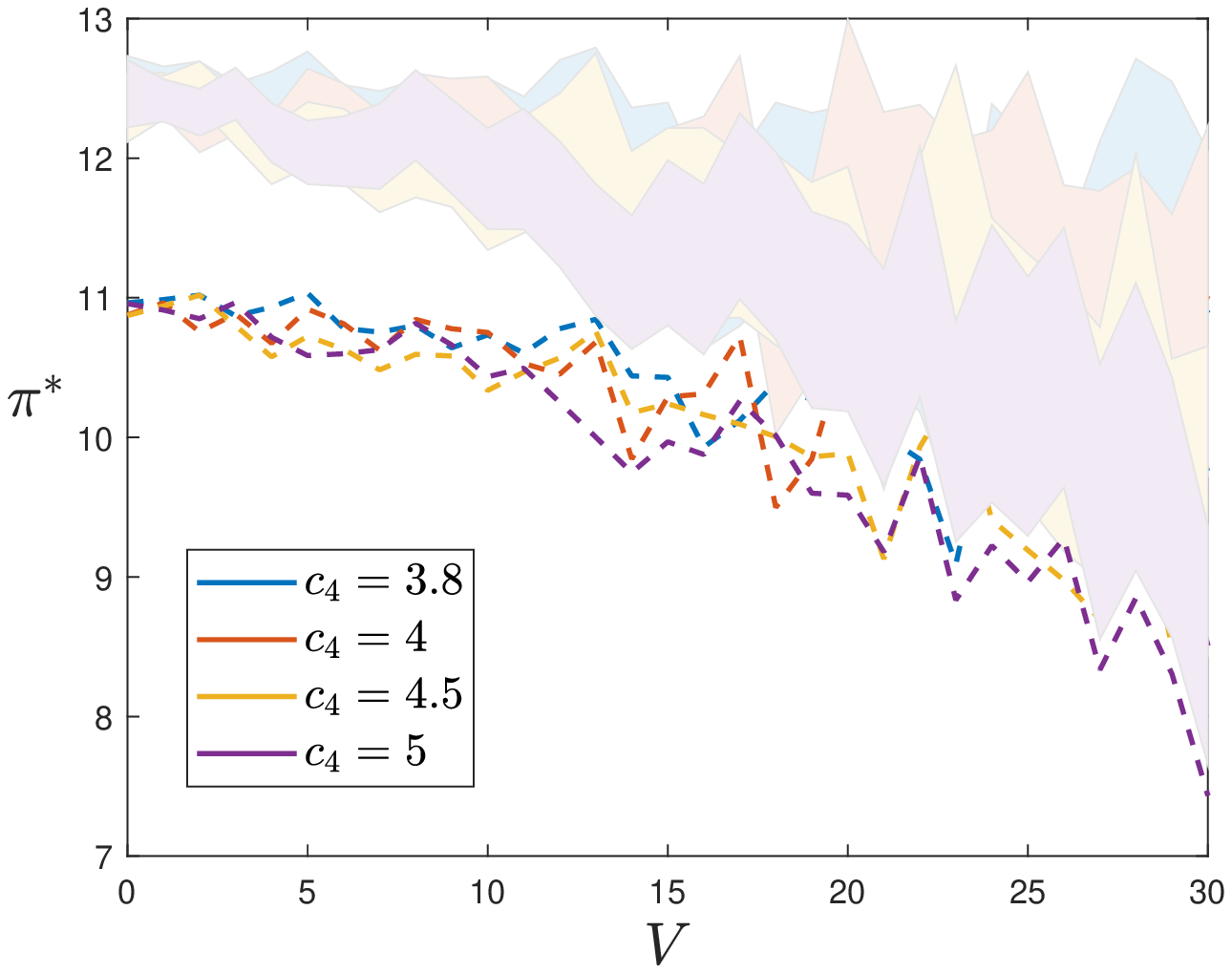}
        \caption{Fatal epidemic, premium minimisation}
        \label{fig:11c}
    \end{subfigure}
    \begin{subfigure}[t]{.45\linewidth}
        \includegraphics[width = \linewidth]{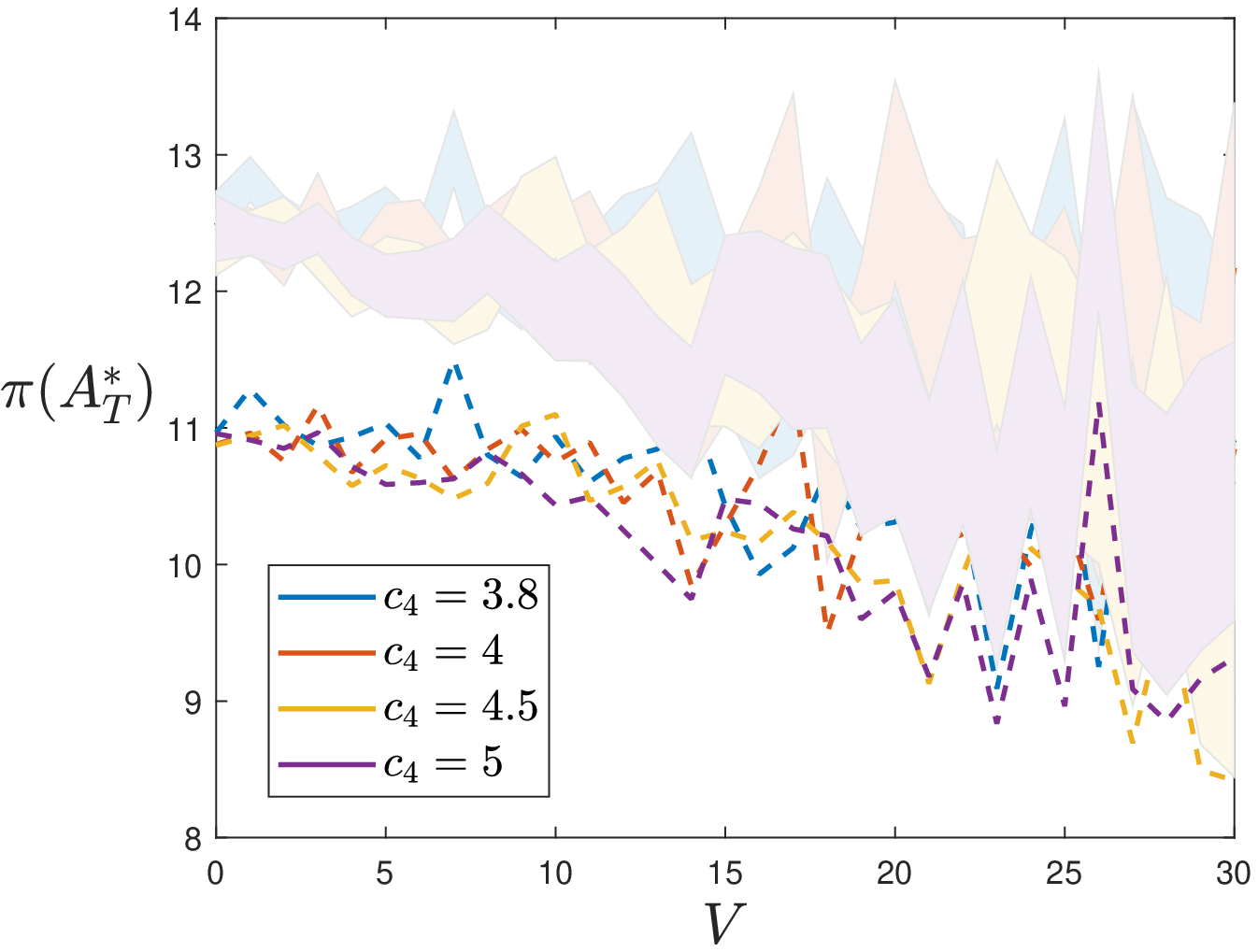}
        \caption{Fatal epidemic, $A_T$ minimisation}
        \label{fig:11d}
    \end{subfigure}
    
    \caption{Premium level for general and fatal epidemics in stochastic health-care scenario for diphtheria, mumps, polio, smallpox. Dashed line -- corresponding premium with discounting factor $\delta = \ln 1.01$. Shaded area -- $95\%$ confidence interval of $\pi$. Basic reproduction number $R_0 = 6$.}
    \label{fig:11}
\end{figure}

\newpage
For harmful diseases (see Fig.~\ref{fig:11}) in non-fatal case we see a more or less distinct area $10 \leq V \leq 12$, where the vaccine is directed to the second centre (vaccinating all susceptibles), and some  persons from the first centre receive vaccine too. Here we still see that both ``optimality'' options provide the same premium levels. From Fig.~\ref{fig:11a} and \ref{fig:11b} we see that it is not important which selling price to choose, the optimal premium will remain almost the same, while we have small number of vaccine. The difference only comes to play when we start to divide vaccine between two centres.

A table of extinct epidemics in stochastic health-centre scenario for severe diseases is omitted due to extremely rare escaping events. Even in non-fatal epidemic case for $V = 30$ the suppression of the infection happens only in $1$ out of $100$ simulations.

\begin{figure}[ht]
    \centering
    
    \begin{subfigure}[t]{.45\linewidth}
        \includegraphics[width = \linewidth]{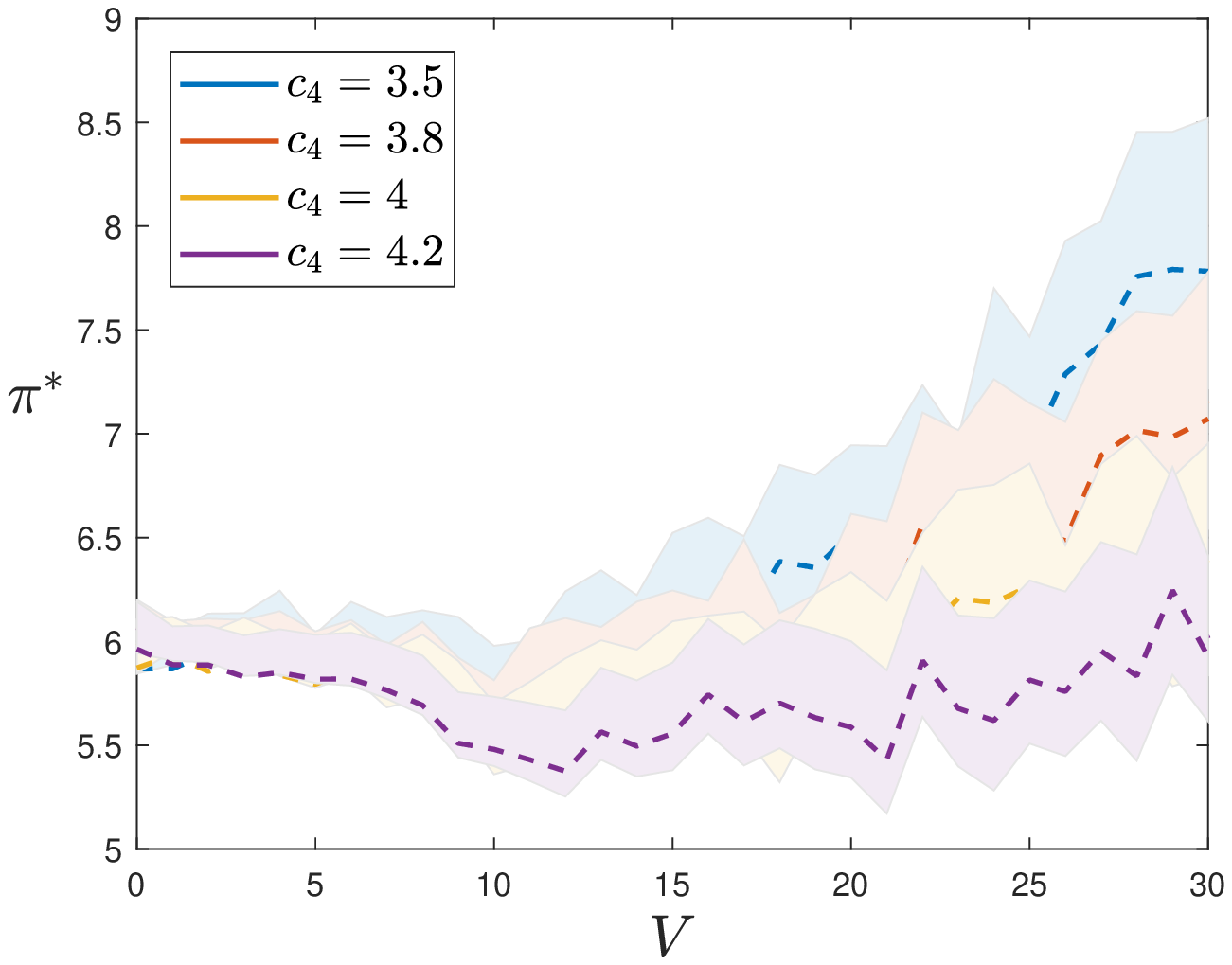}
        \caption{General epidemic, premium minimisation}
        \label{fig:12a}
    \end{subfigure}
    \begin{subfigure}[t]{.45\linewidth}
        \includegraphics[width = \linewidth]{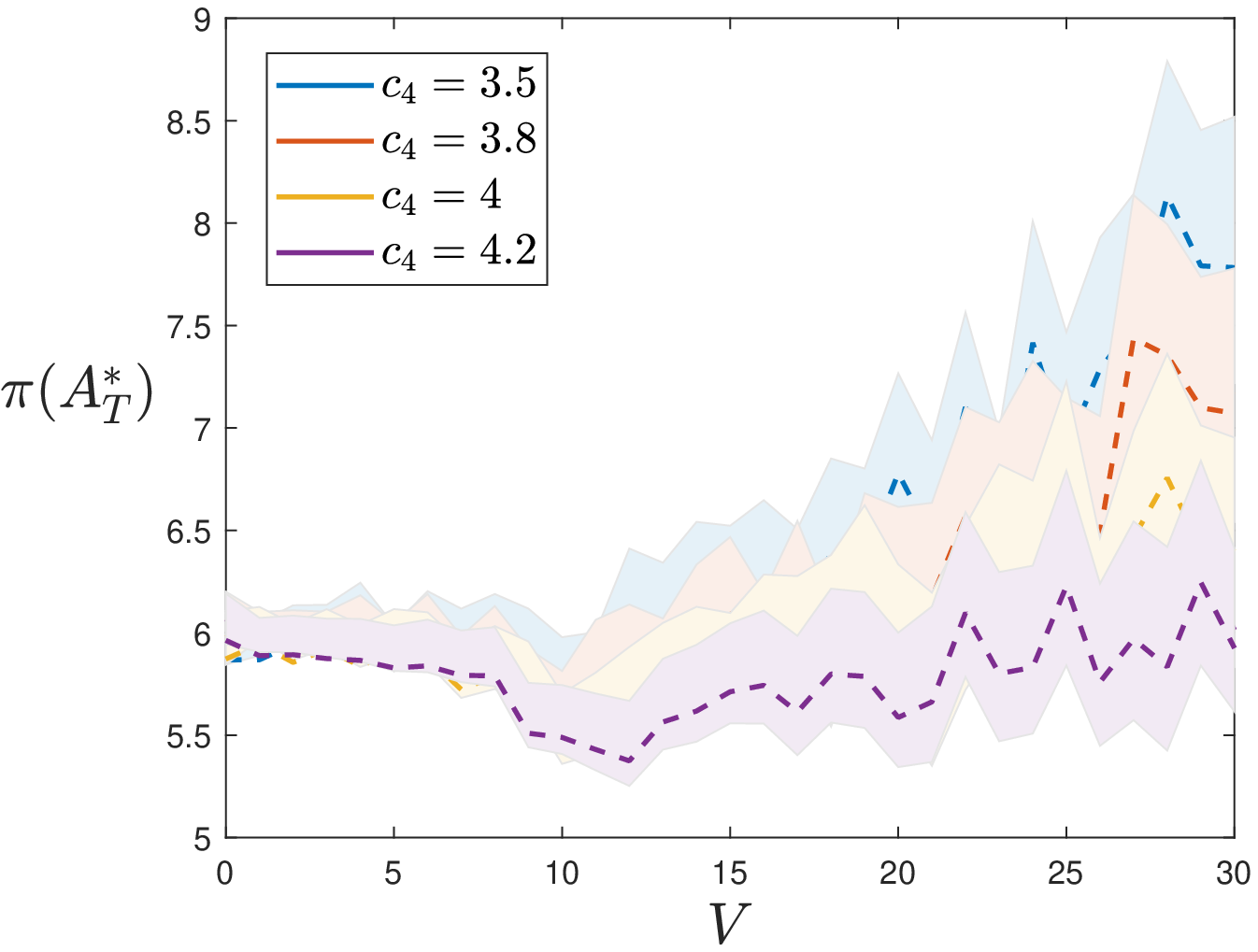}
        \caption{General epidemic, $A_T$ minimisation}
        \label{fig:12b}
    \end{subfigure}
    \begin{subfigure}[t]{.45\linewidth}
        \includegraphics[width = \linewidth]{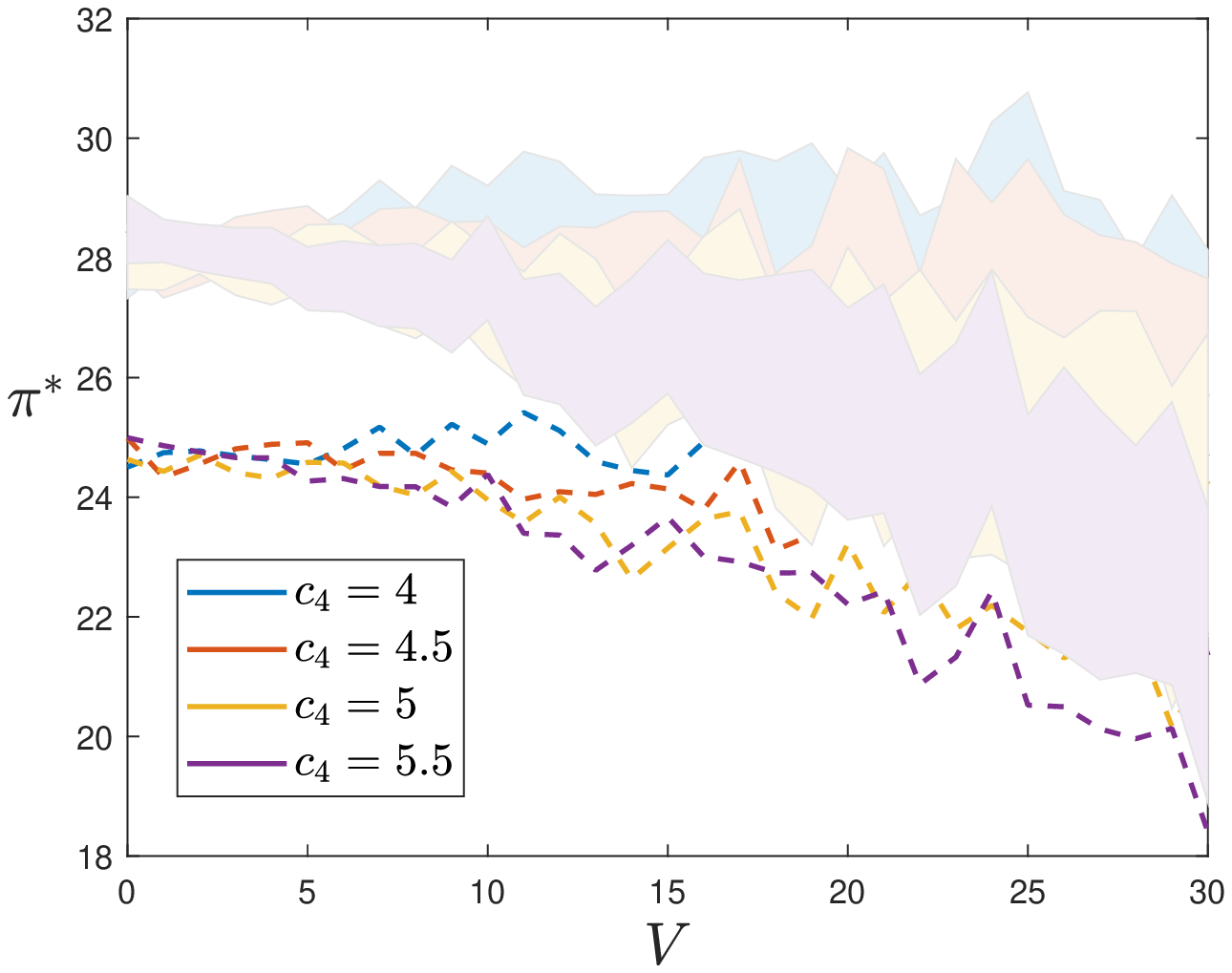}
        \caption{Fatal epidemic, premium minimisation}
        \label{fig:12c}
    \end{subfigure}
    \begin{subfigure}[t]{.45\linewidth}
        \includegraphics[width = \linewidth]{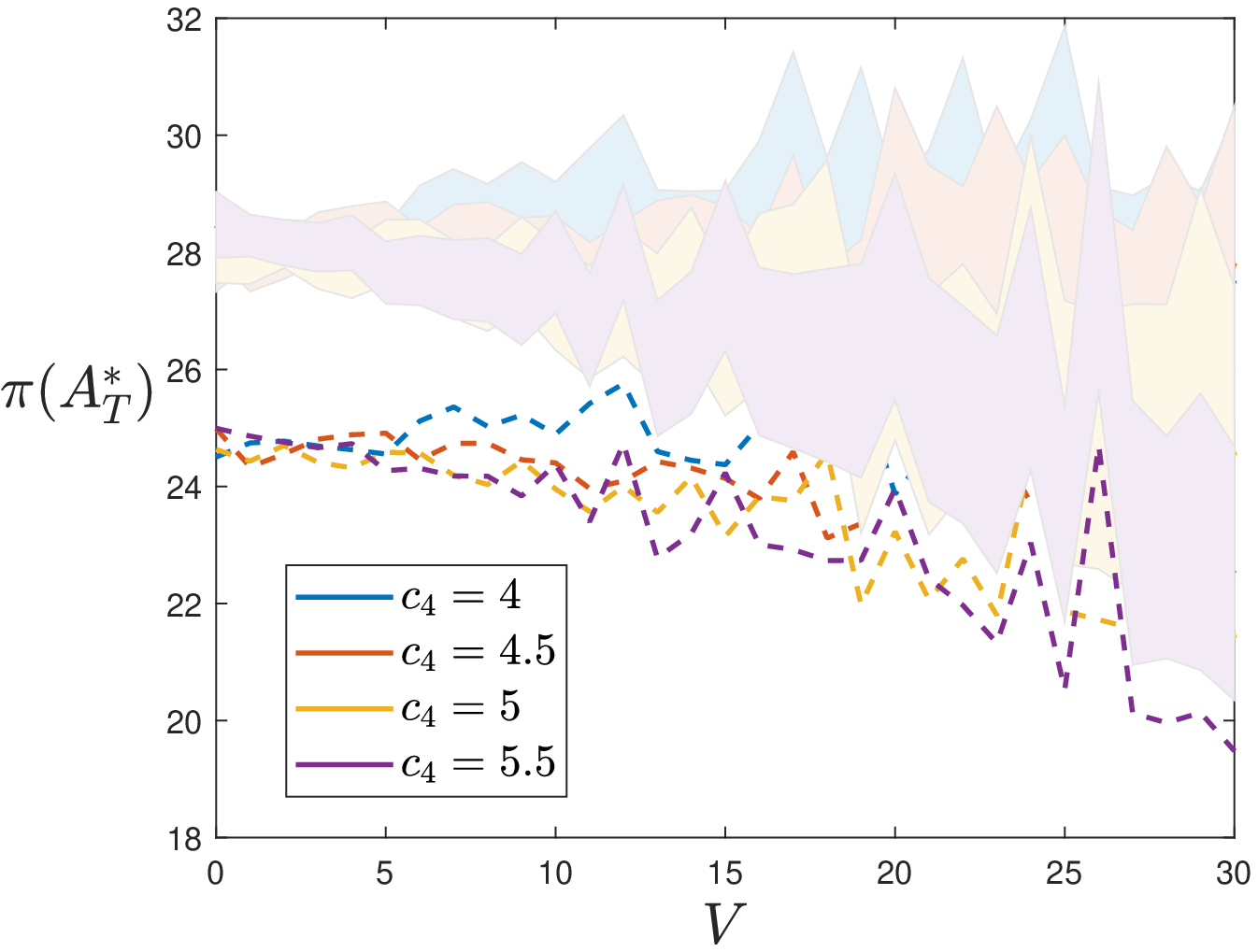}
        \caption{Fatal epidemic, $A_T$ minimisation}
        \label{fig:12d}
    \end{subfigure}
    
    \caption{Premium level for general and fatal epidemics in stochastic health-centre scenario for malaria and measles. Dashed line -- corresponding premium with discounting factor $\delta = \ln 1.01$. Shaded area -- $95\%$ confidence interval of $\pi$. Basic reproduction number $R_0 = 12$.}
    \label{fig:12}
\end{figure}

Fig.~\ref{fig:12} for severe diseases show very similar results (accurate to constant) compared to harmful diseases, see Fig.~\ref{fig:11}. In particular, we have the area $10 \leq V \leq 12$ when both ``optimal'' premiums are minimal, which corresponds to full vaccination of the second centre and (if there is vaccine left) the remaining part is directed to the first centre. For low vaccine amount there is almost no difference between premium levels with respect to selling price $c_4$.

\clearpage
\section{Conclusions}

There is a major theory that supports all models, including analysis of the solutions and dependencies from initial conditions. In general, all models are supposed to study individuals with certain probabilities of passing/receiving the infection, migration, immunisation, etc. However, the theory shows that solution of stochastic model with a lot of time steps is approaching the solution of differential problem with relatively large number of initially infected persons. At the same time, when the number of infected individuals is small, the epidemic has high probability to be naturally suppressed, which is only possible in stochastic model.

In this work we concentrate on the simplest SIR model with two centres and migration fluxes. We study both differential and stochastic models in different "what-if" scenarios. We assume that there is a vaccine that should be optimally distributed among the centres. Individuals who did not get the vaccine must buy a mandatory health-care policy that covers all expenses of treatment in case of getting infected. Taking into account the dynamics of epidemic, we calculate the health-care premium according to equivalence principle.

We propose two measures of ``optimality'' -- one being financially driven, while the second aims to maintain the population healthy. Generally, both approaches result in the same premium level, however, vaccine allocation strategies are different. If both measures  gave different results, it would necessary to consider a linear combination of two with proper coefficients.

In the SIR model we also studied fatal and general (non-fatal) epidemics. In the model after being infected a person is transformed into removed (recovered) group. While it is natural in case of non-fatal epidemic, this assumption drastically restricts the number of infections that could be described by fatal SIR model. Therefore, the results of fatal epidemics are very narrow-focused. Moreover, the proposed minimisation of functional $A_T$, see~\eqref{eq:A_minimisation}, is not the best option for fatal epidemics, because it does not account for number of deaths. Hence, an improvement should be considered, for example, in a form
\begin{equation}
\label{eq:A_T_min_fatal}
    A_T^* = \min_{\substack{(w_1,\ldots,w_n): w_i \geq 0, \\ w_1 + \ldots + w_n = 1}} \sum_{i=1}^n \int_0^T [\alpha_1 I_i(t) + \alpha_2 R_i(t)] \, \mrm{d}t, \quad \alpha_i \in \mbb{R}_+.
\end{equation}
That way, not only we try to reduce the number of lost working days, but also the total exposure to death.

The results of numerical integration~\ref{sect:numerical_deterministic} and stochastic simulations~\ref{sect:numerical_stochastic} show some interesting results of vaccine allocation strategies and resulting premiums. The calculations are made for infections with different hostility levels. The main concern here is to investigate strategies under low vaccine stock.

Future works could consider more detailed cases, like 
\begin{itemize}
    \item non-fatal epidemic, individuals may choose to vaccinate depending on the vaccine selling price $c_4$;
    \item fatal epidemic, SIRS model, minimisation of enhanced functional~\eqref{eq:A_T_min_fatal};
    \item non-instant vaccination (i.e. it take time to vaccinate people, increasing exposure to a disease).
\end{itemize}
Also, most of the models' parameters should be specified by experts, providing more accurate and applicable results.



\bibliographystyle{abbrvnat}


\end{document}